\documentclass[a4paper,fleqn,usenatbib]{mn2e}
\usepackage[T1]{fontenc}
\usepackage{ae,aecompl}
\usepackage{graphicx}
\usepackage{amssymb}
\usepackage{amsmath}
\usepackage{setspace}
\usepackage{hyperref}
\usepackage[utf8]{inputenc}
\usepackage[pass]{geometry}
\usepackage{enumerate}
\usepackage{caption}
\usepackage{csquotes}
\usepackage{color}
\usepackage{placeins}
\usepackage{longtable}
\usepackage{supertabular}
\usepackage{pdflscape}
\usepackage{pifont}
\usepackage{multirow}
\usepackage{booktabs}
\usepackage{units}
\usepackage[subrefformat=parens,labelformat=parens]{subfig}

\newcommand{\cmark}{\ding{51}}
\newcommand{\xmark}{\ding{55}}
\newcommand\blfootnote[1]{
\begingroup
\renewcommand\thefootnote{}\footnote{#1}
\addtocounter{footnote}{-1}
\endgroup
}

\title[Emission-line stars in M31]{Emission-line stars in M31 from the SPLASH and PHAT surveys}

\author[L. J. Prichard et al.] {Laura J. Prichard$^{1\star}$, Puragra Guhathakurta$^{2}$, Katherine M. Hamren$^{2}$, \newauthor Julianne J. Dalcanton$^{3}$, Claire E. Dorman$^{2}$, Anil C. Seth$^{4}$, Benjamin F. Williams$^{3}$,
\newauthor Gabriel A. Damon$^{5}$, Anita Ilango$^{6}$, Megha Ilango$^{6}$\\
$^{1}${\small Sub-department of Astrophysics, Department of Physics, University of Oxford, Denys Wilkinson Building, Keble Road, Oxford OX1 3RH, UK}\\
$^{2}${\small UCO/Lick Observatory, University of California Santa Cruz, 1156 High Street, Santa Cruz, CA 95064, USA}\\
$^{3}${\small Department of Astronomy, Box 351580, University of Washington, Seattle, WA 98195, USA}\\
$^{4}${\small Department of Physics $\&$ Astronomy, University of Utah, Salt Lake City, UT 84112, USA}\\
$^{5}${\small Santa Cruz High School, 415 Walnut Avenue, Santa Cruz, CA 95060, USA}\\
$^{6}${\small Cupertino High School, 10100 Finch Avenue, Cupertino, CA 95014, USA}\\
}

\date{Accepted 2016 November 15. Received 2016 November 11; in original form 2016 September 8}
\pubyear{2016}

\begin{document}
\label{firstpage}
\maketitle

\begin{abstract}
We present a sample of 224 stars that emit H$\alpha$ (H$\alpha$ stars) in the Andromeda galaxy (M31). The stars were selected from $\sim$~5000 spectra, collected as part of the Spectroscopic and Photometric Landscape of Andromeda's Stellar Halo survey using Keck II/DEIMOS. We used six-filter \textit{Hubble Space Telescope} photometry from the Panchromatic Hubble Andromeda Treasury survey to classify and investigate the properties of the H$\alpha$ stars. We identified five distinct categories of H$\alpha$ star: B-type main sequence (MS) stars, `transitioning'-MS (T-MS) stars, red core He burning (RHeB) stars, non-C-rich asymptotic giant branch (AGB) stars, and C-rich AGB stars. We found $\sim$~12 per cent of B-type stars exhibit H$\alpha$ emission (Be stars). The frequency of Be to all B stars is known to vary with the metallicity of their environment. Comparing this proportion of Be stars with other environments around the Local Group, the result could indicate that M31 is more metal rich than the Milky Way. We predict that the 17 T-MS H$\alpha$ stars are Be stars evolving off the MS with fading H$\alpha$ emission. We separated RHeB from AGB H$\alpha$ stars. We conclude that the 61 RHeB and AGB stars are likely to be Long Period Variables. We found that $\sim$~14 per cent of C-rich AGB stars (C stars) emit H$\alpha$, which is an upper limit for the ratio of C-rich Miras to C stars. This catalogue of H$\alpha$ stars will be useful to constrain stellar evolutionary models, calibrate distance indicators for intermediate age populations, and investigate the properties of M31.
\end{abstract}

\section{Introduction}
\label{sec:intro}
\vspace{0 mm}

Studying stellar populations in different environments in the Local Group (LG) is essential for advancing our knowledge of stellar evolution. Understanding the different stages of evolution for massive and intermediate mass stars under a variety of initial and environmental conditions is crucial for constraining stellar evolutionary models. Emission lines arising from different types of stars are powerful diagnostic tools for determining their physical conditions. The H$\alpha$ line, an \blfootnote{\vspace{-4 mm}$^{\star}$Email: \href{mailto:Laura.Prichard@physics.ox.ac.uk}{Laura.Prichard$@$physics.ox.ac.uk}}emission line in the Balmer series of hydrogen at 6563 $\text{\AA}$, is among the most useful due to the abundance of hydrogen in the universe. This emission line can arise from multiple processes and in many types of stars. Conducting a survey of stars that exhibit H$\alpha$ emission (H$\alpha$ stars) can give us insights in to some of these lesser-known stages of stellar evolution. The analysis of H$\alpha$ stars can also tell us about the properties of their environment, galaxy evolution, and even be used as distance indicators.

H$\alpha$ emission is typically associated with the brightest stars such as B-type main sequence (MS) stars. B stars have the maximum rotational velocity compared to other stellar classes, with typical values of $\sim$~200~km~s$^{-1}$ \citep{McNally1965}. They have temperatures of 10,000--30,000 K and a mass range of 3 M$_{\odot} <$ M $<$ 20 M$_{\odot}$. They are short lived and because of this, they are often associated with a parent cloud or nebulous regions of the interstellar medium (ISM), so-called HII regions. Beyond temperatures of around 10,000 K, hydrogen is ionized so H$\alpha$ absorption and emission should not be seen. However, studies show that for rapidly rotating massive stars, H$\alpha$ emission arises from material in a rotationally flattened envelope that creates a circumstellar disc \citep[e.g.][]{Struve1931, Dougherty1992} or accreted material that forms a disc. B-type stars that have exhibited H$\alpha$ emission at least once are classified as Be stars \citep{Jaschek1980}. As discussed in the review by \cite{Porter2003}, whether the Be phase is an evolutionary stage of all B-type stars is still contested. 

Surveys of Be stars have shown that their environment can affect their frequency; lower metallicity environments have a higher proportion of Be stars \citep[e.g.][]{Maeder1999, Martayan2007}. It is thought that stars rotate faster in lower metallicity environments, perhaps due to weaker magnetic coupling of the star with its surroundings from formation, causing more of them to exhibit emission lines originating from their rotationally flattened envelope \citep{Maeder1999, Wisniewski2006, Martayan2007}. As well as the metallicity dependence on the proportion of Be stars, it was also suggested that there could be some evolutionary dependence on their frequency \citep[e.g.][]{Schild1976, Tarasov2012}. The age of star clusters was found to affect the frequency of Be stars, and it was concluded that the Be phase was an evolutionary stage of all B stars that occurs at $\sim$~10 Myr \citep{Fabregat2000}. This discovery could mean that there is an age-metallicity degeneracy for the environments of Be stars. More studies of the properties of Be stars and their environments are needed to better understand the evolution of massive MS stars and put constraints on the processes that occur within them.

The evolutionary path a star takes off the MS is dependent on its mass \citep[e.g.][and references therein]{Hayashi1962}. For intermediate- to high-mass stars ($\sim$~3.5--15 M$_{\odot}$), He burning (HeB) begins in a non-degenerate core \citep[e.g.][and references therein]{Gallart2005}. The star becomes more luminous and red, and populates the red core HeB (RHeB) branch in colour-magnitude space \citep[e.g.][]{Dohm-Palmer2002, McQuinn2011}. Stars traverse the colour-magnitude diagram (CMD) blue-ward during their HeB lifetime, completing a `blue loop' before moving red-ward again towards the RHeB sequence \citep[e.g.][]{Stothers1968}. The bluest point in a star's `blue loop' phase defines the blue core HeB (BHeB) branch on a CMD \citep[e.g.][]{Bertelli1994}. Low- to intermediate-mass stars ($\sim$ 0.8 to $\lesssim$ 3.5 M$_{\odot}$) at the end of their red giant branch (RGB) phase, ignite He in their degenerate core in a He flash, moving on to the horizontal branch on a CMD. Once core He is depleted, both low- and high-mass stars (up to $\sim$ 8 M$_{\odot}$), move up their Hayashi line with increasingly convective envelopes \citep{Hayashi1961}, and a star is then in the asymptotic giant branch (AGB) phase of evolution.

Some AGB stars can exhibit H$\alpha$ emission. AGB stars have masses 0.8 M$_{\odot} <$ M $<$ 8 M$_{\odot}$, they have C-O cores and are burning H/He in their shells. Despite the different evolutionary tracks taken by stars of different masses, both low- to high-mass MS stars evolve on to the AGB. It is during the AGB phase that a star loses most of its mass, transforming in to a planetary nebula \citep[e.g.][]{Iben1974,Iben1983}. Towards the end of the AGB phase, a star will experience a series of thermal pulses (TP-AGB phase) that arise from instabilities in the H/He burning shells \citep{Schwarzschild1965}. These stars have convective envelopes and atmospheres of atomic and molecular gas, and dust \citep[e.g.][]{Schwarzschild1967, Iben1974}. Studying the mass loss in these stars helps us to understand the metal enrichment of the ISM \citep[e.g.][]{Gehrz1989, Blum2006}.

In general, AGB stars are categorised as being oxygen rich (C/O~$<$~1), M-type stars (M stars), or carbon rich (C/O~$>$~1), C-type stars (C~stars). This carbon excess is caused by the `third dredge-up' that occurs during the TP-AGB phase, bringing carbon to the surface \citep[see][and references therein]{Iben1983}. The transition from oxygen to carbon rich is known to depend on the metallicity of the environment and stellar mass \citep[e.g.][]{Iben1974, Karakas2002, Battinelli2005}. C~stars have increased opacity due to the presence of CN in their atmospheres, this leads to a substantial increase in their mass loss \citep[e.g.][]{Marigo2002, Mattsson2010, Sandin2010}. C-rich stars also tend to be more luminous at redder wavelengths than their non-C-rich counterparts \citep[e.g.][]{Frogel1980,Frogel1990}. Studying populations of C~stars is useful to constrain models of AGB stars \citep{Karakas2014}, understand their role in the enrichment of the ISM \citep[e.g.][]{Boyer2011}, and study the metallicity gradients of their environments \citep[e.g.][]{Cioni2008, Feast2010}.

Long Period Variables (LPVs) are AGB stars that undergo long-period pulsations and exhibit H$\alpha$ emission \citep[e.g.][]{Merrill1921,Merrill1952}. LPVs typically exhibit strong Balmer emission for around 90 per cent of their cycle, however these line strengths can vary between cycles \citep[e.g.][]{Joy1954}. The pulsations in LPVs arise from shock waves in the lower photosphere of the star and result in shock-excited H$\alpha$ emission \citep[e.g.][]{Bowen1988}. LPVs can be split in to two main categories, semiregular variables (SRs) and Mira variables (Miras)\footnote{We use the term Mira variables to mean AGB stars with long-period, large-amplitude pulsations (see \citealt{Whitelock1991}).}. These stars have pulsation periods of $\sim$~100--1000s of days, with the average period of Miras being longer than SRs by a factor of two, although there is significant overlap \citep[e.g.][]{Wood1977, Hughes1990}. 

SRs and Miras share many common characteristics and it is now accepted that SRs are the progenitors of Miras \citep[e.g.][]{LloydEvans1973, Whitelock1986}. In many cases the SR and Mira classifications overlap and they are often misidentified due to their restrictive classical definition \citep[e.g.][]{Kerschbaum1992}. Studies of Miras and SRs usually involve multi-epoch observations of their photometric properties to distinguish between the two types. Miras tend to have larger amplitude magnitude fluctuations ($\Delta$I~$\sim$~2.5) compared to SRs \citep[$\Delta$I~$\sim$~0.5; e.g.][]{Hughes1989,Wood1996}. Miras typically have a regular pulsation period, while SRs have intermittent regularity as their name implies \citep[e.g.][]{Feast1996, Wood2000}. The two types of LPV also occupy distinct regions of \textit{K}-band and period (logP) space \citep[e.g.][]{Feast1989, Hughes1990}. To correctly classify Miras from other LPVs, \textit{K} magnitude light curves are essential \citep[e.g.][]{Battinelli2014}. 

Miras have a period-luminosity relation that makes them ideal distance indicators \citep[e.g.][]{Glass1981, Glass1982}, particularly when observed in the infrared \citep{Whitelock2013}. These older stars are in the haloes of galaxies and can be more easily resolved than those in the discs of spirals. Observing them in the infrared avoids contamination from ISM extinction. Miras are the ideal candidates for distance measurements, but so far, large spectroscopically confirmed samples of Mira variables have been difficult to obtain in external galaxies due to instrument limitations. Identifying large samples of Miras spectroscopically will help us to calibrate them for use as standard candles of intermediate-age stars. As Miras are AGB stars, they too can be defined as either C or O rich. Models of C-rich Mira variables (C-Miras) are well constrained \citep{Arndt1997, Wachter2002, Sandin2010}. In addition to having well developed models, this population of stars possesses both the environmental dependence of C~stars \cite[e.g.][]{Karakas2002} with the standard candle benefits of Mira variables \citep{Glass1981}, making them valuable probes of galactic environments.

Investigating populations of C-Miras in different environments has been the focus of a recent study \citep[see][and references therein]{Battinelli2014}. C-Miras presented in the study were identified by photometrically monitoring variations in spectroscopically confirmed C~stars. To identify C-Miras in previous studies, C star catalogues were cross-matched with variable star surveys. Compiling data from the literature, it was shown that the proportion of C-Miras to all C~stars (C-Mira/C~star) was around 12 per cent in many extra-galactic systems (NGC 6822, NGC 147, NGC 185, Fornax dwarf galaxy, and Leo 1). Looking at data for the Large Magellanic Cloud \citep[LMC;][]{Cioni2001, Cioni2003}, a C-Mira/C~star value of $<$ 22 per cent was found \citep{Battinelli2014}. This upper limit for the LMC arose due to limitations of the sample and the value could be down to 12 per cent as seen in other systems. However, the Milky Way (MW) halo has a significantly larger C-Mira/C~star fraction of 40 per cent \citep{Mauron2014}, and this value was recalibrated and expected to be up to 50 per cent \citep{Battinelli2014}. When comparing this C-Mira/C~star proportion with the Sagittarius dwarf spheroidal galaxy (Sgr dSph), a tidally disrupted galaxy in the MW halo \citep{Ibata1994}, a similar proportion of 48 per cent was found \citep{Battinelli2013}. Why the C-Mira/C~star proportion in the halo of our galaxy and Sgr dSph differs appreciably from the 12 per cent found in other galaxies is not clear. Currently there is no comparative C-Mira/C~star value for the Andromeda galaxy (M31), which is similar to the MW in morphology and mass, to investigate this difference further.

Given its relative proximity, around 785 $\pm$ 25 kpc \citep{McConnachie2005}, M31 provides an ideal test bed for investigating the properties of stars with H$\alpha$ emission. With the advantage of an external view of a spiral galaxy with a vast range of environments and stellar types to survey, a range of astrophysical research is possible. Many large surveys of H$\alpha$ stars have been performed throughout the LG. These surveys included regions of the Small Magellanic Cloud (SMC) and LMC \citep[e.g.][]{Henize1956, Bohannan1974, Azzopardi1986, Pereira2001, Martayan2011, Reid2012}, Galactic environments \citep[e.g.][]{Downes1988, Kun1992, Miszalski2013} and some larger extragalactic systems such as M33 \citep[e.g.][]{Sholukhova1999}. Extensive narrowband photometric coverage of M31 and M33 detected only the most massive H$\alpha$ stars \citep{Massey2006}. Thus, there remains a distinct lack of data and comparative statistics of the full H$\alpha$ star population of M31.

With spectra from the Keck II telescope collected as part of the Spectroscopic and Photometric Landscape of Andromeda's Stellar Halo (SPLASH) project \citep{Dorman2012}, we have been able to spectroscopically identify individual H$\alpha$ stars in M31. Using supporting six-filter \textit{Hubble Space Telescope} (HST) photometry from the Panchromatic Hubble Andromeda Treasury (PHAT) survey \citep{Dalcanton2012}, we have analysed and classified H$\alpha$ stars in M31. Details of the two data sets used in this paper are given in \S \ref{sec:data}. The algorithm and selection conditions to obtain our sample of H$\alpha$ stars is outlined in \S \ref{sec:samp}. The classification and analysis of the H$\alpha$ star sample using photometry, emission-line properties, and distribution is described in \S \ref{sec:analysis}. A discussion of the scientific results and their significance is covered in \S \ref{sec:disc}. We present a summary of our findings in \S \ref{sec:conc}.

\section{Data}
\label{sec:data}

To investigate emission-line stars in M31, we used two complementary data sets. To classify and analyse our selected H$\alpha$ stars we used photometric data from the PHAT survey, detailed in \S \ref{sec:photdata}. To select the emission-line stars and determine their properties, we analysed a spectroscopic sample from the SPLASH survey, described in \S \ref{sec:specdata}.

\subsection{Photometric data}
\label{sec:photdata}

The PHAT survey is a multi-cycle HST program. The survey covers around a third of M31's disc in six filters that span from ultraviolet (UV) to near-infrared (NIR). With HST's observing capabilities, the disc of M31, out to 20 kpc, is resolved in to more than 100 million stars. \cite{Dalcanton2012} chose to observe the northeast quadrant of M31 as it has the least obscured active star forming regions and least extinction. The spatial coverage selected by the PHAT team also maximised the range of environments and thus stellar evolutionary stages observed. They used two UV filters (F275W and F336W), one optical filter (F475W), one far-red filter (F814W), and two NIR filters (F110W and F160W). The filters were chosen to constrain extinction and effective temperature of a range of stellar populations. Different age stellar populations were identified by comparison with artificial CMDs generated using the six filters. Details of the observations, data reduction, and production of six-filter photometric catalogues of stars with accurate astrometry are given in \cite{Dalcanton2012}, and \cite{Williams2014}. The PHAT survey provides a large and multidimensional catalogue of data that enables the classification of objects based on colour. 

\subsection{Spectroscopic data}
\label{sec:specdata}

To select H$\alpha$ stars, we used a spectroscopic sample that was collected as part of the SPLASH survey. Although the initial aim of the SPLASH project was to characterise the halo of M31, in recent years it has focussed on the disc and inner regions. In this latest phase of the survey, the crowded disc-dominated region of M31 observed by the SPLASH team was chosen to overlap with the PHAT survey coverage. The data analysed in this paper comes from this crowded, northeast quadrant of M31 where the two surveys overlap. The spectroscopic targets were selected from the PHAT catalogue to span a range of stellar ages including MS, AGB, and RGB stars \citep{Dorman2015}. All stars had apparent magnitudes greater than F814W = 22 or F475W = 24 to ensure good quality spectra could be obtained of the fainter targets. The isolation of each star was taken into account in order to reduce possible contamination from neighbouring stars \citep[see][]{Dorman2012}. 

The spectra were reduced using the \textsc{spec2d} and \textsc{spec1d} software \citep{Cooper2012, Newman2013} that were modified for the SPLASH survey. \textsc{spec2d} extracted a one-dimensional spectrum from the two-dimensional data, \textsc{spec1d} then calculated redshifts for each spectrum by comparing it to spectral templates. \cite{Dorman2012} describe the full data reduction of the spectroscopic sample.

\cite{Hamren2015} used this spectroscopic sample to identify C~stars in order to investigate the C-star to M-star ratio in M31. To identify AGB C~stars, they used a $\chi^{2}$ comparison, position in CN~$-$~TiO, and F475W~$-$~F814W colour-colour space. Within the 5295, 600~line~mm$^{-1}$ SPLASH spectra, they identified 83 C~stars. We later use this sample to identify any C~stars among our sample of H$\alpha$ stars and for comparison to other studies in the discussion of AGB H$\alpha$ stars in \S \ref{sec:AGB}.

\section{Sample selection}
\label{sec:samp}

To select a sample of H$\alpha$ stars, we used the 1D SPLASH spectra described in \S \ref{sec:specdata}. The 2D spectra from the SPLASH survey were used in this study only to verify the result of our detection algorithm performed on the 1D spectra. The list of emission-line sources that we used to check the efficiency of the algorithm used to find H$\alpha$ stars, was determined from visual inspection of $\sim \nicefrac{1}{3}$ of the total 5295, 600~line~mm$^{-1}$ 2D spectra. The 2D spectra can be useful for determining the nature of the emission, i.e. if from the star or a nebulous region of ionised gas. An automated detection algorithm for the 2D spectra will therefore be the focus of future work using the SPLASH sample but beyond the scope of this paper. For an example of 2D SPLASH spectra of four stars, two classified as H$\alpha$ stars and two with nebulous H$\alpha$ emission, see Appendix \ref{sec:appB}.

\subsection{Selecting the data subset}
\label{sec:projdata}

From the sample of 5295 SPLASH spectra, we wanted to ensure that each H$\alpha$ star was selected in the same way in order to produce a uniform sample. To do this we used a subset of data from which to select our sample. We ensured each spectrum in the subset was a stellar object, based on a previous classification, and had data in all of the sections that the selection algorithm used to identify H$\alpha$ stars. 

Some of the spectra in the sample were previously identified as X-ray sources. Some LPVs can be `symbiotic stars', evolved giants that are accreting on to a compact companion, typically a white dwarf \citep[e.g.][]{Warner1972, Allen1984}. Symbiotic stars cannot be compared easily alongside isolated LPVs due to their interaction, as emission may arise from both the LPV and compact companion. As white dwarfs are UV bright and often emit in the X-ray, we decided that where possible we would try to avoid symbiotic stars by not including these X-ray sources. Finally, some objects were previously identified by PHAT as star clusters, and we were only interested in individual stars. For consistency in the data subset, we removed the previously identified X-ray sources and PHAT star clusters, leaving 5121 stellar spectra.

As well as the H$\alpha$ line, some of the spectra also exhibited other emission lines including some forbidden lines of nitrogen and sulphur: [NII]a,b $\lambda$6548,6583 and [SII]a,b $\lambda$6716,6731. These lines are typically associated with diffuse HII regions that are markers for massive star formation. They are also found in planetary nebulae (PNe), and much less frequently in some hot MS stars that experience strong stellar winds. B-type stars with forbidden emission lines are called B[e] stars. Although B[e] stars can exhibit these lines, they are hard to distinguish from hot stars surrounded by ionised ISM. To minimise the chance of selecting a star in an HII region, rather than an H$\alpha$ star, or an object that emits [NII] and [SII] lines such as a PNe, these lines were tested for. To calculate root-mean-squared (RMS) noise values of the spectra, we used a region of continuum red-ward of the H$\alpha$ line between 6600 and 6640 $\text{\AA}$. Fitting a straight line and taking the standard deviation (STDEV) of the data to the line gave the RMS value, and thus a measurement of the noise for each spectrum. To ensure that every spectrum in our sample had undergone a similar selection process, we removed spectra that had data missing in any of the key areas we analysed. If any spectrum did not have data in the H$\alpha$/[NII] region, the [SII] region, or the region to measure continuum, these spectra were omitted. This left a subset of 4686 spectra from which we selected our H$\alpha$ star sample.

\subsection{Selection algorithm}
\label{sec:sampHa}

We developed a selection algorithm for finding emission-line stars from the 1D spectroscopic SPLASH data. The spectra had been shifted to the rest frame and their respective velocities were given by \textsc{spec1d}. To begin our analysis of each spectrum, we normalised each of the 4686 subset of spectra by dividing by the continuum level local to the H$\alpha$ line. To guide the selection process and to verify the efficiency of the selection algorithm, we used the list of emission-line sources visually identified from a subsample of the 5295 2D SPLASH spectra. 

To measure the properties of spectral features, we fitted Gaussian curves to each 1D spectrum of the 4686 spectra in the data subset. A double Gaussian curve was used to fit the H$\alpha$ $\lambda$6563 and [NII]b $\lambda$6583 lines. In the narrow region of the spectrum between $\sim$~6540--6590 $\text{\AA}$, there are three emission lines, H$\alpha$ and the two [NII] lines. We adopted the fixed ratio of [NII]a,b lines as 1:3 \citep{Osterbrock1989}. As the [NII]b line is stronger, we opted to fit only that and the H$\alpha$ using a double Gaussian to reduce the number of model parameters the program was trying to optimise. We fitted for both the H$\alpha$ and [NII]b lines, and the slope of the local continuum simultaneously using Eq. \ref{eq:dg}. We also used this method for fitting the [SII]a,b lines.
\begin{gather} \label{eq:dg}
f_{dG}(x)=\frac{A_{1}}{\sigma_{1}\sqrt{2\pi}}\exp{\left(-\frac{\left(x-\lambda_{1}\right)^2}{2{\sigma_{1}}^2}\right)}\\
\hspace{13 mm}+\frac{A_{2}}{\sigma_{2}\sqrt{2\pi}}\exp{\left(-\frac{\left(x-\left(\lambda_{1}+\delta\lambda\right)\right)^2}{2{\sigma_{2}}^2}\right)}\nonumber\\
\hspace{13 mm}+\left(mx+c\right).\nonumber
\end{gather}
Here $A_{1}$ and $A_{2}$ are the peak amplitudes, and $\sigma_{1}$ and $\sigma_{2}$ are the line widths for the two peaks, either H$\alpha$ and [NII]b or [SII]a,b. $\lambda_{1}$ is the central wavelength of the first line which was not fixed, but we used a constant wavelength separation ($\delta\lambda$) between the two lines being measured, this allowed for small shifts in velocity. The slope of the continuum was given by the gradient ($m$), and y-intercept ($c$). We found a double Gaussian was more reliable at fitting the data as it had less independent parameters than fitting individual Gaussians to each emission line. To account for lines of different widths, the ranges of the fits were chosen to cover wavelengths of 6553~-~6593~$\text{\AA}$ for the H$\alpha$ and [NII]b lines, and 6706~-~6741~$\text{\AA}$ for the [SII]a,b lines. 

To find the best-fitting curve, the \textsc{mpfitfun} routine\footnote{Developed by C. Markwardt for Interactive Data Language (IDL).}, which utilises the Levenberg--Marquardt method was used. By visually inspecting the fits of many spectra, a set of initial parameters was given to \textsc{mpfitfun} to ensure the best-fitting curve of this region. We did not fix any of the input values in Eq. \ref{eq:dg} but we did find that the program was sensitive to initial values for each of the seven input model parameters. Specifically, we set a positive peak height and line width to encourage the program to find a line, we found the program could more easily recover emission this way if it was present.

To calculate emission-line strengths, we used equivalent widths (EWs); the strength of the line relative to the continuum. We summed over each of the emission lines separately. Using Eq. \ref{eq:ewcal}, an emission line has a negative EW value:
\begin{equation} \label{eq:ewcal}
EW \approx \sum\limits_{i=1}^N\left({ 1 - \frac{f_{dG}(x_{i})}{F_0}}\right)\cdot\Delta \lambda.
\end{equation}
Here, N is the number of wavelength intervals, $f_{dG}(x_{i})$ is the value of the double Gaussian fit for each wavelength channel, and $\Delta \lambda$ is the resolution of each wavelength channel in Angstroms. The constant continuum level $F_0$ was obtained from the average continuum level over the wavelength range of the fit. To quantify the goodness of the fit for each channel, the reduced $\chi^{2}$ was calculated as follows:
\begin{equation}
\frac{\chi^{2}}{N} = \frac{1}{N}\sum\limits_{i=1}^N\frac{(F_{i} - f_{dG}(x_{i}))^{2}}{\sigma^{2}}.
\label{eq:ewerr}
\end{equation}
Here, $F_{i}$ is the flux in each wavelength channel and $\sigma^{2}$ is the variance of the data. We found this a useful test of how well the model fit the data and thus the reliability of the measured parameters.

\subsection{Selection criteria}
\label{sec:selcrit}

Using the set of parameters determined in \S \ref{sec:sampHa}, we wanted to identify stars with significant H$\alpha$ detection. To ensure we selected a clean H$\alpha$ star sample, we also wanted to omit objects with significant [NII]b or [SII]a,b emission. To identify H$\alpha$ stars reliably and automatically from the data subset, we chose selection criteria. To find the optimum set of criteria, the objects selected by the algorithm were compared to the visually selected subsample of H$\alpha$ stars, as determined from the 2D spectra, to determine its ability to find H$\alpha$ stars from the 1D spectra. We chose the following set of criteria (numbers 1--5) to be lenient (around 2.5$\sigma$ accuracy). This combination of constraints produced a list of H$\alpha$ stars with $\sim$~5 per cent contamination. The final criterion was a visual inspection, if the algorithm had falsely identified H$\alpha$ or had failed to detect [NII] or [SII] emission, we removed the star from our final sample. The criteria used, their order, and their efficiency of removing non-H$\alpha$ stars is detailed below.
\vspace{4mm}
\begin{enumerate}[1. ]
\setlength\itemsep{4mm}
\item H$\alpha$ EW value $<$ $-2.5$ $\times$ $\Delta$EW $\longrightarrow$ removed 2816 stars
\vspace{-3mm}
\begin{figure*}
\centering
\subfloat[][]{
	\includegraphics[width=0.49\textwidth, trim=170 745 430 185,clip]{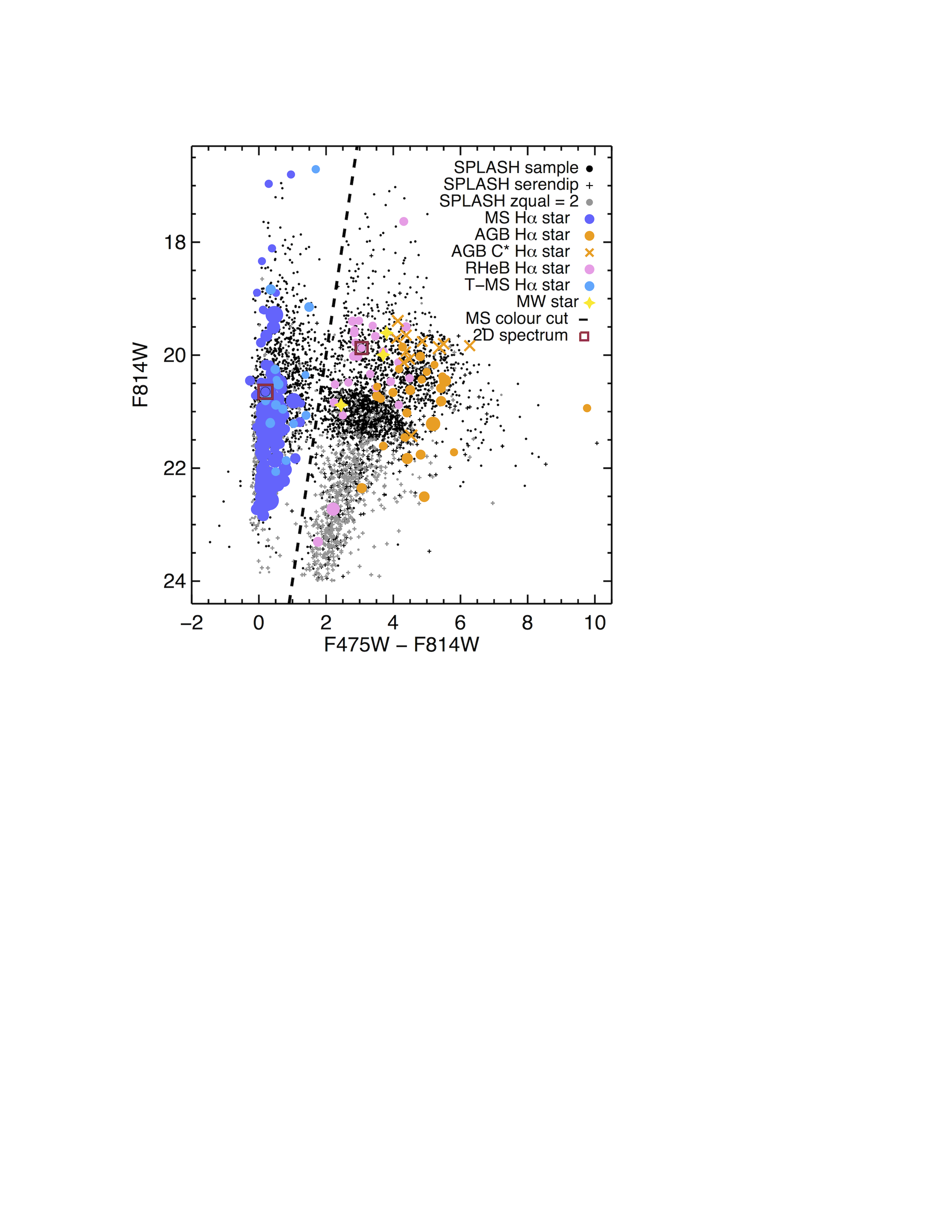}\label{opcmd}
	}
\subfloat[][]{
	\includegraphics[width=0.49\textwidth, trim=170 745 430 185,clip]{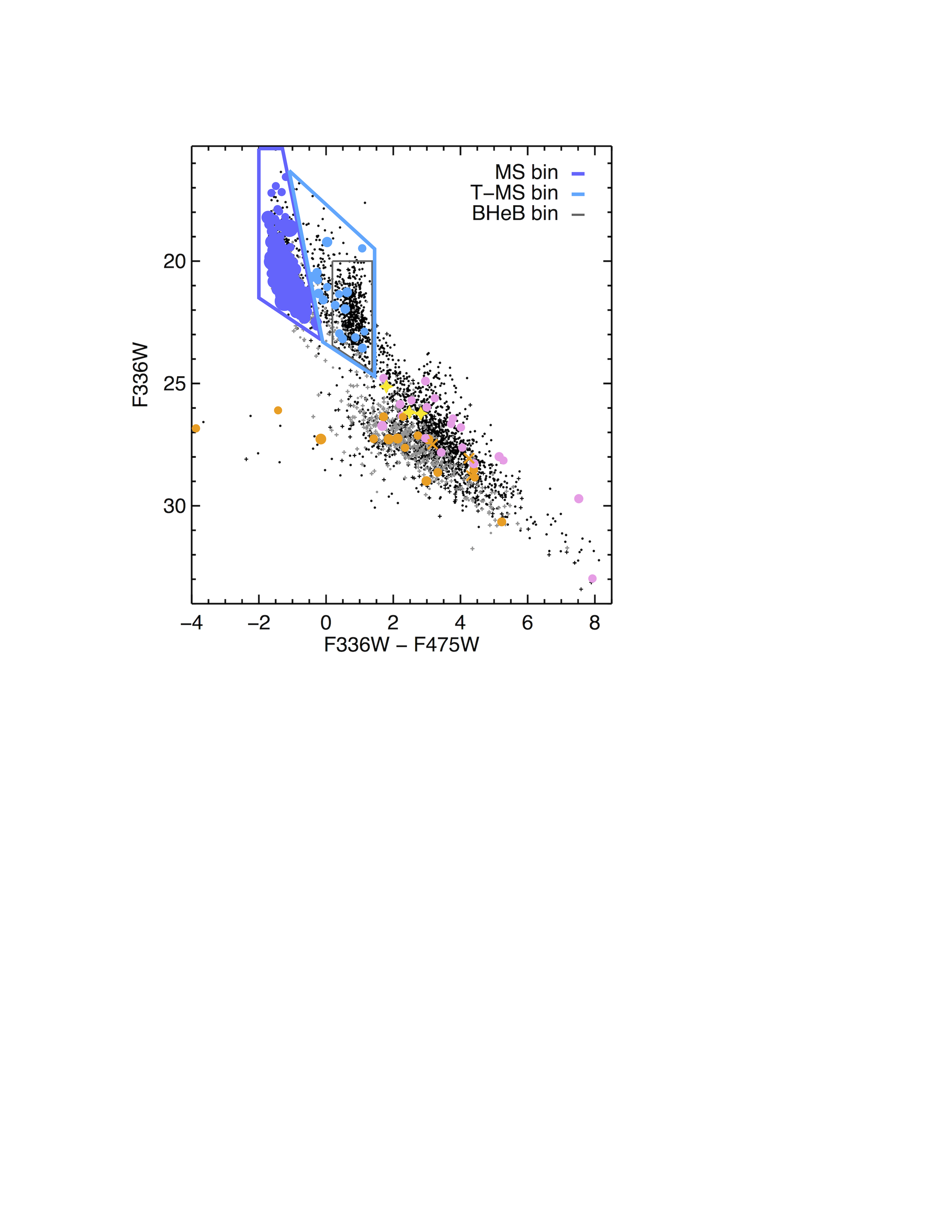}\label{uvcmd}
	}\\
\subfloat[][]{
	\includegraphics[width=0.49\textwidth, trim=170 745 430 185,clip]{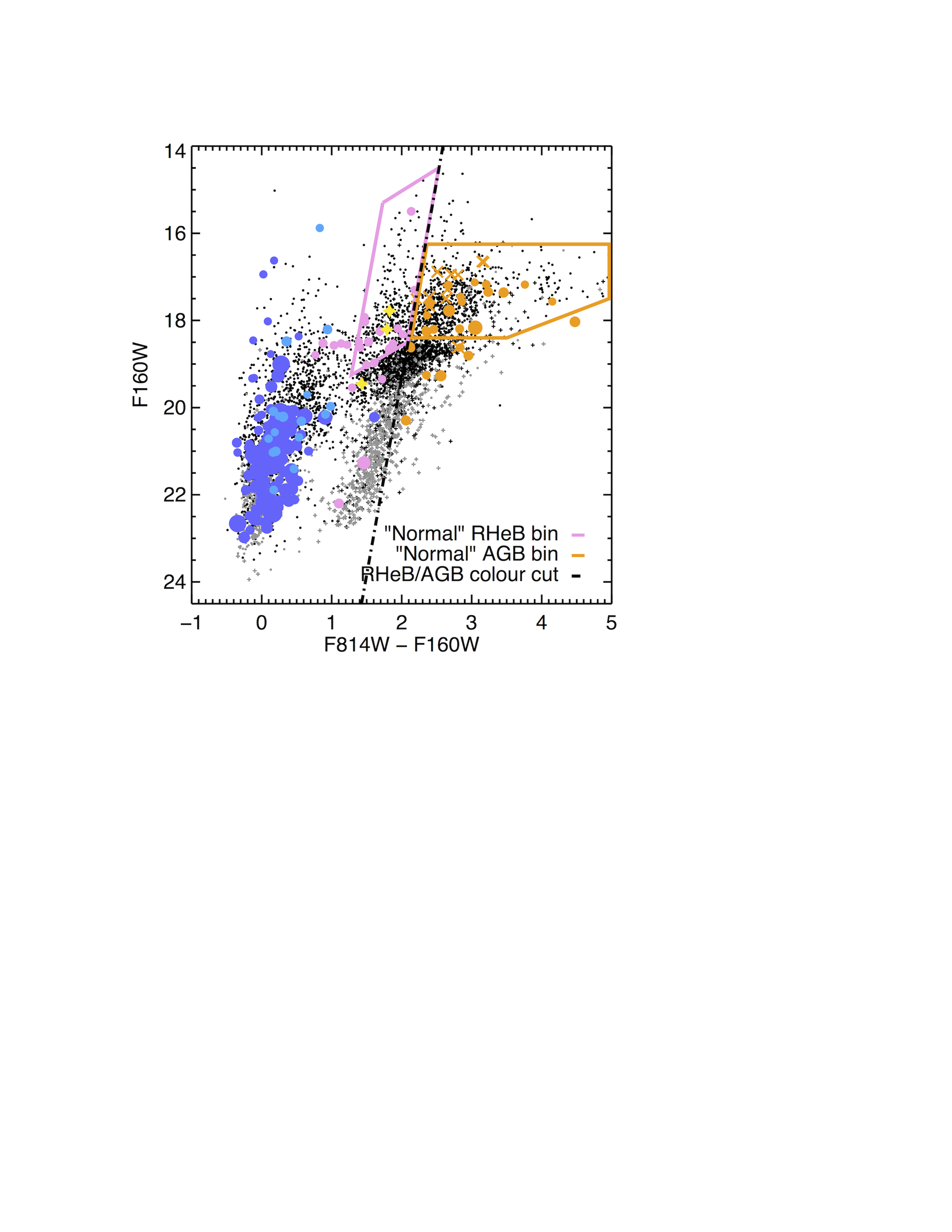}\label{opnircmd}
	}
\subfloat[][]{
	\includegraphics[width=0.49\textwidth, trim=170 745 430 185,clip]{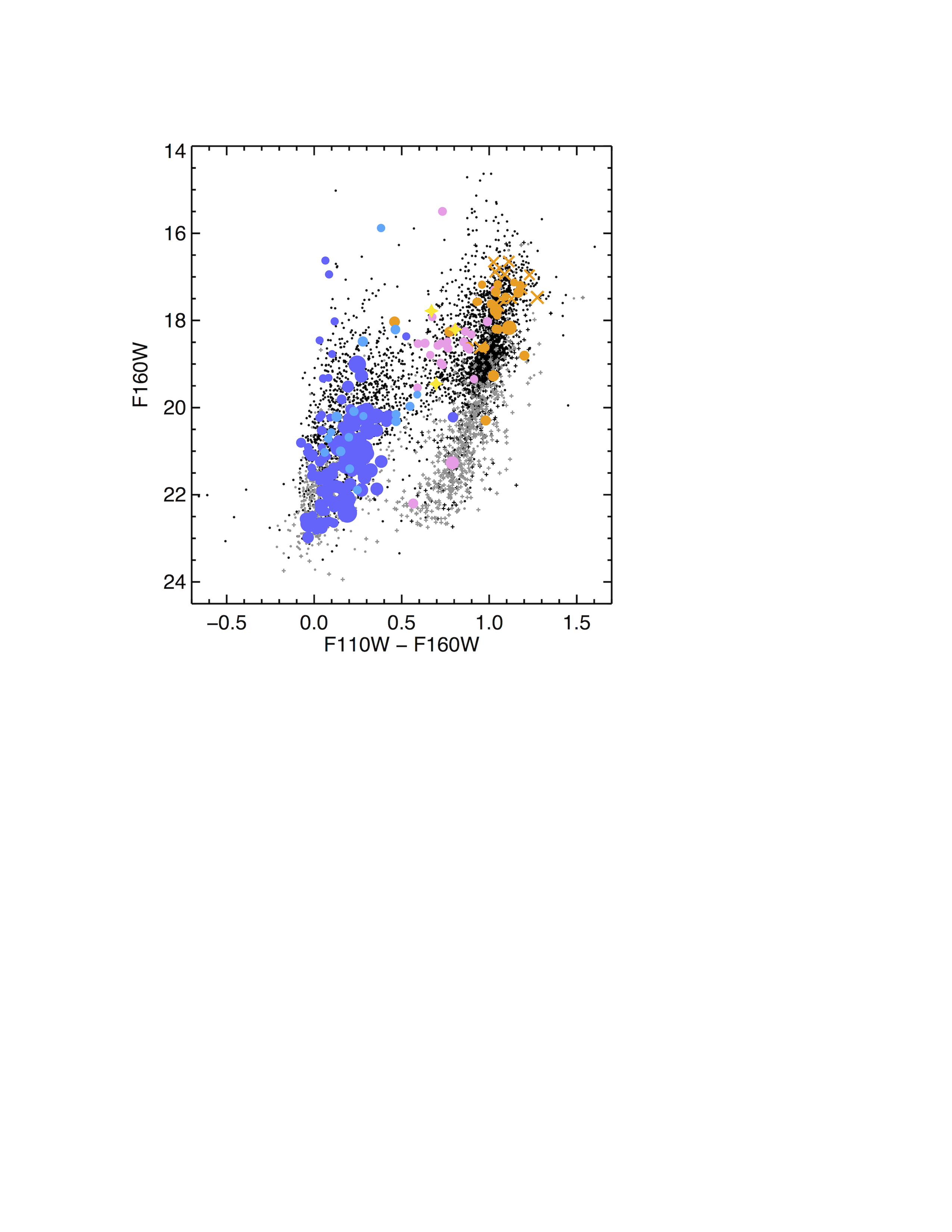}\label{nircmd}
	}
\caption[]{CMDs of PHAT photometry. Small points are the 5295 SPLASH objects, including serendipitously observed sources `serendips' (small crosses), and sources with poor spectral quality (zqual = 2; grey points). Larger points are H$\alpha$ stars; their size indicates H$\alpha$ emission-line strength (|EW|), their `normal' population is defined by their respective polygons. Yellow points, not scaled to line strength, are foreground MW H$\alpha$ stars.
\protect\subref{opcmd} The optical colour cut (black dashed line) separates the blue and red H$\alpha$ stars. Examples of the 2D SPLASH spectra of H$\alpha$ stars in red boxes are shown in Appendix \ref{sec:appB}.
\protect\subref{uvcmd} We distinguish regular MS stars (dark blue polygon), from the BHeB stars (thin grey polygon), and define T-MS H$\alpha$ stars (pale blue polygon).
\protect\subref{opnircmd} Redder H$\alpha$ stars are separated in to RHeB and AGBs using a colour cut (\citealt{Rosenfield2016}; dot-dash line).\small}
\label{fig:cmd}
\end{figure*}
\item H$\alpha$ peak height $>$ 2.5 $\times$ RMS $\longrightarrow$ removed 357 stars
\item $[$NII$]$b peak height $<$ 2.5 $\times$ RMS $\longrightarrow$ removed 84 stars
\item H$\alpha$ $\&$ $[$NII$]$b Gaussian $\frac{\chi^{2}}{N}$ $<$ 0.62 $\longrightarrow$ removed 38 stars
\item $[$SII$]$b peak height $<$ 2.5 $\times$ RMS, for $[$SII$]$a,b Gaussian $\frac{\chi^{2}}{N}$ $<$ 0.62 \hspace{4mm} OR \hspace{4mm} 
$[$SII$]$b peak height $<$ 4 $\times$ RMS, for $[$SII$]$a,b Gaussian $\frac{\chi^{2}}{N}$ $>$ 0.62 $\longrightarrow$ removed 15 stars
\item Visual inspection $\longrightarrow$ removed 12 stars
\end{enumerate}
\vspace{4mm} 

\begin{figure*} 
\centering
\subfloat[][]{
	\includegraphics[width=0.49\textwidth , trim=170 745 430 185,clip]{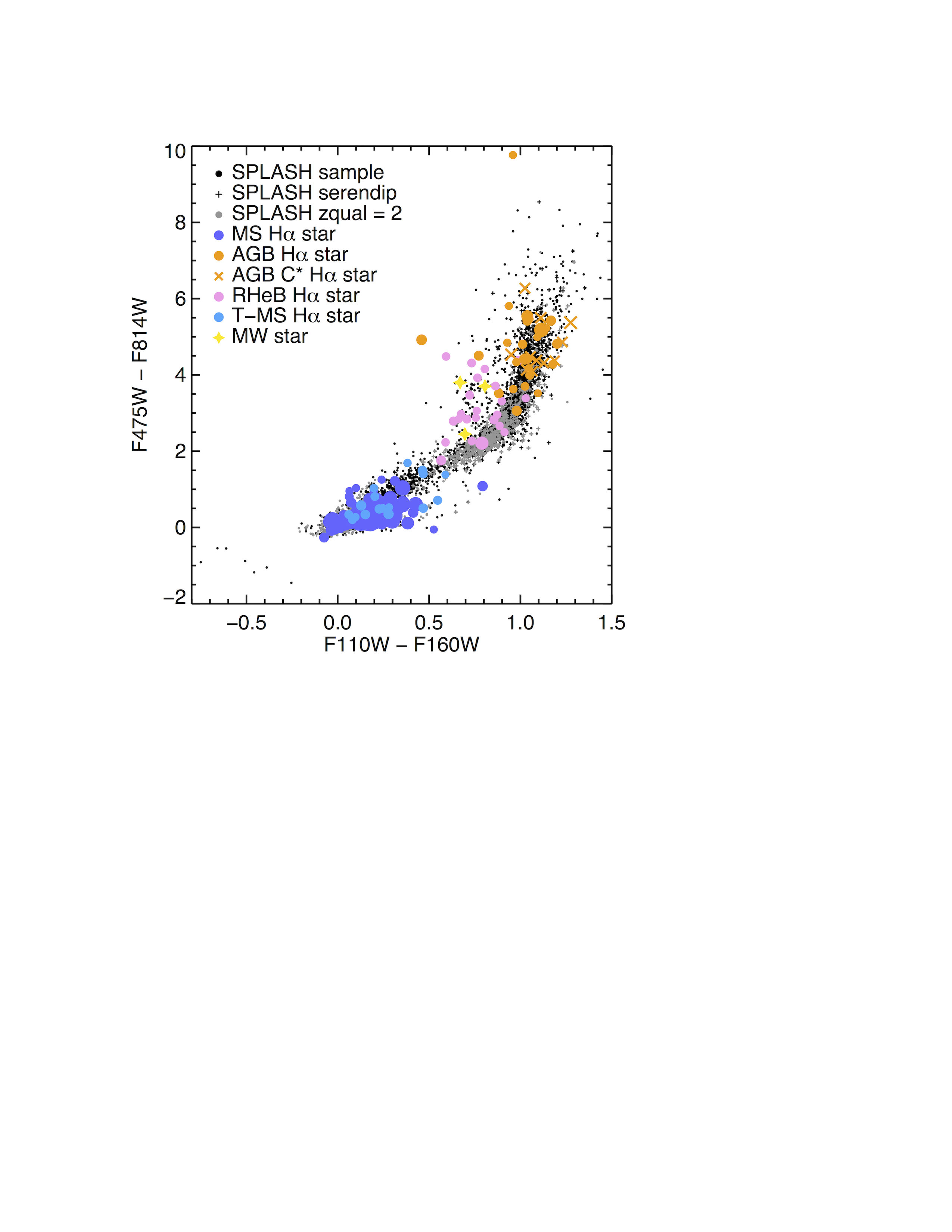}\label{opnir}
	}
\subfloat[][]{
	\includegraphics[width=0.49\textwidth , trim=170 745 430 185,clip]{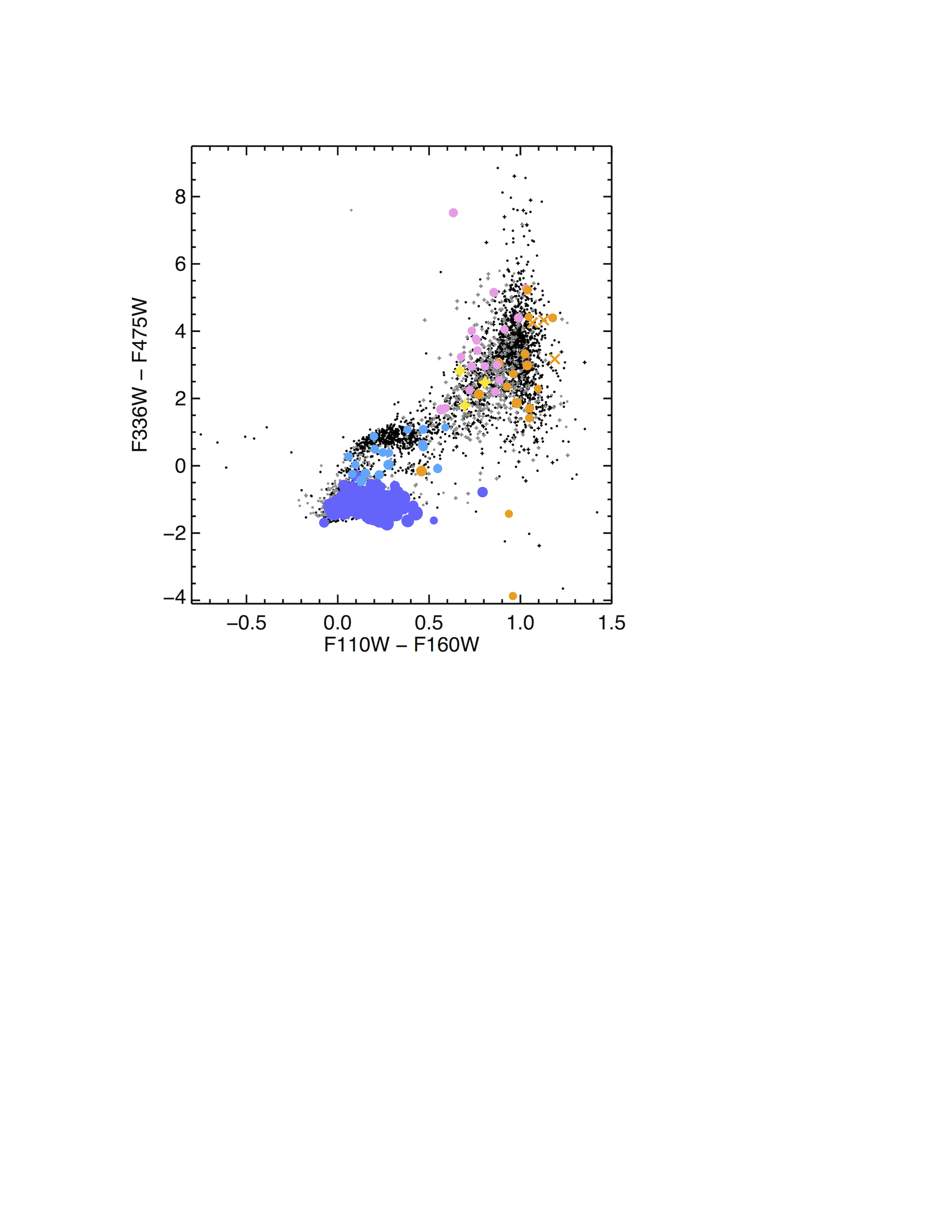}\label{uvop_nir}
	}
\caption[]{2CDs of the PHAT photometry follow the same labelling as Fig.~\ref{fig:cmd}. 
\protect\subref{opnir} 2CD of optical~$-$~far-red vs. NIR colours.
\protect\subref{uvop_nir} 2CD of UV~$-$~optical vs. NIR colours. Fewer redder stars appear as only $\sim$~89 per cent of the H$\alpha$ stars have F336W data. See \S \ref{sec:photom}.\small}
\label{fig:colc}
\end{figure*}

This set of criteria produced a sample of 229 H$\alpha$ stars, $\sim$~5.5 per cent of the total sample. By comparing our selected sample with the visually selected list we were able to verify the algorithm's efficiency of selecting H$\alpha$ stars. For a full list of selected H$\alpha$ stars, their coordinates, and determined properties, see the catalogue in Appendix A. 

\section{Data analysis}
\label{sec:analysis}

With a sample of 229 H$\alpha$ stars selected, we used the six-filter HST photometry and their measured parameters from the selection algorithm to classify them and investigate their properties.

\subsection{H$\boldsymbol{\alpha}$ star photometry}
\label{sec:photom}

CMDs of the PHAT photometry are shown in Fig.~\ref{fig:cmd}. The small points are of the objects in the sample of 5295 SPLASH spectra that have data in the relevant HST filters. We have separated them in to four categories. The small crosses are serendipitously observed sources (`serendips'), the grey points are all sources with poor spectral quality (i.e. zqual~=~2) and thus unreliable velocity measurements, the small grey crosses are both poor quality and serendips. The black points are the remaining objects in the SPLASH sample. The larger points represent the H$\alpha$ stars that we selected, their symbol size is a measure of their H$\alpha$ line-strengths (|EW|). In Fig.~\subref*{opcmd} are two H$\alpha$ stars in red boxes. These stars correspond to the two H$\alpha$ stars for which we present their 2D SPLASH spectra in Appendix \ref{sec:appB}. 

\begin{figure} 
\includegraphics[width=0.49\textwidth , trim=85 370 70 95,clip]{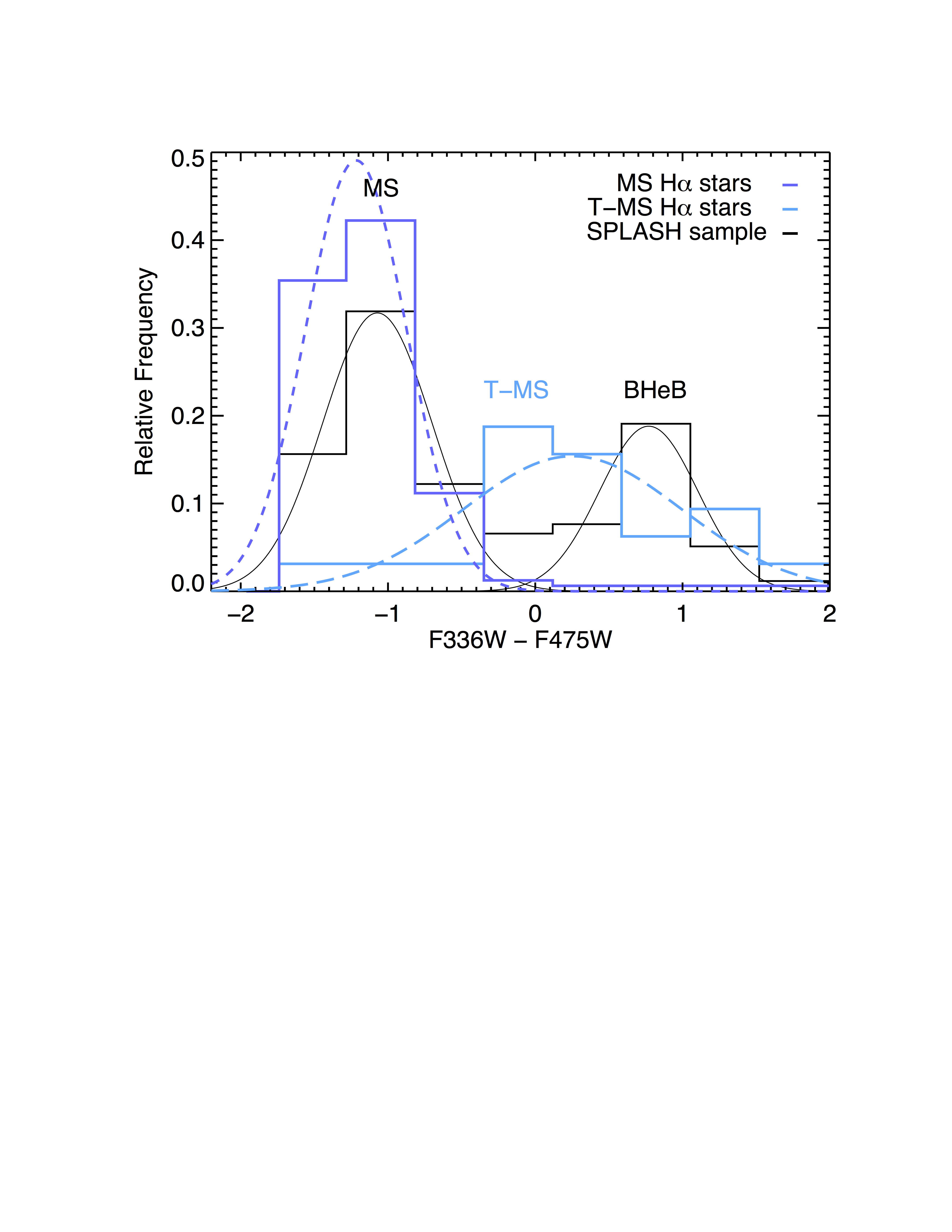}
\caption[]{Histogram of relative frequencies of stars blue-ward of the optical colour cut in Fig.~\subref*{opcmd} vs. the UV~$-$~optical colour. H$\alpha$ T-MS stars (light blue long-dashed Gaussian) are not associated with the MS or BHeB sequence. See \S \ref{sec:tmsha}.\small}
\label{fig:TPhist}
\end{figure}

We take advantage of the HST photometry by plotting a range of CMDs that allow us to differentiate different stellar populations. Fig.~\subref*{opcmd} is an optical/far-red CMD of the F475W and F814W filters that does a good job of separating the blue and red stars in the SPLASH sample. We use an optical colour cut to define this separation, indicated by the dashed line. The RGB stars form a dense clump of black points between F475W~$-$~F814W $\sim$~2--4 mag and centred on F814W $\sim$~21 mag. The points below the RGB stars, mostly poor spectral quality grey points due to their faint magnitudes, are likely to be both RGB stars, and variable AGB or RHeB stars that are near minima in their cycles. Note that RGB stars do not exhibit H$\alpha$ emission, so any H$\alpha$ stars below the magnitude of the tip of the RGB (TRGB) are either variable AGB or RHeB H$\alpha$ stars or some contamination from symbiotic stars. The UV/optical CMD (Fig.~\subref*{uvcmd}) of the F336W and F475W filters separates the different varieties of bluer stars, while the far-red/NIR CMD Fig.~\subref*{opnircmd} of the F814W and F160W filter is used to separate the redder stars. We show the NIR CMD of the F110W and F160W filters (Fig.~\subref*{nircmd}) and the colour-colour diagrams (2CDs) in Fig.~\ref{fig:colc} to show the distributions of the SPLASH sample and H$\alpha$ stars in different CMD and 2CD spaces.

The H$\alpha$ star photometry is included in the catalogue in Appendix \ref{sec:appA}. Two of our selected H$\alpha$ stars do not have PHAT photometry as these were serendips that were not in the PHAT catalogue. This left 227 H$\alpha$ stars we were able to classify. Some of our selected H$\alpha$ stars did not have data in all six filters. Many stars, mostly the redder RHeB and all AGB stars do not have data in the UV filter F275W. For some sources that were faint in F275W, the detection would have been noise, and so were deemed unreliable. We therefore chose a magnitude cut of F275W $<$ 24.5 mag, leaving only 157 of 227 ($\sim$~69 per cent) H$\alpha$ stars with reliable F275W detections. Due to this large proportion, we decided not to include the F275W filter in our analysis for consistency. For the remaining five filters, the F336W filter has detections for 157 ($\sim$~89 per cent), F475W and F814W have data for all, F110W has data for all but one, and F160W has data for all but two of the 227 H$\alpha$ stars with PHAT photometry. 

In the following section we describe the method for identifying MW H$\alpha$ stars. The parameters for identifying the five classifications of H$\alpha$ stars are described in the subsequent sections. A summary of values determined in these sections and additional average properties for each class are given in Table \ref{tab:avg}.

\subsubsection{Removing foreground MW H$\alpha$ stars}
\label{sec:mw}

Identifying foreground MW stars required multiple diagnostics in order to separate them from the stars in M31. Low-mass old halo stars are most commonly confused with red massive stars at the distance of M31, such as the RHeB and AGB stars. In order to identify which H$\alpha$ stars in our sample were foreground MW stars we used a combination of CMD space and spectral-line strengths. Using a set of visibly identified MW stars as a training sample, we plotted the SPLASH sample and the MW stars on two 2CDs; F336W~$-$~F475W vs. F475W~$-$~F814, and F814W~$-$~F160W vs. F475W~$-$~F814W. A third plot of the line strength (|EW|) of the surface gravity sensitive feature Na~I $\lambda$8190--8900, used to distinguish low and high mass stars \citep{Faber1980}, against the F475W~$-$~F814W colour was also used. Using the known MW stars as a training set, polygons enclosing potential foreground contamination stars were drawn on the three separate plots. Stars that fell in all three of these polygons were classified as definite foreground MW stars. Of our H$\alpha$ star sample, we identified three MW H$\alpha$ stars, likely to be M-dwarf stars. We have marked these on the CMDs and 2CDs in Figs. \ref{fig:cmd} and \ref{fig:colc} respectively (yellow stars). We remove these stars from our sample of H$\alpha$ stars, leaving 224 in total, and do not include them in the analysis.

\subsubsection{Classifying MS H$\alpha$ stars}
\label{sec:msha} 

To identify the MS H$\alpha$ stars we used the UV/optical (F336W and F475W) CMD in Fig.~\subref*{uvcmd}. Using this CMD, we were able to distinguish the MS stars from the BHeB stars in our sample \citep{Dohm-Palmer1997}. We used PAdova and TRieste Stellar Evolution Code (PARSEC) stellar evolutionary tracks \citep{Bressan2012} to confirm the location of the MS stars in this CMD. The stars we identified as MS stars are enclosed in the dark blue polygon, drawn by eye and centred at F336W~$-$~F475W~$\sim -1$ mag. We therefore classify MS H$\alpha$ stars as any H$\alpha$ star within this MS bin, and the `normal' MS population against which to compare the MS H$\alpha$ stars, as any other SPLASH object within this bin. We found 146 MS H$\alpha$ stars, $\sim$~12 per cent of the MS population.

\subsubsection{Classifying T-MS H$\alpha$ stars}
\label{sec:tmsha} 

The bluer stars in the SPLASH sample can be further separated. Again using the PARSEC evolutionary tracks \citep{Bressan2012}, we were able to confirm that the dense clump of black points in the thin grey polygon on the UV/optical CMD (Fig.~\subref*{uvcmd}) are BHeB stars. As can be seen in the histogram in Fig.~\ref{fig:TPhist}, the blue stars in the SPLASH sample separate in to two groups, MS and BHeB stars (solid black Gaussian curves). As can be seen from the UV/optical CMD, there are H$\alpha$ stars that lie outside the MS bin (dark blue), we therefore defined an additional polygon red-ward of the MS polygon by eye to enclose these additional H$\alpha$ stars (pale blue). As can be seen from Fig.~\ref{fig:TPhist}, this population of stars do not line up with either the MS or BHeB stars in colour space but instead falls between them (pale blue long-dash Gaussian). From this we conclude that these stars are `transitioning' from the MS (T-MS) and not associated with the BHeB sequence. These T-MS stars are most likely stars that are evolving off the MS and their H$\alpha$ emission is fading as a result. Alternatively, the T-MS stars could also be reddened MS stars. We classify 17 stars as T-MS H$\alpha$ stars.

To investigate the properties of the T-MS stars, we wanted to define a `normal' T-MS population. We therefore defined any SPLASH object that was enclosed in the T-MS bin but \textit{not} in the BHeB bin as a `normal' T-MS star. Taking a ratio of the H$\alpha$ stars that fall in the `normal' T-MS bin and the total number of SPLASH objects in the same region, we found a proportion of $\sim$~5 per cent of the T-MS stars emit H$\alpha$.

\begin{figure*}
\centering
\includegraphics[width=0.61\textwidth, trim=70 370 200 90,clip]{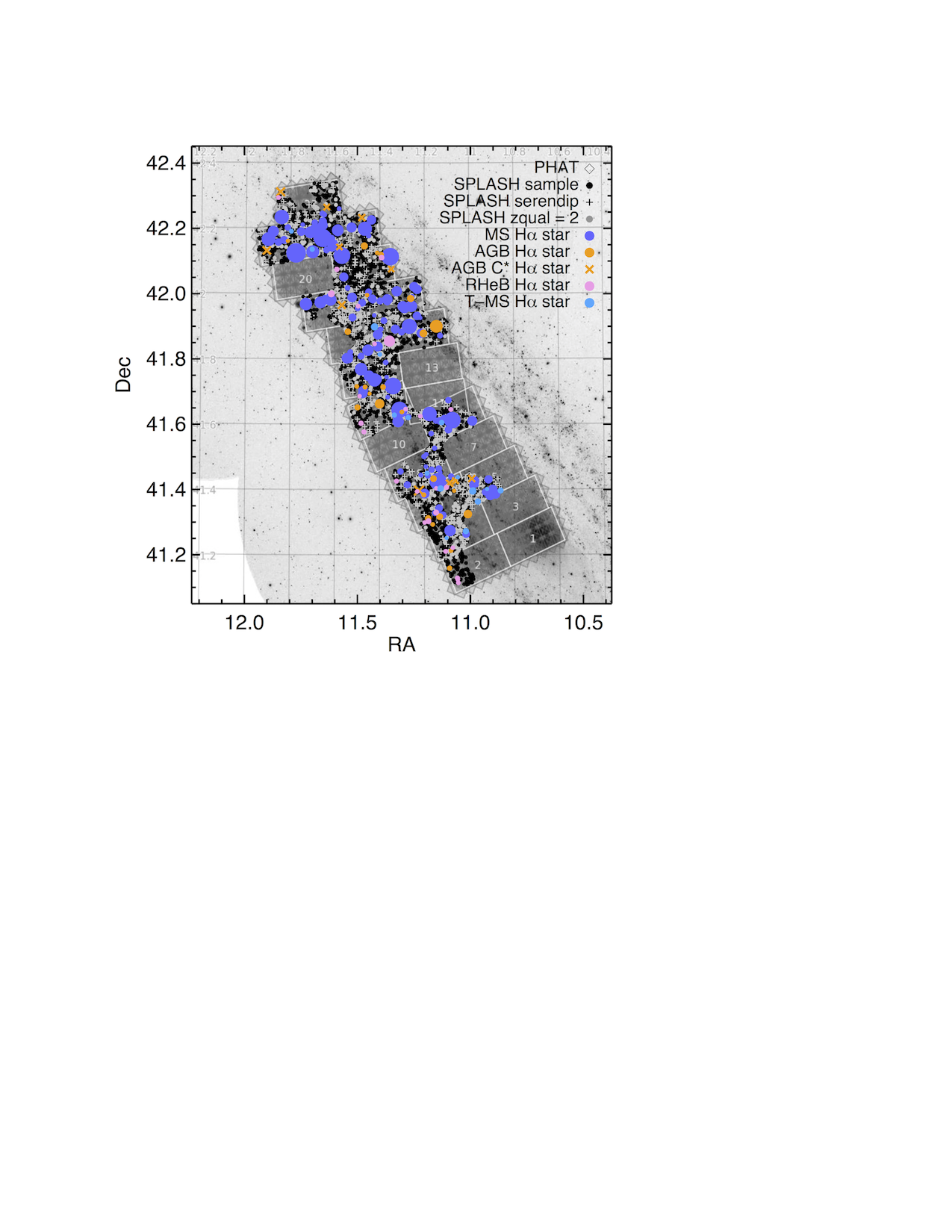}
\caption[]{Spatial distribution of the spectroscopic and photometric samples in M31. The background is a composite of a GALEX image of M31 with the PHAT survey coverage shown by the grey `bricks' (composed by Dustin Lang). Points follow the same labelling as Fig.~\ref{fig:cmd}. See \S \ref{sec:space}.\small}
\label{fig:spatial}
\end{figure*}

\subsubsection{Classifying RHeB H$\alpha$ stars}
\label{sec:rhha} 

For the redder stars in the SPLASH sample, we used the far-red/NIR (F814W/F110W) CMD in Fig.~\subref*{opnircmd} to separate the different populations \citep{Melbourne2012}. We separated RHeB stars from AGB stars using the line defined by \cite{Rosenfield2016}, and scaled to the TRGB in M31, shown by the dot-dash line. We defined RHeB H$\alpha$ stars as those that lie red-ward of the optical~$-$~far-red colour cut shown in Fig.~\subref*{opcmd} but blue-ward of the far-red~$-$~NIR colour cut in Fig.~\subref*{opnircmd}. We classify 25 H$\alpha$ stars as RHeB stars. As discussed in \cite{Rosenfield2016}, this definition works best at brighter magnitudes but close to the TRGB, these populations become increasingly more merged. However, as we are unable to spectroscopically distinguish the RHeB stars from the AGB stars at this resolution \citep{Melbourne2012, Dalcanton2012}, we use this definition to separate RHeB stars from AGBs knowing that there could significant be contamination. \cite{Rosenfield2016} estimate the mean contamination by AGB stars in the RHeB population to be $\sim$~4 per cent.

To define a `normal' RHeB star population against which to compare the RHeB H$\alpha$ stars, we again used the far-red/NIR CMD. We defined a polygon by eye that was positioned blue-ward of the \cite{Rosenfield2016} colour cut but above the TRGB (dense black points centred at F160W $\sim$ 19 mag). Comparing the RHeB H$\alpha$ stars and the other SPLASH objects within this bin, we found a proportion of $\sim$ 4 per cent of RHeB stars emit H$\alpha$. To verify the RHeB H$\alpha$ stars we defined that were not in the `normal' RHeB bin were correctly classified, we inspected their individual spectra. While we were unable to definitively say whether a star was an AGB star or RHeB star, we could check that those outside the `normal' RHeB bin had typical RHeB spectral features, i.e. strong TiO absorption. We found that stars blue-ward of the bin had TiO absorption and those below the bin, in the RGB clump, mostly had strong TiO absorption. The two with weaker TiO absorption that lie below the `normal' RHeB bin, we have defined as `RHeB?' in the H$\alpha$ star catalogue in Appendix \ref{sec:appA} but include in the RHeB class for the analysis. 

\subsubsection{Classifying AGB H$\alpha$ stars}
\label{sec:agbha} 

Following the classification of the RHeB stars from the AGB stars, we therefore defined all AGB stars as being those red-ward of the colour cut shown by the dot-dash line in Fig.~\subref*{opnircmd} (orange points). The AGB H$\alpha$ stars that lie below the TRGB are likely to be variable AGB stars at their minima, or contamination by symbiotic stars. We further separated the AGB stars in to non-C-rich (C/O $<$ 1) AGB stars and C-rich AGB stars (AGB C*) based on spectral features. The AGB C* H$\alpha$ population is discussed in the next section. We found 25 non-C-rich AGB stars that emit H$\alpha$. To define a `normal' AGB population we again drew a polygon by eye above the TRGB but extending red-ward of the RHeB/AGB colour cut (orange polygon). Using this reference sample and focusing just on the non-C-rich AGBs within the polygon, we found a proportion of $\sim$~2 per cent emit H$\alpha$. As with the RHeBs, distinguishing AGBs from RHeBs can result in contamination. \cite{Rosenfield2016} determine the mean contamination by RHeBs in the AGB population to be $\sim$~3 per cent.
 
\subsubsection{Classifying AGB C* H$\alpha$ stars}
\label{sec:csha} 

Those AGB H$\alpha$ stars that are also C stars, as identified by \cite{Hamren2015}, see \S \ref{sec:specdata}, are shown by the orange X symbols in Fig.~\ref{fig:cmd}. We found 11 AGB C* H$\alpha$ stars. From the total 5295 SPLASH spectra, we take all C stars red-ward of the RHeB/AGB colour cut (77 in total) to be our `normal' AGB C* population. We conclude that $\sim$~14 per cent of the C stars show H$\alpha$ emission.

\subsection{Spatial distribution}
\label{sec:space}

A 2D projection of the sample in M31 is shown in Fig.~\ref{fig:spatial}. The sample is plotted over a GALEX image of M31 with the PHAT survey coverage shown by the numbered grey `bricks' (image by Dustin Lang). The spectroscopic sample lies within the region covered by the PHAT survey as a requirement of the SPLASH target selection. The points again follow the labelling convention of the CMDs in Fig.~\ref{fig:cmd}, where the small points indicate the total SPLASH sample of 5295, 600~line~mm$^{-1}$ spectra, and are split in to sub-categories. The larger points are the H$\alpha$ stars separated in to five classifications; MS (dark blue points), T-MS (pale blue points), RHeB (pink points), AGB (orange points), and AGB C* (orange Xs). Their symbol size is a measure of their H$\alpha$ emission-line strength (|EW|).

\begin{figure} 
\includegraphics[width=0.49\textwidth , trim=0 65 60 100,clip]{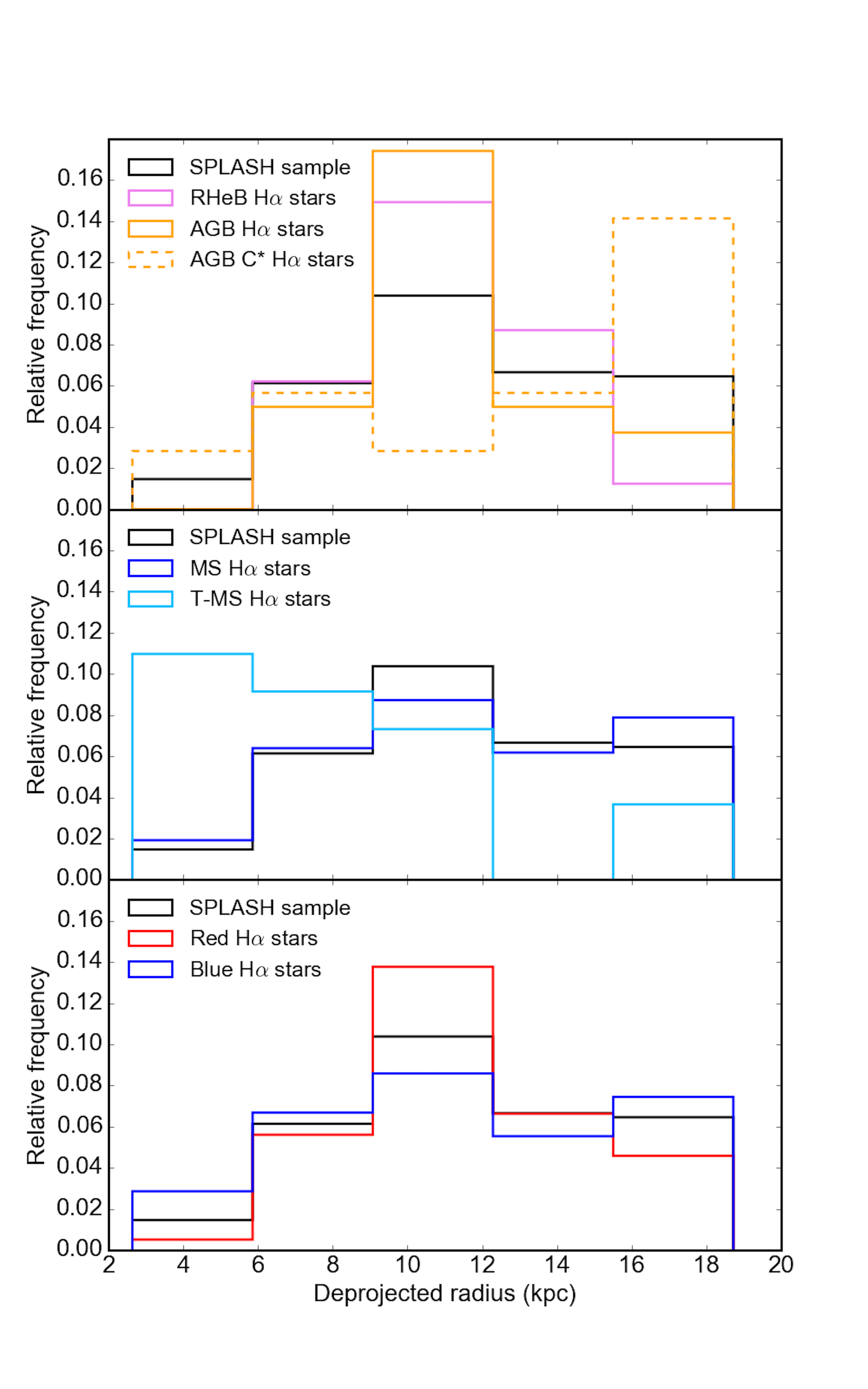}
\caption[]{Histograms of the relative frequencies of the SPLASH sample (black) and H$\alpha$ stars vs. R$_{deproj}$. Top panel shows red H$\alpha$ stars. Middle panel shows blue H$\alpha$ stars. Bottom panel shows the combined red and blue H$\alpha$ stars. See \S \ref{sec:space}.\small}
\label{fig:Rhist}
\end{figure}

The richest sections of M31 covered by the PHAT survey were targeted by SPLASH in order to maximise the use of the DEIMOS masks. Therefore, the dense star forming rings were prioritised. To better understand the spatial distribution of the sample, we determined deprojected radii from the centre of M31 (R$_{deproj}$) of the stars. We used inclination angle $i~=~70^{\circ}$, position angle $43^{\circ}$, and centre $\alpha(2000) = 10\fdg68473$ and $\delta(2000) = 41\fdg26905$ \citep{Dalcanton2012} to calculate R$_{deproj}$. The main features seen in this northeast quadrant of M31 are the 5-, 10-, and 15-kpc star forming rings \citep{Lewis2015}. These three main features, which are covered by the SPLASH survey, are at approximately $\alpha(2000) = 11\fdg1$ and $\delta(2000) = 41\fdg6$, $\alpha(2000) = 11\fdg3$ and $\delta(2000) = 41\fdg9$, and $\alpha(2000) = 11\fdg7$ and $\delta(2000) = 42\fdg2$ respectively on the 2D projection in Fig. \ref{fig:spatial}. Given the targeted selection the SPLASH sample, we therefore look at the distribution of H$\alpha$ stars relative to the total 600 line mm$^{-1}$ SPLASH sample of 5295 objects, which peaks at the 10-kpc ring.

\begin{figure*}
\centering
\includegraphics[width=0.7\textwidth, trim=80 365 125 165,clip]{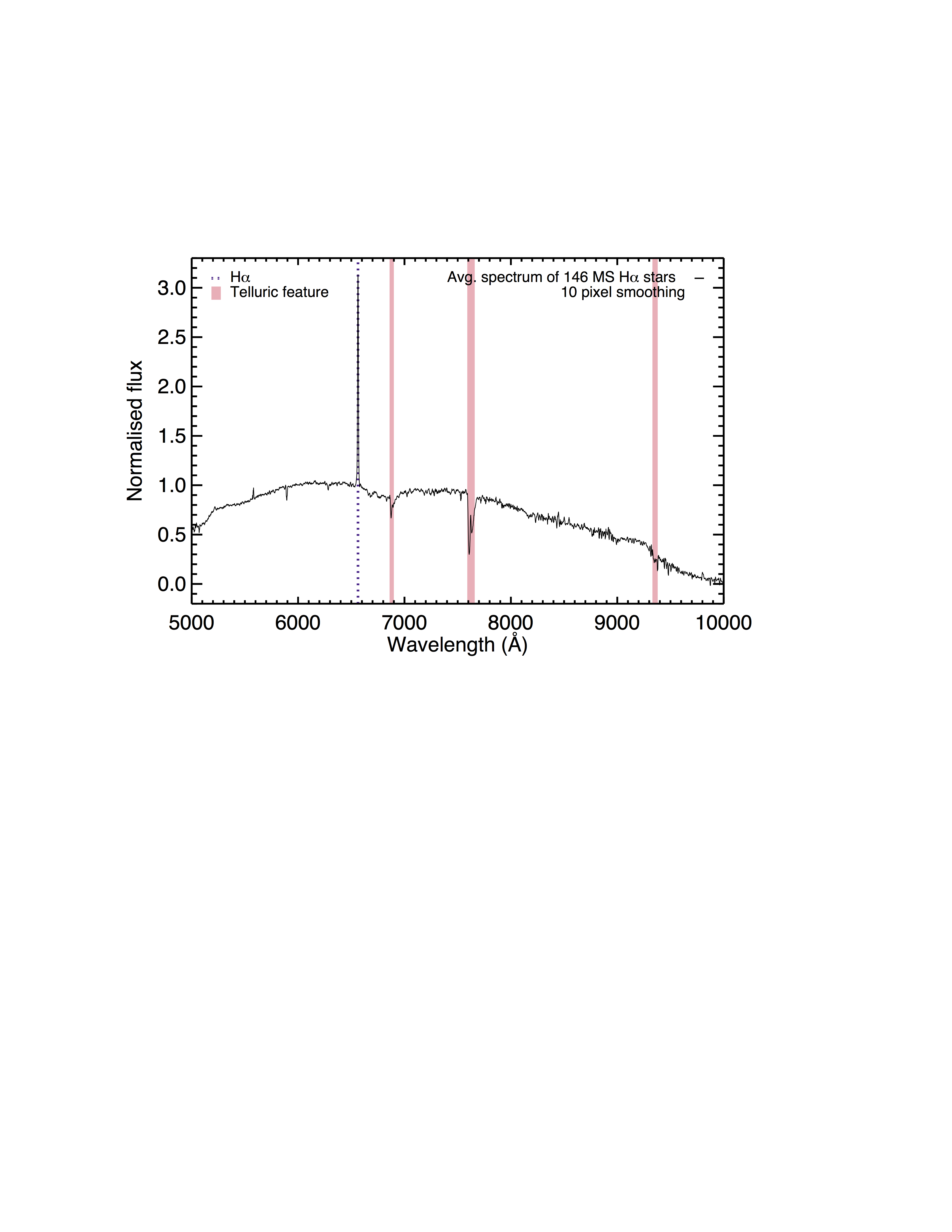}
\caption[]{Composite spectrum of the 146 bright MS H$\alpha$ stars, with 10-pixel smoothing. The continuum, normalised to the H$\alpha$ region, is relatively smooth with emission and absorption lines. The H$\alpha$ emission line (vertical dotted line) is prominent, there is broad telluric absorption lines at $\sim$~6880 $\text{\AA}$ and $\sim$~7620 $\text{\AA}$ shown by the magenta bands, and there is some bad sky subtraction seen in the spectrum. \small}
\label{fig:MS}
\end{figure*}

\begin{figure*}
\centering
\includegraphics[width=0.7\textwidth, trim=80 365 125 165,clip]{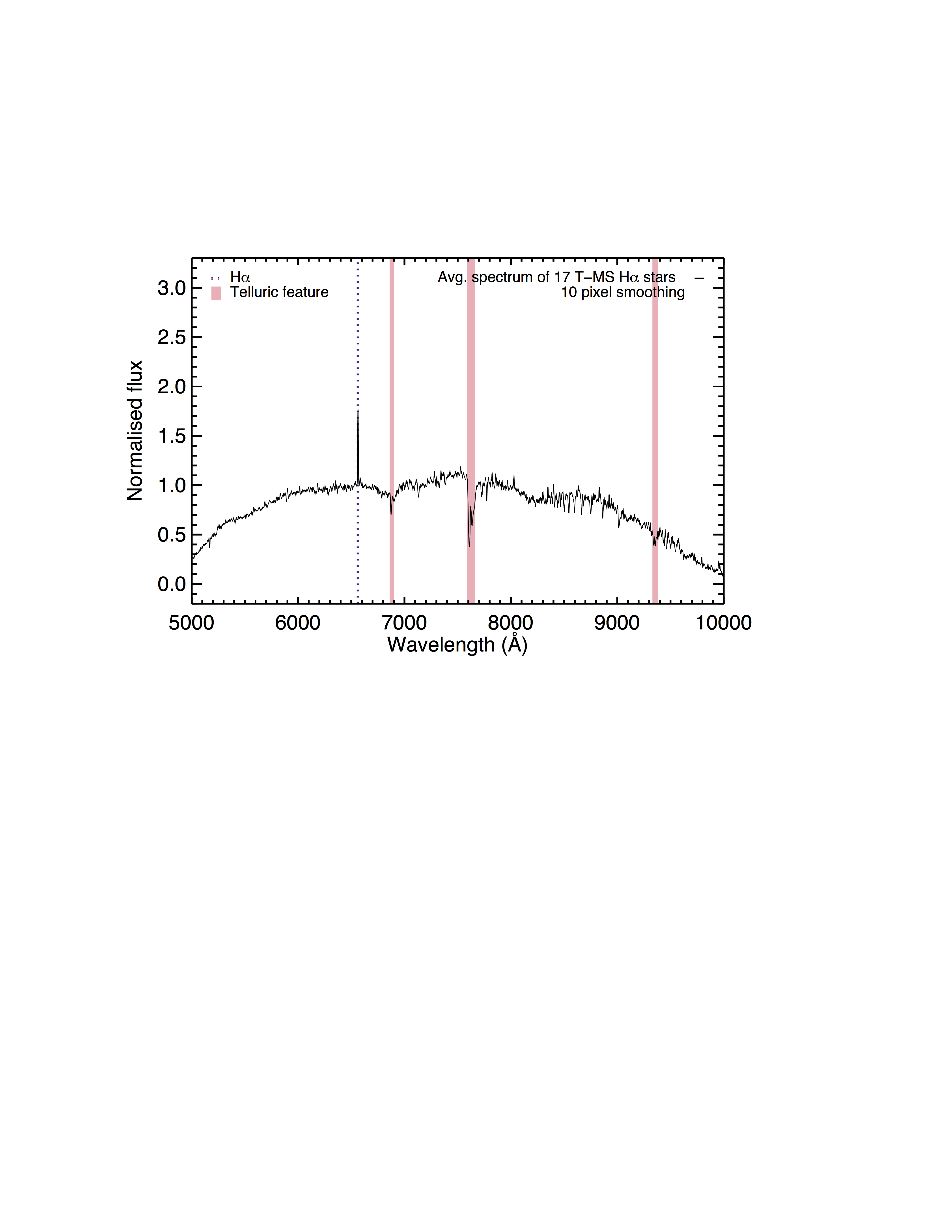}
\caption[]{Composite spectrum of the 17 T-MS H$\alpha$ stars, normalised to the H$\alpha$ region, and with 10-pixel smoothing. The H$\alpha$ emission line (vertical dotted line) and telluric absorption (magenta bands) are shown. The T-MS stars have similar spectral properties to the MS star spectrum in Fig.~\ref{fig:MS}. However, they seem to show weaker H$\alpha$ emission and have a bumpier continuum.\small}
\label{fig:TMS}
\end{figure*} 

\begin{figure*}
\centering
\includegraphics[width=0.7\textwidth, trim=80 365 125 165,clip]{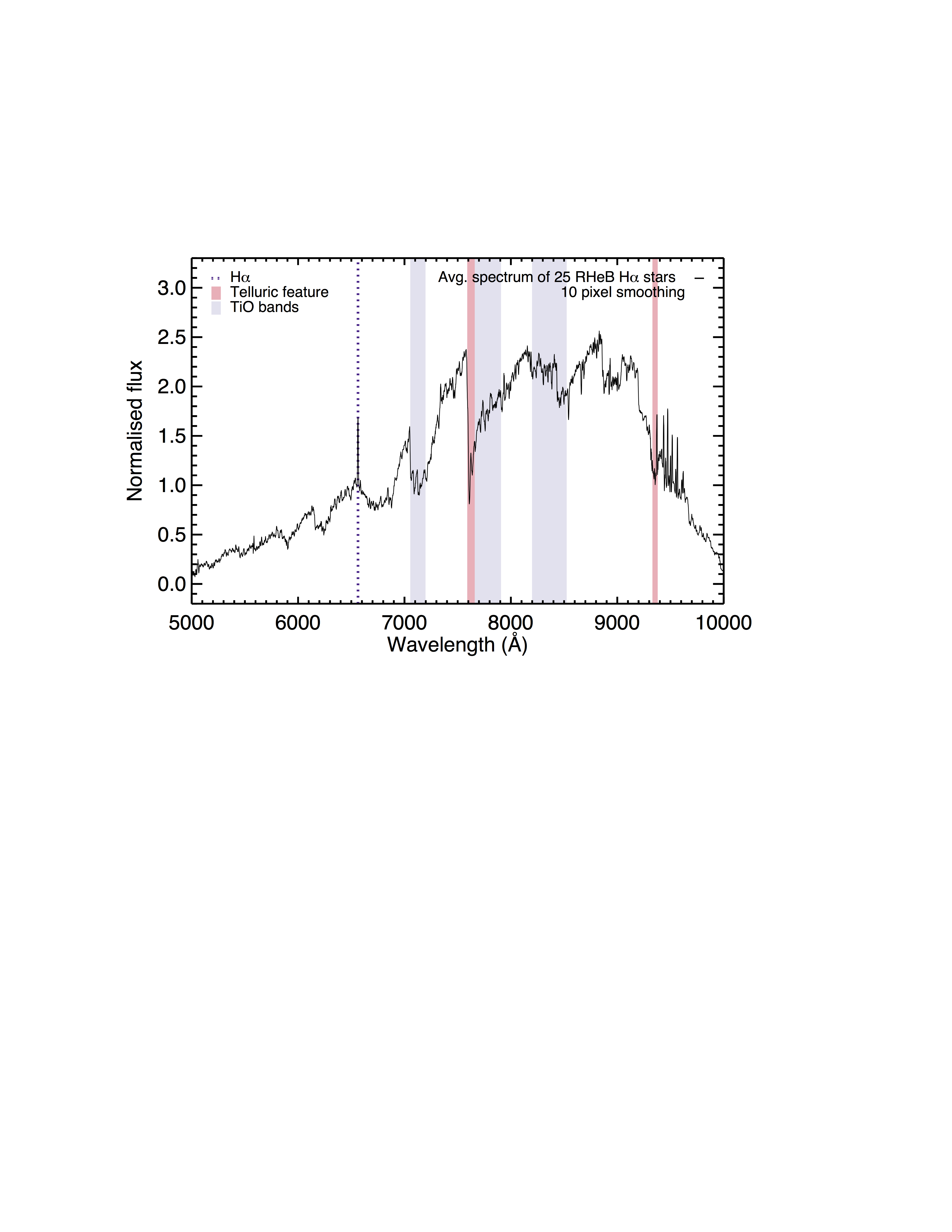}
\caption[]{Composite spectrum of the 25 RHeB H$\alpha$ stars, with continuum normalised to the H$\alpha$ region and 10-pixel smoothing. The H$\alpha$ emission line (vertical dotted line) and telluric absorption (magenta bands) are shown. The RHeB spectrum also shows strong TiO absorption features as indicated by the pale bands. \small}
\label{fig:RH}
\end{figure*}

\begin{figure*}
\centering
\includegraphics[width=0.7\textwidth, trim=80 365 125 165,clip]{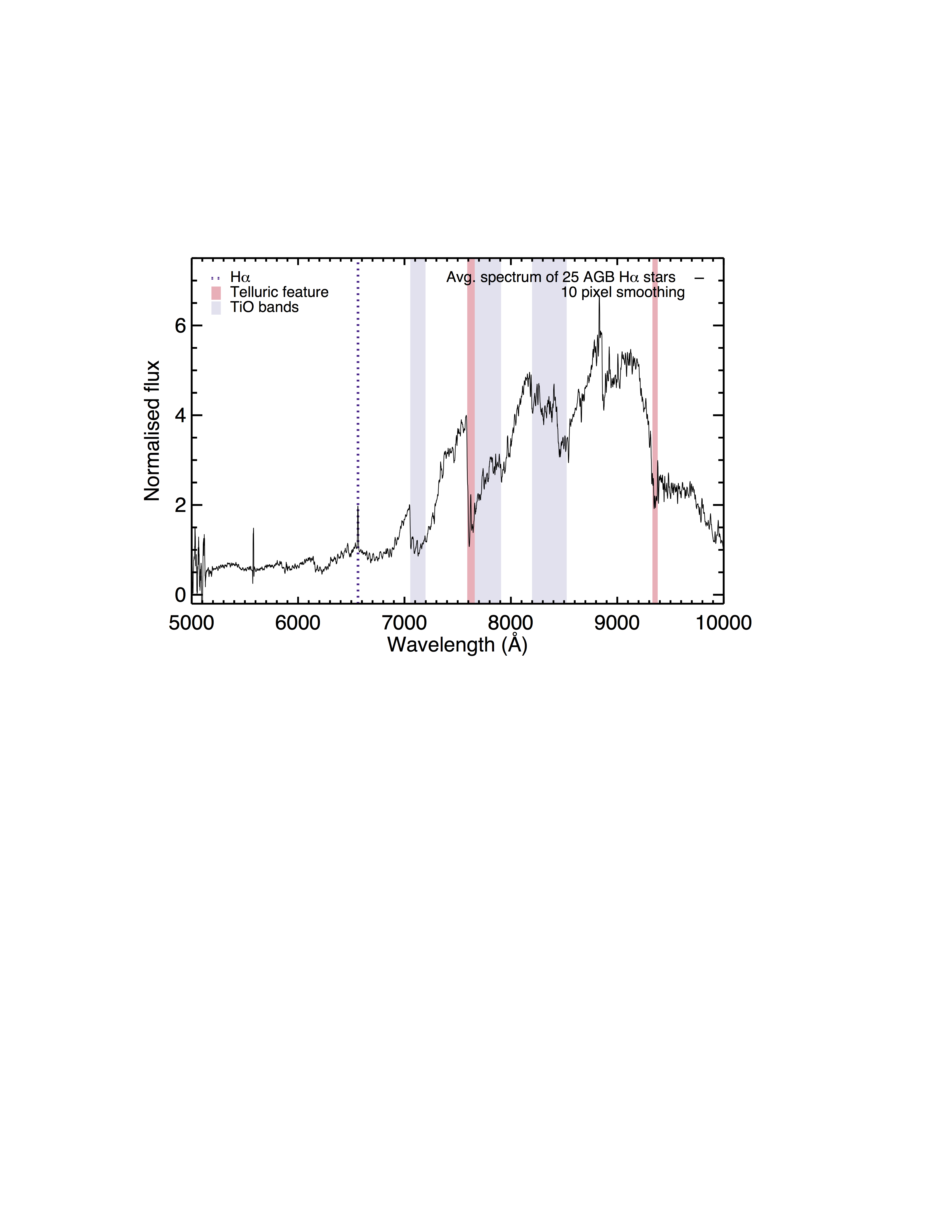}
\caption[]{Composite spectrum of the 25 non-C-rich AGB stars with H$\alpha$ emission, normalised to the H$\alpha$ region continuum, and with 10-pixel smoothing. The H$\alpha$ emission line (vertical dotted line), telluric absorption (magenta bands), and TiO absorption (pale bands) are shown. The AGB composite spectrum is very similar to the RHeB spectrum (Fig.~\ref{fig:RH}) despite their different evolutionary paths.\small}
\label{fig:AGB}
\end{figure*}

\begin{figure*}
\centering
\includegraphics[width=0.7\textwidth, trim=80 365 125 165,clip]{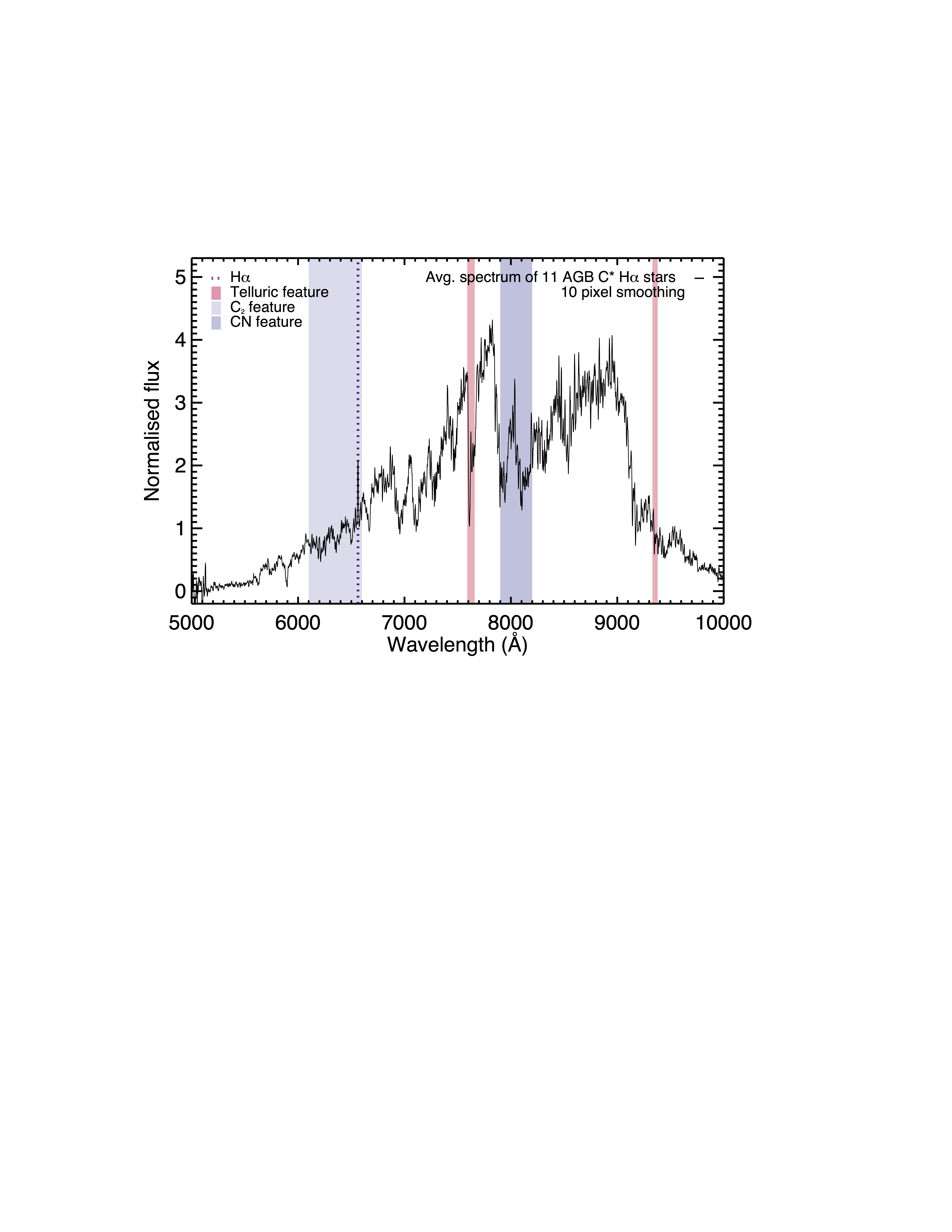}
\caption[]{Composite spectrum of the 11 C-rich AGB stars with H$\alpha$ emission, normalised to the H$\alpha$ region, and with 10-pixel smoothing. The H$\alpha$ emission line (vertical dotted line) and telluric absorption (magenta bands) are shown. The darker bands show the C$_2$ ($\sim$~6100--6600 $\text{\AA}$) and CN ($\sim$~7900--8200 $\text{\AA}$) features. \small}
\label{fig:CS}
\end{figure*} 

Histograms of the different types of H$\alpha$ star as a function R$_{deproj}$ are shown in Fig. \ref{fig:Rhist}. When splitting the red (all AGBs and RHeBs) from the blue stars (MS and T-MS) in the sample (bottom panel of Fig. \ref{fig:Rhist}), we find a relative increase in the number of red H$\alpha$ stars in the 10-kpc ring. This is consistent with the prolonged star formation (SF), spanning $\sim$ 0.5--1 Gyrs, predicted to have occurred in this long-lived feature \citep[e.g.][]{Dalcanton2012, Lewis2015}. The blue H$\alpha$ stars seem to show a relative increase compared to the SPLASH sample at the inner and outer most regions, consistent with the 5-kpc ring 15-kpc ring respectively. It has been shown the 15-kpc ring has had a relatively recent SF burst, starting $\sim$ 80 Myrs ago \citep{Lewis2015}, consistent with the increased number of younger H$\alpha$ stars. 

Due to the small numbers of individual H$\alpha$ star within their classes, excluding MS stars, it is difficult to describe trends in their distribution with certainty. In the middle panel of Fig. \ref{fig:Rhist}, we do see that the majority of the T-MS stars lie at R$_{deproj}$ $<$ 12 kpc. This could be due to more evolved MS stars existing in this region, consistent with slower and more extended SF \citep{Lewis2015}, but could also relate to increased dust that reddens the MS stars. As shown in the top panel of Fig. \ref{fig:Rhist}, the AGB and RHeB stars have a similar distribution across M31 with the relative peak in the 10-kpc ring appearing in both classes. As the AGB C*s are the smallest sub-group of H$\alpha$ stars, drawing conclusions on their distribution is done so with caution. There appears to be a relative increase in the number of AGB C* H$\alpha$ stars around the 15-kpc ring, as compared to the total number of SPLASH objects. As there is thought to have been a relatively recent SF burst, the presence of this older population of lower mass stars is less consistent. However, when looking at the AGB C* distribution in the 2D projection (Fig. \ref{fig:spatial}), the AGB C*s are evenly spread throughout the outer region. This could mean these are halo stars projected onto the disc, or they could be the remnants of an older and more sparse population formed over a longer period of more quiescent SF, still in agreement with SF history (SFH) maps of the PHAT sample \citep{Lewis2015}.

\subsection{Composite spectra}
\label{sec:comp}

To inspect the spectral qualities of the five H$\alpha$ star classifications we identified, we made composite spectra (i.e. mean spectra) of all those we selected in each group. The composite spectra shown in Figs. \ref{fig:MS} to \ref{fig:CS} are normalised to the continuum local to the H$\alpha$ emission line and smoothed by 10 pixels. We identify the H$\alpha$ emission line (dotted line) and show the broad telluric absorption features (magenta bands). For the SPLASH spectra, beyond the telluric feature at $\sim$ 9300 $\text{\AA}$ the spectrum is contaminated by 2nd order effects but we do not use any of this region in our analysis.

A composite spectrum of all the 146 MS H$\alpha$ stars is shown in Fig.~\ref{fig:MS}. The prominent H$\alpha$ emission line at 6563 $\text{\AA}$, shown by the dotted line, is the strongest spectral feature. The continuum is relatively smooth with broad sky absorption and some bad sky subtraction features. The composite spectrum of the 17 T-MS H$\alpha$ stars in Fig.~\ref{fig:TMS} has many similar properties to the MS stars. It has a slightly bumpier continuum but this could well be an effect of the small numbers of T-MS stars. However, the H$\alpha$ line is noticeably weaker in the T-MS stars. 

The composite spectrum for the 25 RHeB H$\alpha$ stars is shown in Fig.~\ref{fig:RH} and for the 25 non-C-rich AGB is in Fig.~\ref{fig:AGB}. Despite their different evolutionary paths, these stars share similar properties and at the resolution of the SPLASH sample, cannot be spectroscopically distinguished \citep{Melbourne2012, Dalcanton2012}. The spectra of RHeB and AGB stars show strong TiO absorption features (faint bands). A composite spectrum of the 11 AGB C* H$\alpha$ stars is shown in Fig.~\ref{fig:CS}. Two carbon features are present in the spectrum; C$_2$ at $\sim$~6100--6600 $\text{\AA}$, and CN at $\sim$~7900--8200 $\text{\AA}$, shown by the darker bands.

\begin{table*}
\caption{Summary of properties and median values for the five classifications of H$\alpha$ stars in our catalogue: MS, T-MS, RHeB, non-C-rich (C/O $<$ 1), and C-rich (C/O $>$ 1) AGB stars. The rows contain: number of H$\alpha$ stars, proportion of their classification to emit H$\alpha$, median H$\alpha$ EW and $\sigma_{v}$ values, median magnitudes in the six HST filters (where data exists) and errors.\small}
\centering
\label{tab:avg}
\resizebox{\textwidth}{!}{%
\begin{tabular}{lrllrllrllrllrll}
\toprule
\multirow{2}{*}{Classification} & \multicolumn{3}{c}{\multirow{2}{*}{MS}} & \multicolumn{3}{c}{\multirow{2}{*}{T-MS}} & \multicolumn{3}{c}{\multirow{2}{*}{RHeB}} & \multicolumn{6}{c}{AGB}  \\ \cmidrule{11-16}
  & \multicolumn{3}{c}{}& \multicolumn{3}{c}{}  & \multicolumn{3}{c}{}  & \multicolumn{3}{c}{C/O $<$ 1}  & \multicolumn{3}{c}{C/O $>$ 1}\\\hline
Number& \multicolumn{3}{c}{$146$} & \multicolumn{3}{c}{$17$}  & \multicolumn{3}{c}{$25$}  & \multicolumn{3}{c}{$25$} & \multicolumn{3}{c}{$11$}  \\\hline
Proportion  & \multicolumn{3}{c}{$\sim 12\%$}& \multicolumn{3}{c}{$\sim 5\%$}  & \multicolumn{3}{c}{$\sim 4\%$}  & \multicolumn{3}{c}{$\sim 2\%$} & \multicolumn{3}{c}{$\sim 14\%$} \\\hline
H$\alpha$ EW ($\text{\AA}$) & $-14.4$ & $\pm$& $1.1$& $-5.2$ & $\pm$ & $0.9$  & $-3.8$  & $\pm$ & $1.0$ & $-6.2$ & $\pm$& $1.5$& $-3.6$& $\pm$& $1.3$\\\hline
H$\alpha$ $\sigma_v$ (km s$^{-1}$)& $112$ & $\pm$ & $5$ & $74$ & $\pm$ & $11$ & $55$ & $\pm$ & $5$ & $65$ & $\pm$ & $9$ & $59$ & $\pm$ & $12$\\\hline
F275W (mag) & $20.4$  & $\pm$& $0.1$& $22.3$ & $\pm$ & $4.6$  & \multicolumn{3}{c}{-} & \multicolumn{3}{c}{-}& \multicolumn{3}{c}{-} \\\hline
F336W (mag)& $20.4$  & $\pm$& $0.6$& $21.3$ & $\pm$ & $0.3$  & $27.6$  & $\pm$ & $6.4$ & $28.6$ & $\pm$& $7.1$& \multicolumn{3}{c}{-} \\\hline
F475W (mag)& $21.6$  & $\pm$& $0.1$& $21.4$ & $\pm$ & $0.3$  & $23.1$  & $\pm$ & $0.2$ & $25.4$ & $\pm$& $0.3$& $24.5$& $\pm$& $0.3$\\\hline
F814W (mag)& $21.3$  & $\pm$& $0.1$& $20.8$ & $\pm$ & $0.3$  & $20.0$  & $\pm$ & $0.2$ & $20.7$ & $\pm$& $0.1$& $19.8$& $\pm$& $0.2$\\\hline
F110W (mag)& $21.3$  & $\pm$& $0.1$& $20.6$ & $\pm$ & $0.3$  & $19.3$  & $\pm$ & $3.2$ & $18.9$ & $\pm$& $0.2$& $18.5$& $\pm$& $0.2$\\\hline
F160W (mag)& $21.1$  & $\pm$& $0.5$& $20.2$ & $\pm$ & $0.3$  & $18.6$  & $\pm$ & $3.3$ & $18.0$ & $\pm$& $0.2$& $17.3$& $\pm$& $0.2$ \\
\bottomrule
\end{tabular}}
\end{table*}

\subsection{H$\boldsymbol{\alpha}$ EW and velocity dispersion distribution}
\label{sec:vel}

To investigate the physical properties of our sample of H$\alpha$ stars from their spectra, velocity dispersion ($\sigma_{v}$) values were calculated from the width of the H$\alpha$ emission line. The width of the H$\alpha$ lines were calculated from the double Gaussian fit described in \S \ref{sec:sampHa}. The $\sigma_{v}$ of the H$\alpha$ line was then calculated by
\begin{equation} \label{eq:vel}
\sigma_{v} = \left(\frac{\sigma_{H\alpha}}{\lambda_{H\alpha}}\right)\cdot c.
\end{equation} 
Here, $\sigma_{H\alpha}$ is the H$\alpha$ line width, given by the STDEV from \textsc{mpfitfun}, $\lambda_{H\alpha}$ is the wavelength of the H$\alpha$ line, and c is the speed of light. The width of the emission line is a measure of the Doppler broadening that occurs when moving material emits photons, i.e. a stellar wind or the pulsating outer layers of a star. Line broadening also depends on the spectral resolution of the instrument used for observations. 

\begin{figure*}
\centering
\subfloat[][]{
	\includegraphics[width=0.75\textwidth , trim=70 370 63 90,clip]{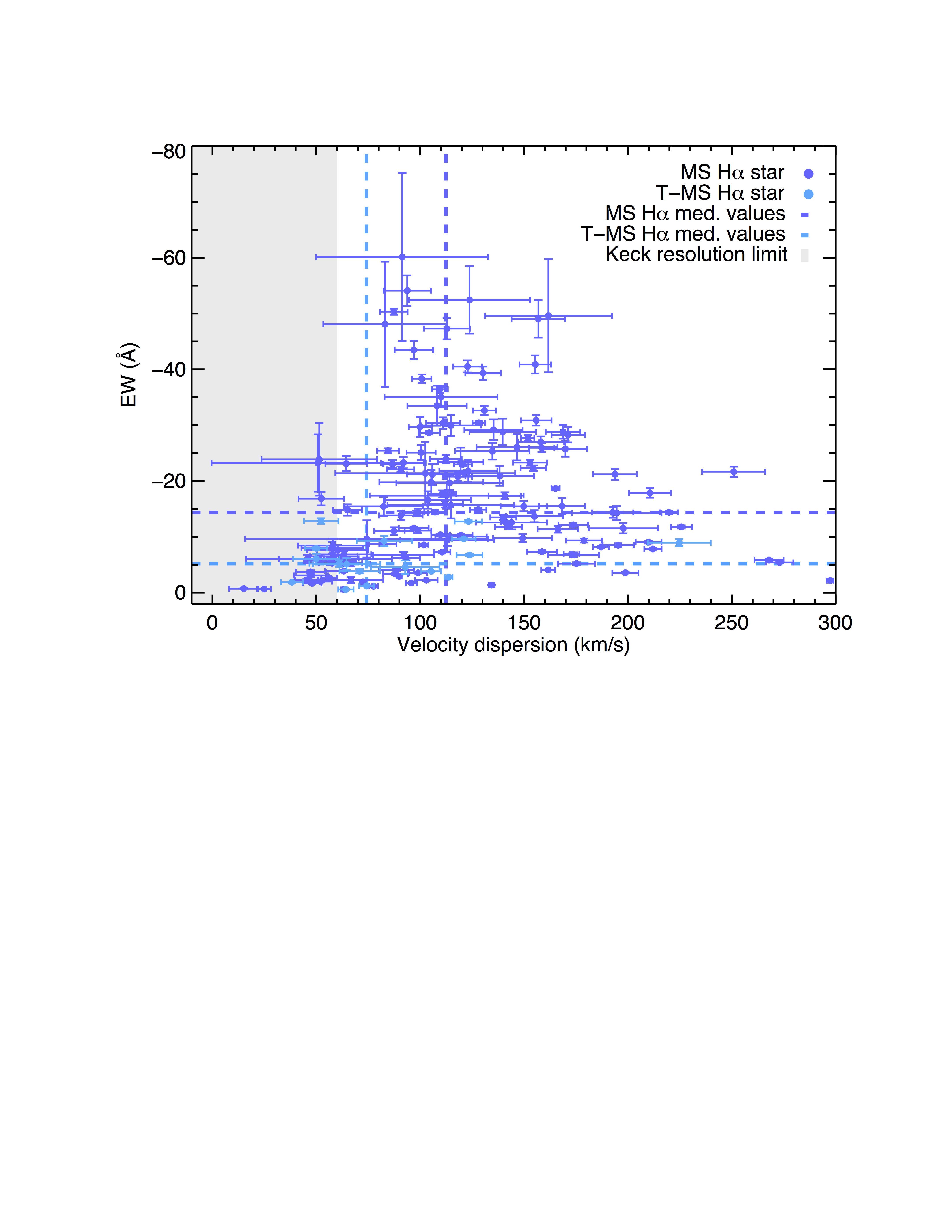}\label{ewvel_ms}
	}\\
\subfloat[][]{
	\includegraphics[width=0.75\textwidth , trim=70 370 63 90,clip]{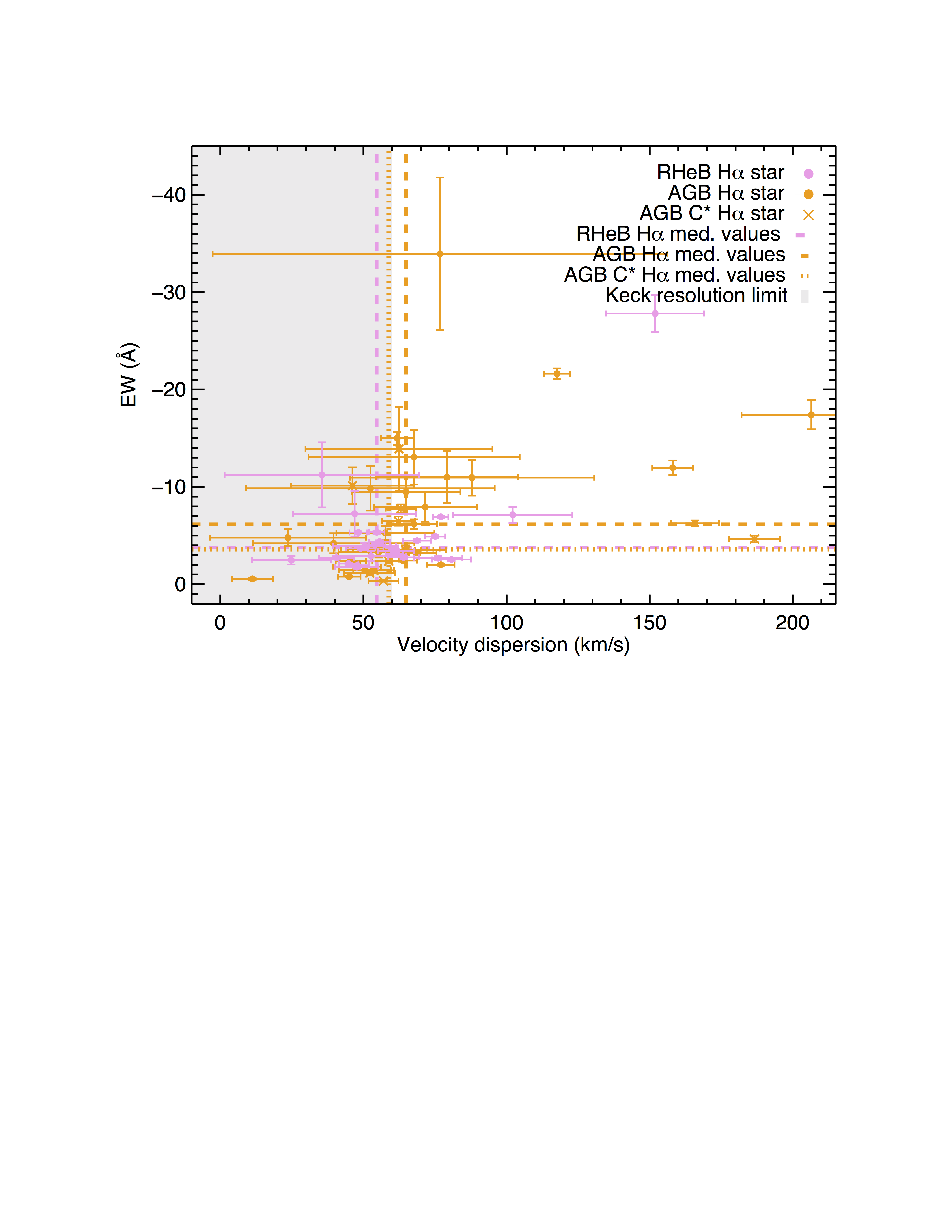}\label{ewvel_rs}
	}
\caption[]{H$\alpha$ line strength (EW) vs. $\sigma_{v}$ of H$\alpha$ stars. Shaded region in both panels is approximate spectral resolution limit for Keck II/DEIMOS ($\sim$ 60 km s$^{-1}$), however this can vary so some $\sigma_{v}$ values below this limit are resolved. Median EW and velocity values for the H$\alpha$ stars are indicated by the dashed horizontal and vertical lines respectively.
\protect\subref{ewvel_ms} MS H$\alpha$ stars (dark blue) and T-MS H$\alpha$ stars (pale blue).
\protect\subref{ewvel_rs} RHeB H$\alpha$ stars (pink), non-C-rich AGB H$\alpha$ stars (orange points), and C-rich AGB H$\alpha$ stars (orange Xs). The majority of these stars have $\sigma_{v}$ values that are unresolved. See \S \ref{sec:vel} and Table \ref{tab:avg}.\small}
\label{fig:ewvel}
\end{figure*} 

\begin{figure*}
\subfloat[][]{
	\includegraphics[width=0.46\textwidth,  trim=45 35 82 75,clip]{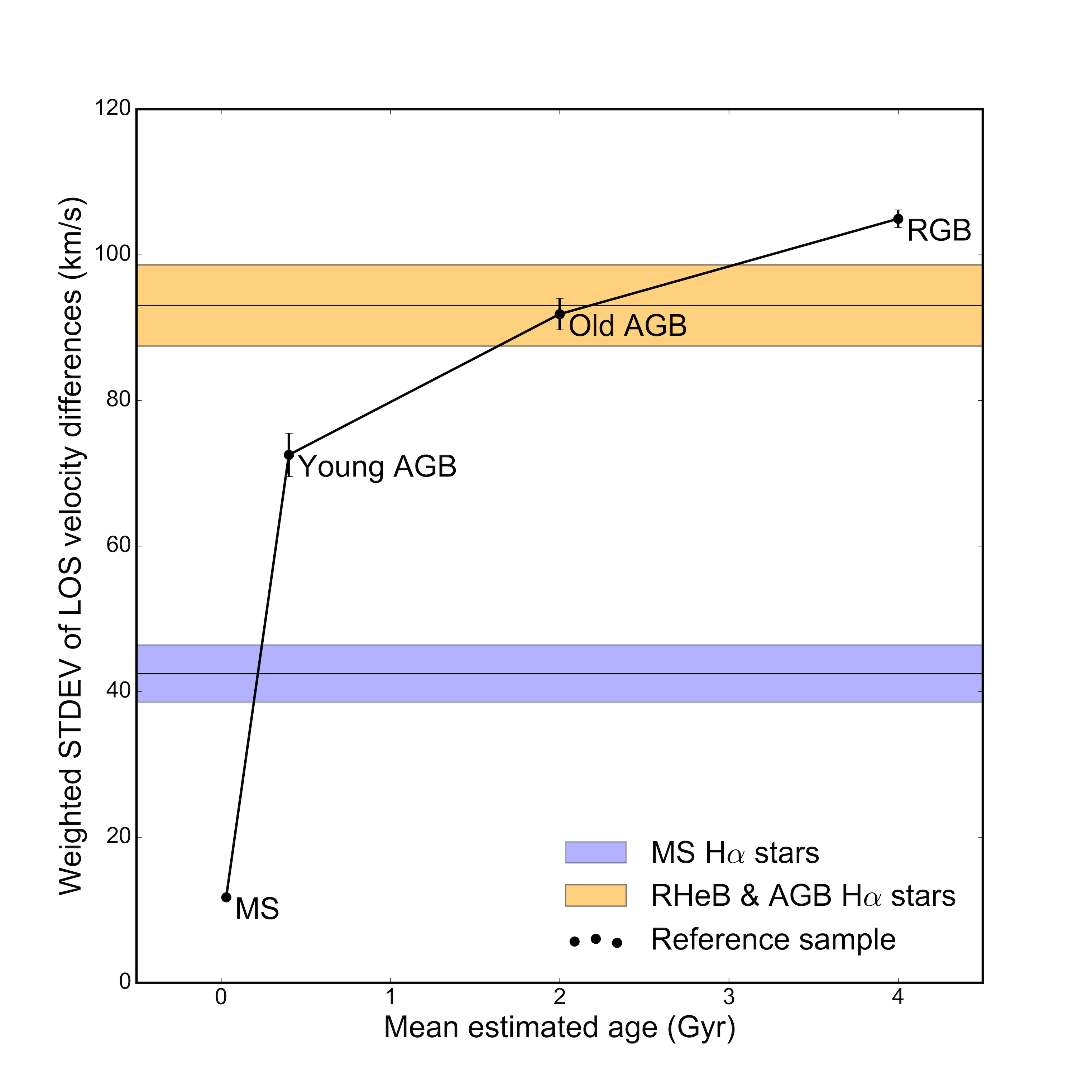}\label{stdev}
	}
\subfloat[][]{
	\includegraphics[width=0.46\textwidth,  trim=45 35 82 75,clip]{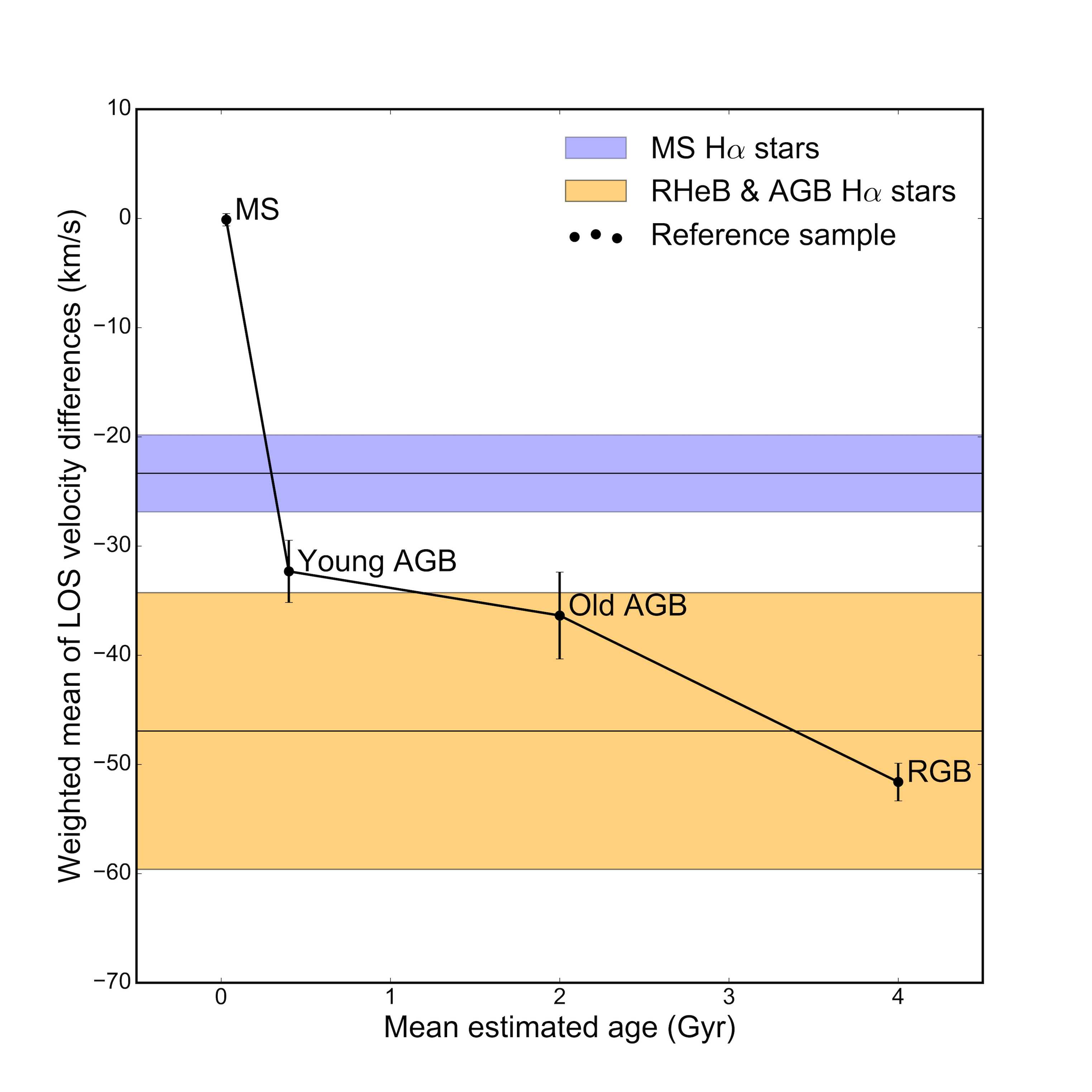}\label{mean}
	}
\caption[]{\protect\subref{stdev} LOSVD vs. mean estimated age for reference sample of normal MS, bright/young AGB, faint/old AGB, and RGB stars (black points) and MS H$\alpha$ stars (blue band), and RHeB and AGB H$\alpha$ stars (orange band). LOSVD calculated as the weighted STDEV of LOS velocity differences and is an indicator of age.
\protect\subref{mean} Same as left panel, except y-axis is the weighted mean (rather than weighted STDEV) of the LOS velocity differences. Similar trends are observed. See \S \ref{sec:kin}.\small}
\label{fig:sipkin}
\end{figure*}

The distribution of $\sigma_{v}$s and EWs for the different classes of H$\alpha$ stars is shown in Fig.~\ref{fig:ewvel}. Fig.~\subref*{ewvel_ms} shows MS (dark blue) and T-MS (pale blue) H$\alpha$ stars, and the RHeB (pink points), AGB (orange points), and AGB C* (orange crosses) H$\alpha$ stars are shown in Fig.~\subref*{ewvel_rs}. In both panels, the shaded region represents the approximate limit of Keck II/DEIMOS resolution. This value, around $\sim$~60~km s$^{-1}$, can vary, hence why the $\sigma_{v}$ for some stars is below this limit, while the majority in this shaded region are unresolved. However, in this shaded region, the EW and thus the H$\alpha$ detection is unaffected. In both plots, the horizontal and vertical dashed lines represent the median values of EW and $\sigma_{v}$ respectively for each class of H$\alpha$ star. A summary of these median values is given in Table \ref{tab:avg} along with additional average values of each class.
 
The MS stars show a relatively broad range of $\sigma_{v}$ values that go up to $\sim$~300~km~s$^{-1}$, with a median of 112~$\pm$~5~km~s$^{-1}$ (dark blue dashed vertical line). The MS H$\alpha$ stars show a large spread of EW values going up to $\sim$~$-$60~$\text{\AA}$. The median value of the MS H$\alpha$ star EW (dark blue dashed horizontal line) is $\sim$~$-$14.4~$\pm$~1.1~$\text{\AA}$. The T-MS H$\alpha$ stars occupy a much smaller region of the plot with lower median $\sigma_{v}$ values (pale blue dashed vertical line) of 74~$\pm$~11~km~s$^{-1}$ and median EW $\sim$~$-$5.2~$\pm$~0.9~$\text{\AA}$ (horizontal line). As could also be seen from the composite spectrum in Fig.~\ref{fig:TMS}, the T-MS H$\alpha$ stars show weaker emission.

In Fig.~\subref*{ewvel_rs} the AGB stars show the largest spread of EWs and $\sigma_{v}$ values. The red H$\alpha$ stars have similar median $\sigma_{v}$ values (vertical lines); RHeBs at 55~$\pm$~5~km~s$^{-1}$ (pink dashed line), AGBs at 65~$\pm$~9~km~s$^{-1}$ (orange dashed line), and AGB C*s at 59~$\pm$~12~km~s$^{-1}$ (orange dotted line). However, all of these median $\sigma_{v}$ values are on the cusp of the approximate limit of Keck II/DEIMOS resolution and so are mostly unresolved. The non-C-rich AGB stars have a larger median |EW| value of $\sim$~$-$6.2~$\pm$~1.5~$\text{\AA}$ (orange dashed horizontal line). The RHeBs and AGB C*s both have similar and weaker median H$\alpha$ emission at $\sim$~$-$3.8~$\pm$~1.0~$\text{\AA}$ and $\sim$~$-$3.6~$\pm$~1.3~$\text{\AA}$ respectively.

\subsection{H$\boldsymbol{\alpha}$ star kinematics}
\label{sec:kin}

A recent study using the SPLASH and PHAT samples investigated the line-of-sight stellar velocity dispersion (LOSVD) of stars in M31 as a function of their age and metallicity \citep{Dorman2015}. Splitting stellar populations in to four categories based on their position in the optical PHAT CMD, they then calculated the ages of these populations based on simulated CMDs. The four categories were young MS, bright/young AGB, faint/old AGB, and RGB stars, with mean ages were 30 Myrs, 0.2 Gyrs, 0.4 Gyrs, and 2 Gyrs respectively. The LOSVD for each of the four age bins was determined by measuring the dispersion of each population within radial bins (of 200$\arcsec$ up to 275$\arcsec$). Stars with less than 15 neighbours within the radial bin were dropped from the analysis as their calculated LOSVD and error were deemed unreliable. They found that the older populations were dynamically hotter than the younger populations, ranging from 30 km s$^{-1}$ for MS stars to 90 km s$^{-1}$ for RGB stars outward from R$_{deproj}$ $\sim$ 10 kpc. This linear increase in LOSVD indicated a continuous or recurring process contributed to the evolution of the disc, whereby the older populations have had more time to be perturbed from their original velocity. 

\begin{figure*} 
\subfloat[][]{
	\includegraphics[width=0.49\textwidth , trim=85 370 70 95,clip]{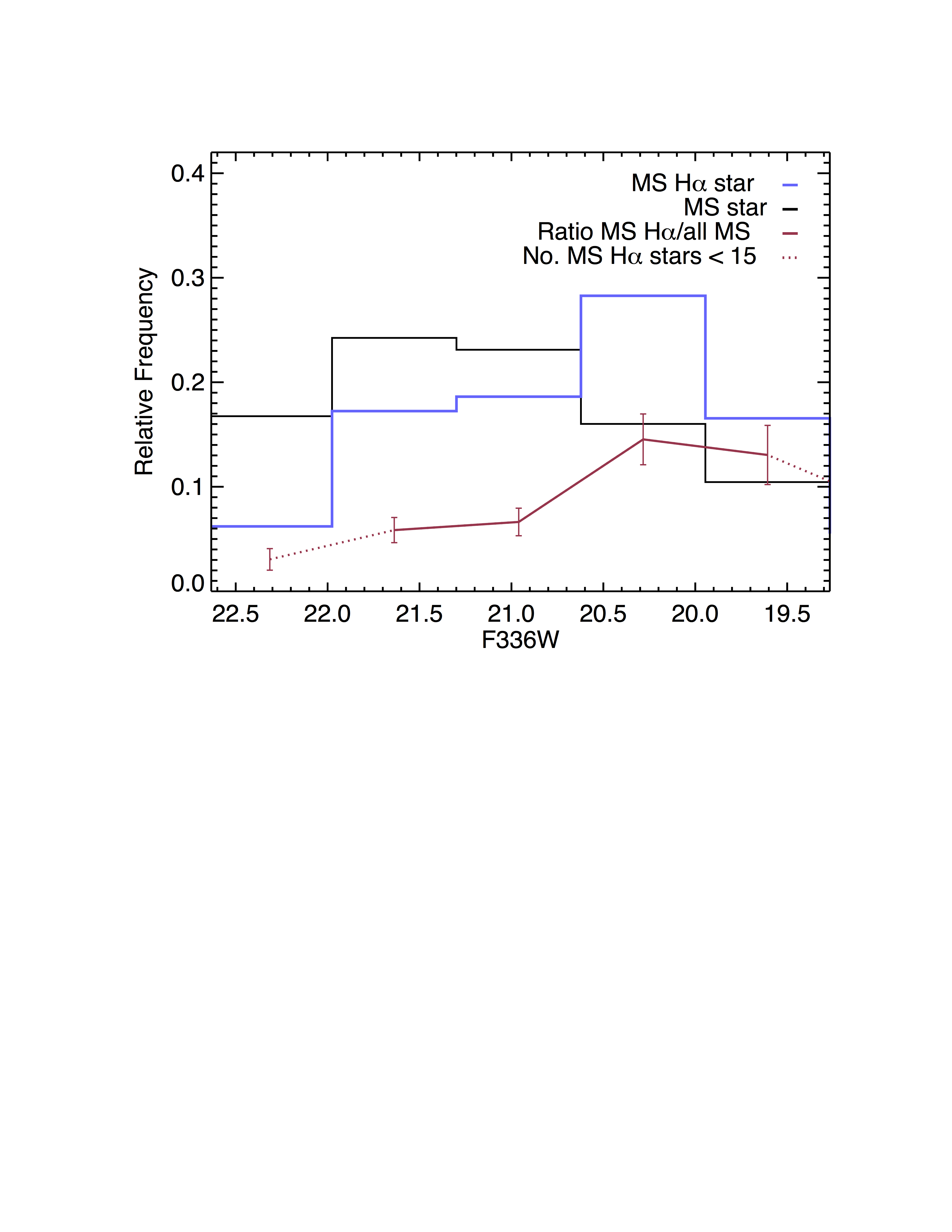}\label{Behist_336}
	}
\subfloat[][]{
	\includegraphics[width=0.49\textwidth , trim=85 370 70 95,clip]{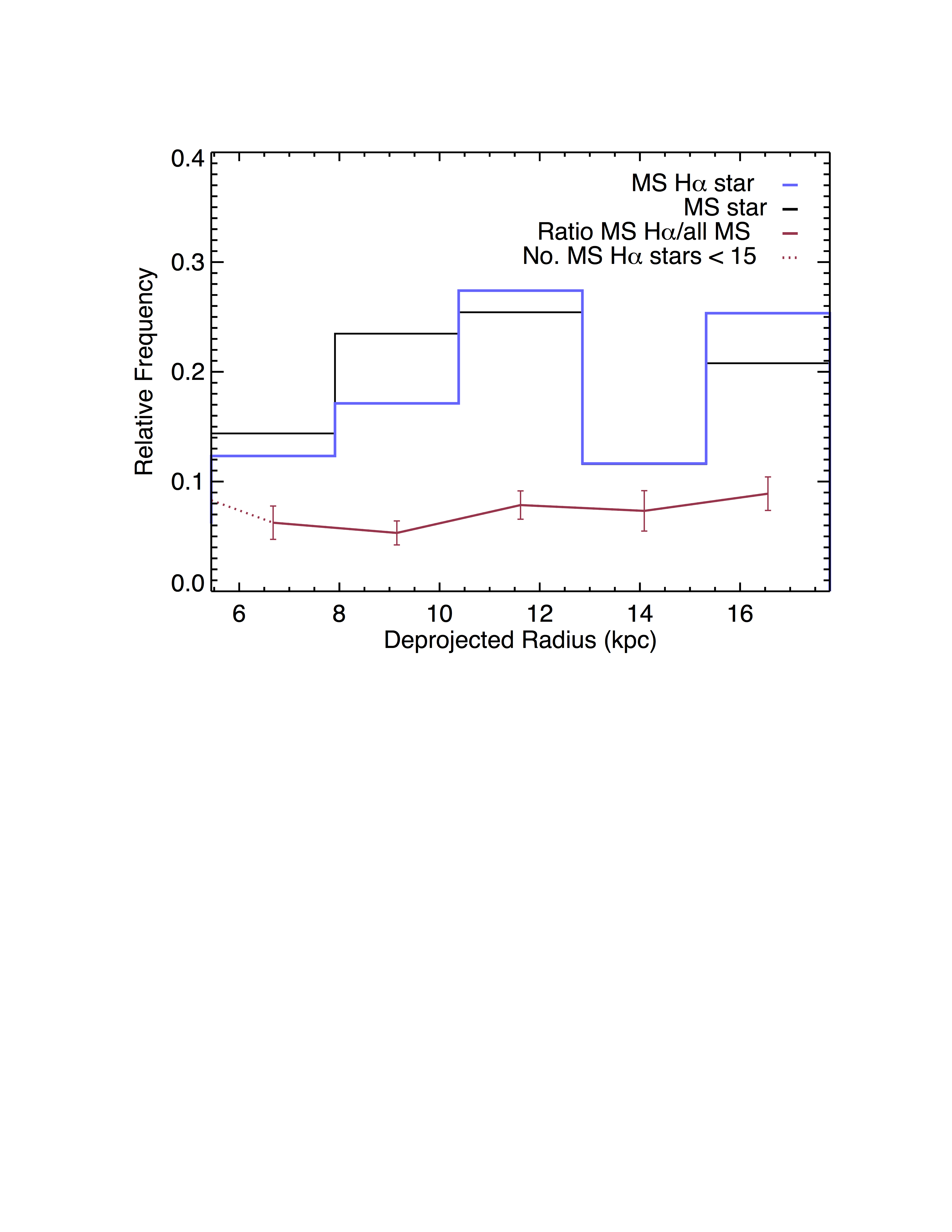}\label{Behist_dpr}
	}
\caption[]{Histograms of the MS H$\alpha$ stars (blue) to all MS stars (black) ratio (i.e. Be/(Be~+~B)) as a function of F336W magnitude \protect\subref{Behist_336} and R$_{deproj}$ \protect\subref{Behist_dpr}. Ratio of the two shown by the red line. Bins with $>$ 15 MS H$\alpha$ stars are shown by the solid red line. Values of the Be/(Be~+~B) ratio are given for these bins in Table \ref{table:Be}. See \S \ref{sec:rat}.\small}
\label{fig:hist}
\end{figure*}

We used a modified version of the kinematic analysis method done by \cite{Dorman2015} in order to predict the ages of the H$\alpha$ MS (not including the T-MS) stars and the red H$\alpha$ stars. Fig.~\ref{fig:sipkin} shows the reference SPLASH sample separated in to the four age bins (black points), and the MS and red H$\alpha$ stars represented by the horizontal blue and orange bands respectively. Here we have combined the RHeBs, AGBs, and AGB C*s in to the red stars category due to their limited numbers, we do note that this will then increase the error margins on the calculated ages. Because H$\alpha$ stars are relatively sparse, the LOSVDs were calculated as LOS velocity differences between the weighted mean of MS star neighbours (within r $<$ 275$\arcsec$) and the LOS velocity of the H$\alpha$ stars, instead of the weighted STDEV of H$\alpha$ neighbours. The MS population is used to define the velocity reference frame because it is the most well-ordered (dynamically cold) of M31's stellar populations. The mean was weighted inversely proportional to the square of the velocity measurement errors. LOSVDs were calculated for MS, bright/young AGB, faint/old AGB, and RGB age bins using the reference SPLASH sample. In accordance with the results from \cite{Dorman2015}, LOSVD increased with the age of the stellar population. H$\alpha$ MS stars (blue bands) had LOSVDs similar to bright/young AGB stars, and the redder H$\alpha$ stars (RHeBs, AGBs, and AGB C*s; orange bands) had LOSVDs similar to faint/old AGB stars. The ages of H$\alpha$ stars appear to correspond with that of their non-emission counterparts.

\begin{table}
\centering
\caption{Be/(Be~+~B) values at different values of F336W magnitude and R$_{deproj}$ (see Fig.~\ref{fig:hist}). Shown in the table are only those bins that contain $>$ 15 H$\alpha$ stars, $\sim$~10 per cent of the MS H$\alpha$ sample.}
\begin{tabular}{|c|c|rll|}
\toprule
\multicolumn{2}{c}{Varied Quantity}  & \multicolumn{3}{c}{Be/(B~+~B)} \\\midrule
\multirow{4}{*}{F336W}	& $21.6$ & $6\%$& $\pm$  & $1\%$  \\\cmidrule{2-5}
						& $21.0$ & $7\%$& $\pm$  & $1\%$  \\\cmidrule{2-5}
						& $20.3$ & $15\%$& $\pm$  & $2\%$  \\\cmidrule{2-5}
						& $19.6$ & $13\%$& $\pm$  & $3\%$  \\\midrule
\multirow{5}{*}{Deprojected Radius (kpc)} & $6.7$  & $6\%$& $\pm$  & $2\%$  \\\cmidrule{2-5}
						& $9.1$  & $5\%$& $\pm$  & $1\%$  \\\cmidrule{2-5}
						& $11.6$ & $8\%$& $\pm$  & $1\%$  \\\cmidrule{2-5}
						& $14.1$ & $7\%$& $\pm$  & $2\%$  \\\cmidrule{2-5}
						& $16.6$ & $9\%$& $\pm$  & $2\%$  \\\bottomrule
\label{table:Be}
\end{tabular}
\end{table}

\subsection{Be/(Be~+~B) variations}
\label{sec:rat}

Studies of the ratio of Be stars to all B stars (Be/(Be~+~B)) have shown it is dependent on ages of stars \citep{Fabregat2000} and the metallicity of their environment \citep{Maeder1999, Martayan2007}. Due to the size of the sample of MS H$\alpha$ stars we have selected, the range of different environments, and spatial extent covered by the data sets, we were able to investigate how this ratio changes in M31. Fig.~\ref{fig:hist} shows histograms of the relative frequencies of Be stars and all B stars, and how their ratio varies as a function of F336W magnitude and R$_{deproj}$. The solid line red line shows how the Be/(Be~+~B) ratio varies as a function of the x-axis for bins with $>$ 15 ($\sim$~10 per cent of all MS H$\alpha$ stars) Be stars in a bin. The values of the Be/(Be~+~B) ratio for these different histograms are given for all those bins with $>$ 15 Be stars in Table \ref{table:Be}.

Fig.~\subref*{Behist_336} shows how the Be/(Be~+~B) ratio varies with F336W magnitude, which can also be interpreted as stellar mass but that also has a metallicity dependence. The ratio significantly increases between the second and third bin ($\sim$ 21.0--20.3 mag). This could be a metallicity dependence on the ratio, as seen in other studies, or a dependence of this ratio on stellar mass, i.e. higher mass MS stars are more likely to emit H$\alpha$.

To investigate the dependence of the Be/(Be~+~B) ratio on environment, we investigate how it varies as a function of R$_{deproj}$ from the centre of M31, which spans a variety of different metallicity regions. Although there is no obvious overall trend of the ratio with radius, there is a slight increase of the ratio at larger radii than the centre. This would follow the metallicity gradient of M31 \citep[e.g.][]{Zurita2012, Gregersen2015}, as the Be/(Be~+~B) fraction is higher in lower metallicity environments.

\section{Discussion}
\label{sec:disc}

The PHAT and SPLASH data show the H$\alpha$ stars fall in to five classes: bright MS, T-MS, RHeB, AGB, and AGB C* stars. Despite the fact that these stars all exhibit H$\alpha$ emission, their evolution, properties, and mechanisms of emission are very different. In this section we will discuss the five classifications of H$\alpha$ stars separately in \S \ref{sec:MS} to \S \ref{sec:CS} respectively.

\subsection{MS H$\boldsymbol{\alpha}$ stars}
\label{sec:MS}

Of the bright MS stars in the sample, $\sim$~12 per cent of them were classified as H$\alpha$ stars. Due to the rarity of O-type stars, we assume that the MS part of the sample contains solely B-type stars for simplicity. Looking at the spatial distribution of the Be stars, we found that there was a relative increase in their number compared to the SPLASH sample at R$_{deproj}$ of $\sim$ 15 kpc (see Figs. \ref{fig:spatial} and \ref{fig:Rhist}). This presence of these young massive stars is consistent with the 15-kpc star forming ring of M31 that was found to have had a relatively recent SF burst, starting $\sim$ 80 Myrs ago \citep{Lewis2015}.

Due to the metallicity dependence of the Be/(Be~+~B) ratio, a previous study investigated a range of environments in the LG with different metallicities. Be/(Be~+~B) ratios of 11, 19, 23, and 39 per cent were determined for star clusters that decreased in metallicity, in the inner and outer Galaxy, LMC and SMC respectively. The `inner MW' was defined as star clusters at galactocentric radii of $<$~8~kpc and `outer MW' as $>$~8~kpc \citep{Maeder1999}. We found an average Be/(Be~+~B) proportion of $\sim$~12 per cent in M31, i.e. both disc and halo stars, out to R$_{deproj}$~$\sim$~20 kpc from the centre of M31. To ensure that we were comparing the samples accurately, we compared the absolute magnitude range used by \cite{Maeder1999} to our sample, and we found that we were probing the same stellar population at the distance of M31. This average value for Be/(Be~+~B) in the inner and outer regions of M31 is consistent with the $\sim$~11 per cent found in just the metal-rich inner region of the MW. When looking at the histogram in Fig.~\subref*{Behist_dpr}, and the values in Table \ref{table:Be} for the bins with $>$ 15 MS H$\alpha$ stars, the values range from 6 $\pm$ 2 per cent to 9 $\pm$ 2 per cent, for R$_{deproj}$ of 6.7 and 16.6 kpc respectively. The average of these values between those radii would give a Be/(Be~+~B) ratio of 7 $\pm$ 2 per cent. This ratio of Be/(Be~+~B) stars in M31 could indicate that it has a higher metallicity than the MW. This would agree with other studies that suggest M31 is likely to be more metal rich than the MW, and may have had a more recent and active accretion history of satellites galaxies \citep[e.g.][]{Kalirai2006, Brown2006}. A more turbulent evolution could have catalysed more star formation in M31, increasing its overall metallicity.

Given the diverse range of environments our Be sample spans in M31, these stars are ideal candidates for follow up observations. Monitoring the variability of the H$\alpha$ emission seen in these stars across these environments will allow us to better understand its nature. Cross matching the sample of Be and B stars to the SFH maps from \cite{Lewis2015}, and metallicity maps of M31, may break the degeneracy in the age-metallicity dependence of the Be phase.

\subsection{T-MS H$\boldsymbol{\alpha}$ stars}
\label{sec:TMS}

We identified T-MS H$\alpha$ stars as those not within the MS stellar bin on the UV/optical CMD in Fig.~\subref*{uvcmd} but not associated with the BHeB phase of evolution as shown by the histogram in Fig.~\ref{fig:TPhist}. We found 17 T-MS H$\alpha$ stars, and show that this makes up $\sim$ 5 per cent of this `transitioning' population. It is likely that these are Be stars evolving off the MS. High mass MS stars evolve rapidly after H burning due to the quick contraction in the core and expansion in the envelope. Stars can traverse the CMD from MS to massive red stars in $\sim$ 1 Myr \citep{Hayashi1962}, hence why the T-MS bin is sparsely populated. Their expansion causes the rotation of their envelope to slow as they conserve angular momentum. As the Be stars could emit H$\alpha$ via a rotationally flattened disc \citep[e.g.][]{Struve1931, Dougherty1992}, if stellar rotation slows, the H$\alpha$ emission from this circumstellar disc would fade as a star evolves off the MS. However, these stars could also be reddened MS H$\alpha$ stars that appear to have left the MS in UV/optical colour space.

As seen in our sample of H$\alpha$ stars, these T-MS stars have characteristically weaker H$\alpha$ emission as shown by their composite spectra (Fig.~\ref{fig:TMS}), distribution of EW values (Fig.~\subref*{ewvel_ms}), and median population values of EW and $\sigma_v$ (Table \ref{tab:avg}), as compared with the MS H$\alpha$ stars. Being cautious not to draw any strong conclusions from their spatial distribution due to their small number, we do see that the majority of T-MS stars lie within R$_{deproj}$ = 12 kpc (Figs. \ref{fig:spatial} and \ref{fig:Rhist}). The presence of these perhaps older more evolved MS stars is then consistent with the prolonged SF in this region \citep{Dalcanton2012, Lewis2015}, but could also be more reddened by an increase in the dust within this region. Given our multi-faceted data set, separating the MS and BHeB stars was possible in the UV/optical CMD \citep{Dohm-Palmer1997, Dohm-Palmer2002}, and the Keck/II DEIMOS spectra allowed us to resolve H$\alpha$ emission for these stars in M31. As a result of combining the PHAT and SPLASH data we have been able to identify this population but do not have many comparative samples against which to compare our T-MS H$\alpha$ stars. However, for the first time we may have captured the evolution of Be stars off the MS as they traverse the CMD towards the RHeB sequence.

\subsection{RHeB H$\boldsymbol{\alpha}$ stars}
\label{sec:RH}

We separated the RHeB stars from the AGB stars in our sample using the far-red/NIR CMD \citep{Melbourne2012}, and using the colour cut defined by \cite{Rosenfield2016}. The stars we identified as RHeB H$\alpha$ stars are likely to be RHeB stars that are evolving up their Hayashi lines and becoming AGB stars. Compared to their `normal' population, we found $\sim$ 4 per cent of the RHeB stars show H$\alpha$ emission. The H$\alpha$ emission we detect in the RHeB stars is weaker than those of the AGB stars, as can be seen by their distribution of EW values (Fig.~\subref*{ewvel_rs}), and corresponding median population values (Table \ref{tab:avg}). As seen in Figs. \ref{fig:spatial} and \ref{fig:Rhist}, the RHeB stars show a relative increase in their numbers as compared to the total SPLASH sample at the 10-kpc star forming ring of M31. This increase in these older stars support the prolonged SF activity, up to 1 Gyr, in the 10-kpc ring \citep{Dalcanton2012, Lewis2015}. It is likely that many RHeB stars are misclassified as AGB stars in other studies without the far-red/NIR CMD to separate them. However, given the fact that these stars both show emission of H$\alpha$ and that at this evolutionary stage, the lower- and higher-mass stars go through qualitatively similar evolution, we predict that these RHeB stars with H$\alpha$ emission also fall under the LPV classification, and are either SRs or Miras.

There has been some discrepancy in the literature over what the ratio of BHeB to RHeB stars should be, and what is observed \citep[e.g.][and references therein]{Gallart2005}. The number of BHeB stars we found in our sample, as defined in the thin grey polygon in Fig.~\subref*{uvcmd} is $\sim$~640. By accounting for the density of the `normal' T-MS stars that could be crossing the BHeB sequence, we estimate the total number of BHeBs to be $\sim$~500. The number of RHeB stars in the `normal' RHeB bin defined in Fig.~\subref*{opnircmd} is $\sim$~360. As we know from CMD space and evolutionary tracks, RHeB stars exist outside of our defined `normal' polygon but we avoided the RGB stars to reduce contamination. Using the ratio of RHeB H$\alpha$ stars within and outside of the RHeB bin, we predict there are $\sim$ 560 RHeB stars in total. These numbers are crude approximations but we do find there are similar numbers of RHeBs to BHeBs as is also seen in other studies \citep{McQuinn2011, Melbourne2012}.

\subsection{AGB H$\boldsymbol{\alpha}$ stars}
\label{sec:AGB}

Only $\sim$~2 per cent of the non-C-rich AGB stars in the SPLASH sample exhibit H$\alpha$ emission. Due to their characteristic Balmer emission, the AGB H$\alpha$ stars in the sample are likely to be LPVs, either SRs or Miras. Given that we do not have \textit{K}-band, multi-epoch observations, we are unable to distinguish between Miras and SRs in our selected H$\alpha$ sample \citep[e.g.][]{Battinelli2014}. Of the two broad LPV categories, Miras are more useful for studying stellar populations and galactic properties. Due to their period-luminosity relationship \citep[e.g.][]{Glass1981, Glass1982}, they can be used as standard candles for intermediate age stellar populations \citep{Whitelock2012}. Identifying them in M31 and improving our understanding of this evolutionary stage will help to refine their use as standard candles, and ultimately improve distance measurements to M31 and other systems. For old to intermediate age populations that lack younger stars, where distance indicators such as Cepheid variables are rare, Mira variables are good alternative standard candles. The selected AGB H$\alpha$ stars are ideal for follow-up observations due to their number and shared environment.

AGB and RHeB stars trace a similar spatial distribution in M31, also exhibiting a relative increase in frequency at the 10-kpc ring (Figs. \ref{fig:spatial} and \ref{fig:Rhist}), consistent with prolonged SF activity in this region \citep{Dalcanton2012, Lewis2015}. Stars in the mass range $\sim$ 0.8 to 8 M$_{\odot}$ have qualitatively similar evolutionary paths when burning He/H in their shells and with a degenerate C-O core. We therefore conclude that the H$\alpha$ emission in both the RHeB population (25 stars) and the stars firmly on the AGB (36 stars, including AGB C*s) are all LPVs. The sample of up to 61 potential Mira variables we have found, is the largest spectroscopically discovered sample of LPVs in M31.

\subsection{AGB C* H$\boldsymbol{\alpha}$ stars}
\label{sec:CS} 

A recent study investigating the C-Mira/C~star proportion in systems of a range of masses and metallicities around the LG had a consistent value of $\sim$ 12 per cent \citep{Battinelli2014}. However, a C-Mira/C~star value of up to 50 per cent was found in the MW halo and Sgr dSph \citep{Battinelli2014, Battinelli2013}. Of the 77 C~stars found in the 600~line~mm$^{-1}$ SPLASH spectra \citep[][see \S \ref{sec:specdata}]{Hamren2015} and red-ward of the colour cut shown in Fig.~\subref*{opnircmd}, 11 of these show H$\alpha$ emission. This proportion of 14 per cent of AGB C*s in M31 exhibiting H$\alpha$ emission is an upper limit to the C-Mira/C~star ratio for M31. If there are SRs in our sample, this value will decrease. The upper limit fits with observations in other LG galaxies but differs significantly from the $\sim$~40--50 per cent found in the MW halo and Sgr dSph \citep[][respectively]{Mauron2014, Battinelli2014, Battinelli2013}. To ensure this was a fair comparison, we compared our NIR filter magnitudes, F110W (\textit{J}-band equivalent) and F160W (\textit{H}-band equivalent), and found that we are probing the same population of stars at the distance of M31 as observed in the MW halo. The cause of this disparity is most likely due to the different environmental effects of each sample. The properties of this population would be interesting to study further.

\section{Conclusion}
\label{sec:conc}

We identified and classified H$\alpha$ stars in M31 using spectroscopy from Keck II/DEIMOS collected for the SPLASH survey and supporting six-filter photometry from HST collected as part of the PHAT survey. We devised a selection algorithm that implemented a set of six selection criteria to find a clean sample of H$\alpha$ stars. From 5295 SPLASH spectra, we selected a sample of 224 H$\alpha$ stars in M31. See Appendix A for a full list of selected H$\alpha$ stars and their properties. We used a combination of CMDs and spectral features in order to classify the H$\alpha$ stars. We have identified five classifications of H$\alpha$ stars; MS, T-MS, RHeB, AGB, and AGB C* stars. The results from the analysis of these five categories of H$\alpha$ star are summarised below.

\begin{itemize}
\item We found 146 MS H$\alpha$ stars, meaning $\sim$~12 per cent of the B-type MS stars were classified as Be stars. The proportion of Be/(Be~+~B) stars has been found to be dependent on the metallicity of their environment. When comparing our results with previous studies, the average proportion of Be to B stars that we found at a range of R$_{deproj}$ $\sim$ 6 to 16 kpc of 7 $\pm$ 2 per cent could indicate that M31 is more metal rich than the MW, supporting the findings of other studies.
\vspace{2mm}

\item We defined a population of H$\alpha$ stars red-ward of the MS on the UV/optical CMD as a T-MS population. These stars had characteristically weaker emission than the MS H$\alpha$ stars, and were not associated with the BHeB star over-density seen in the parent SPLASH sample. We concluded that these 17 T-MS H$\alpha$ stars ($\sim$ 5 per cent of the `normal' T-MS population), are Be stars evolving off the MS and their H$\alpha$ emission fades as the star's outer layers expand and its rotation slows. This could be the first time the evolution of Be stars off the MS has been captured in CMD space. However, we note that these stars could also be reddened MS H$\alpha$ stars. 
\vspace{2mm}

\item Using the far-red/NIR CMD, we were able to separate RHeB stars from AGB stars in our sample. There are approximately a similar number of BHeB to RHeB stars in the SPLASH sample. We found 25 of the RHeB stars showed H$\alpha$ emission, $\sim$ 4 per cent of the RHeB population. Given the similarity in the evolutionary paths of stars in the mass range $\sim$ 0.8 to 8 M$_{\odot}$ beyond the HeB phase, we predict that these RHeB stars are evolving on to the AGB. Due to exhibiting H$\alpha$ emission, we conclude that these stars, that are often misidentified as AGB stars, are undergoing regular dynamic pulsations in their outer layers causing the shock excited Balmer emission. We conclude that the RHeB H$\alpha$ stars are LPVs, and are either Mira or SR variables.
\vspace{2mm}

\item We found a sample of 25 non-C-rich AGB H$\alpha$ stars, making up just $\sim$~2 per cent of the non-C-rich AGB population. The presence of H$\alpha$ emission in the AGB stars means that these stars are LPVs. Of these LPVs, the AGB H$\alpha$ stars could be SRs or Miras. We found a relative excess of AGBs, and all other red stars in our sample as compared to the parent SPLASH sample, around the 10-kpc star forming ring in M31. This supports evidence from other studies of prolonged star formation in this long-lived feature.
\vspace{2mm}

\item We found 11 AGB C* H$\alpha$ stars, which is a proportion of $\sim$ 14 per cent of the C~stars in the SPLASH sample that exhibit H$\alpha$ emission, which is an upper limit for the C-Mira/C~star ratio in M31. This is consistent with many other galaxies in the LG, although this value is likely to be lower than this due to the presence of SRs. Regardless, this value differs significantly from the $\sim$~40--50 per cent C-Mira/C~star proportion in the MW halo and Sgr dSph, and this disparity is worth additional investigation. 
\vspace{2mm}

\item Combining the total number of RHeB, and AGB stars, both O and C rich, we found a sample of 61 LPV stars in our sample. This is the largest spectroscopically identified sample of LPVs in M31. This catalogue can be used to improve the use of Mira variables as distance indicators for intermediate-age populations of stars and the potential C-Miras can be used to investigate the properties of M31. 
\end{itemize}

\section*{Acknowledgements}
\noindent L.J.P. wishes to thank collaborators at the University of California Santa Cruz (UCSC) for use of their data and for their supervision, and thanks the University of Leeds for supporting the collaboration with UCSC for this project. L.J.P. also thanks collaborators at Leeds, UCSC, and Oxford for useful discussion and support. L.J.P is supported by a Hintze Scholarship, awarded by the Oxford Centre for Astrophysical Surveys, which is funded through generous support from the Hintze Family Charitable Foundation. L.J.P. acknowledges financial support from the Royal Astronomical Society Research and Grants Fund, supervisors at UCSC and University of Oxford, and Christ Church, Oxford. P.G., K.M.H., and C.E.D. were supported by NSF grant AST-1412648 and NASA/STScI grant HST-GO-12055. G.A.D., A.I., and M.I. carried out their work as part of the Science Internship Program at UCSC. We are grateful to the anonymous referee for useful comments that helped improve the clarity of the paper.

\bibliography{HaStarsM31bib}

\clearpage

\onecolumn
\appendix
\begin{landscape}
\section{Catalogue of H$\boldsymbol{\alpha}$ Stars}
\label{sec:appA}

{ \footnotesize \textbf{Table A1:} List of selected H$\alpha$ stars. The columns contain: object name, H$\alpha$ star classification, DEIMOS mask name, mask slit number, right ascension (RA), declination (Dec), H$\alpha$ line EW and error, $\sigma_v$ and error, Heliocentric velocity (v$_{Hel}$), R$_{deproj}$, and the PHAT six-filter photometry. The object names sometimes conflict with the classifications as they are based on an earlier misclassification. Object names in a box relate to the objects that we present a section of their 2D SPLASH spectrum in Appendix \ref{sec:appB}.}

\FloatBarrier
\begin{table*}
\label{table:Ha}
\centering
\resizebox{1.35\textwidth}{!}{%
\begin{tabular}{lcccllrllrllcccccccc} 
\toprule
\multicolumn{1}{|c|}{Object} & \multicolumn{1}{c|}{Class.} & \multicolumn{1}{c|}{Mask} & \multicolumn{1}{c|}{Slit No.} & \multicolumn{1}{c|}{RA} & \multicolumn{1}{c|}{Dec} & \multicolumn{3}{c|}{H$\alpha$ EW ($\text{\AA}$)} & \multicolumn{3}{c|}{$\sigma_{v}$ (km s$^{-1}$)} & \multicolumn{1}{c|}{v$_{Hel}$ (km s$^{-1}$)} & \multicolumn{1}{c|}{R$_{deproj}$ (kpc)} & \multicolumn{1}{c|}{F275W} & \multicolumn{1}{c|}{F336W} & \multicolumn{1}{c|}{F475W} & \multicolumn{1}{c|}{F814W} & \multicolumn{1}{c|}{F110W} & \multicolumn{1}{c|}{F160W}\\ \midrule
AGB1800   & RHeB   & mct6B & 14  & $00:44:12.77$ & $+41:06:50.4$ & $-3.11$  & $\pm$ & $0.10$ & $61$  & $\pm$ & $3$  & $-62.8$  & $12.48$ & -       & -       & $23.65$ & $20.33$ & $19.22$ & $18.32$ \\
RGB3554   & RHeB?  & mct6B & 25  & $00:44:13.39$ & $+41:07:27.1$ & $-2.48$  & $\pm$ & $0.44$ & $25$  & $\pm$ & $14$ & $-343.0$ & $12.27$ & -       & $27.62$ & $23.57$ & $21.07$ & $20.26$ & $19.35$ \\
AGB4591   & RHeB   & mct6B & 30  & $00:44:13.75$ & $+41:07:43.1$ & $-4.38$  & $\pm$ & $0.12$ & $56$  & $\pm$ & $4$  & $-112.4$ & $12.19$ & -       & $27.82$ & $24.39$ & $20.47$ & $19.41$ & $18.64$ \\
AGB17874  & AGB    & mct6B & 61  & $00:44:22.05$ & $+41:09:31.6$ & $-5.25$  & $\pm$ & $0.72$ & $58$  & $\pm$ & $17$ & $-333.9$ & $12.19$ & -       & $28.53$ & $24.13$ & $19.86$ & $18.37$ & $17.19$ \\
AGB63507  & RHeB   & mct6B & 104 & $00:44:26.16$ & $+41:12:37.7$ & $-3.78$  & $\pm$ & $0.11$ & $59$  & $\pm$ & $3$  & $-400.2$ & $11.28$ & -       & -       & $24.89$ & $20.41$ & $19.13$ & $18.53$ \\
AGB65520  & AGB    & mct6B & 106 & $00:44:20.56$ & $+41:12:45.1$ & $-2.37$  & $\pm$ & $0.14$ & $46$  & $\pm$ & $5$  & $-268.2$ & $10.70$ & -       & $27.63$ & $25.28$ & $20.43$ & $18.50$ & $17.57$ \\
AGB74155  & RHeB   & mct6B & 117 & $00:44:18.71$ & $+41:13:17.3$ & $-2.68$  & $\pm$ & $0.22$ & $76$  & $\pm$ & $8$  & $80.6$   & $10.30$ & -       & $26.43$ & $22.65$ & $19.78$ & $19.31$ & $18.56$ \\
AGB105172 & T-MS   & mct6B & 147 & $00:44:25.29$ & $+41:15:04.7$ & $-0.54$  & $\pm$ & $0.04$ & $64$  & $\pm$ & $4$  & $-59.1$  & $10.21$ & -       & $22.88$ & $21.74$ & $20.35$ & $20.29$ & $19.70$ \\
AGB147540 & AGB    & mct6B & 186 & $00:44:39.99$ & $+41:17:34.1$ & $-2.44$  & $\pm$ & $0.12$ & $64$  & $\pm$ & $5$  & $-295.8$ & $10.67$ & -       & $26.83$ & $30.70$ & $20.94$ & $18.14$ & $17.18$ \\
\framebox[1.8cm][l]{AGB154104} & RHeB   & mct6B & 193 & $00:44:48.54$ & $+41:17:55.9$ & $-2.12$  & $\pm$ & $0.08$ & $45$  & $\pm$ & $3$  & $-578.9$ & $11.36$ & -       & $26.65$ & $22.93$ & $19.87$ & $19.25$ & $18.49$ \\
AGB158331 & RHeB   & mct6B & 195 & $00:44:44.51$ & $+41:18:09.8$ & $-7.13$  & $\pm$ & $0.83$ & $102$ & $\pm$ & $21$ & $-480.2$ & $10.88$ & -       & $27.99$ & $22.84$ & $20.02$ & $19.35$ & $18.49$ \\
AGB167748 & AGB    & mct6B & 201 & $00:44:44.89$ & $+41:18:41.3$ & $-9.48$  & $\pm$ & $1.76$ & $65$  & $\pm$ & $19$ & $-137.0$ & $10.73$ & -       & $27.29$ & $24.24$ & $20.72$ & $19.49$ & $18.60$ \\
AGB173678 & AGB    & mct6B & 206 & $00:44:32.58$ & $+41:19:01.0$ & $-11.00$ & $\pm$ & $2.69$ & $79$  & $\pm$ & $25$ & $-83.8$  & $9.43$  & -       & $27.25$ & $25.12$ & $20.62$ & $19.03$ & $18.26$ \\
MS178449  & MS     & mct6B & 209 & $00:44:29.27$ & $+41:19:17.4$ & $-13.67$ & $\pm$ & $1.09$ & $155$ & $\pm$ & $14$ & $-340.9$ & $9.02$  & $22.22$ & $22.13$ & $22.91$ & $21.83$ & $21.01$ & $20.22$ \\
AGB187069 & RHeB   & mct6B & 213 & $00:44:36.35$ & $+41:19:46.0$ & $-7.24$  & $\pm$ & $2.33$ & $47$  & $\pm$ & $21$ & $-272.2$ & $9.53$  & -       & $28.28$ & $23.88$ & $19.50$ & $19.02$ & $18.03$ \\
MS128721  & MS     & mct6C & 25  & $00:44:21.57$ & $+41:16:26.3$ & $-28.80$ & $\pm$ & $2.37$ & $140$ & $\pm$ & $16$ & $-255.2$ & $9.33$  & $20.57$ & $20.83$ & $22.38$ & $22.25$ & $22.33$ & $22.15$ \\
\framebox[1.8cm][l]{MS179275}  & MS     & mct6C & 64  & $00:44:31.66$ & $+41:19:19.9$ & $-8.53$  & $\pm$ & $0.13$ & $102$ & $\pm$ & $2$  & $-246.8$ & $9.23$  & $19.17$ & $19.34$ & $20.84$ & $20.65$ & $20.59$ & $20.43$ \\
MS185530  & MS     & mct6C & 72  & $00:44:37.73$ & $+41:19:40.7$ & $-8.16$  & $\pm$ & $0.13$ & $187$ & $\pm$ & $4$  & $-303.4$ & $9.69$  & $19.65$ & $19.74$ & $20.96$ & $20.67$ & $20.96$ & $20.84$ \\
MS201708  & MS     & mct6C & 88  & $00:44:33.28$ & $+41:20:34.2$ & $-14.36$ & $\pm$ & $0.93$ & $193$ & $\pm$ & $10$ & $-92.6$  & $8.96$  & $21.93$ & $21.85$ & $22.80$ & $22.52$ & $22.40$ & $22.29$ \\
MS251784  & AGB    & mct6C & 145 & $00:44:49.48$ & $+41:23:00.5$ & $-4.79$  & $\pm$ & $0.86$ & $24$  & $\pm$ & $27$ & $-208.8$ & $9.69$  & -       & $28.63$ & $25.31$ & $21.61$ & $20.28$ & $19.25$ \\
MS257904  & MS     & mct6C & 149 & $00:44:48.61$ & $+41:23:14.0$ & $-23.26$ & $\pm$ & $1.00$ & $92$  & $\pm$ & $11$ & $-278.9$ & $9.54$  & $20.51$ & $20.71$ & $22.11$ & $22.02$ & $21.96$ & $21.82$ \\
AGB269781 & RHeB   & mct6C & 158 & $00:44:56.46$ & $+41:23:38.1$ & $-2.75$  & $\pm$ & $0.17$ & $64$  & $\pm$ & $5$  & $-490.2$ & $10.15$ & -       & $26.31$ & $24.06$ & $20.58$ & $19.69$ & $18.97$ \\
MS313163  & MS     & mct6C & 180 & $00:45:06.98$ & $+41:24:51.2$ & $-12.60$ & $\pm$ & $0.16$ & $141$ & $\pm$ & $2$  & $-249.0$ & $10.76$ & $18.97$ & $19.18$ & $20.69$ & $20.52$ & $20.65$ & $20.43$ \\
AGB341522 & RHeB   & mct6C & 189 & $00:45:18.46$ & $+41:25:32.3$ & $-4.07$  & $\pm$ & $0.18$ & $54$  & $\pm$ & $4$  & $-271.3$ & $11.62$ & -       & -       & $23.13$ & $19.67$ & $19.25$ & $18.53$ \\
MS420720  & MS     & mct6C & 211 & $00:45:14.27$ & $+41:27:18.8$ & $-9.39$  & $\pm$ & $1.09$ & $113$ & $\pm$ & $23$ & $-190.9$ & $10.72$ & $21.77$ & $21.88$ & $22.75$ & $22.71$ & $22.76$ & $22.65$ \\
AGB275033 & AGB C* & mct6D & 35  & $00:44:54.60$ & $+41:23:48.0$ & $-13.90$ & $\pm$ & $4.30$ & $62$  & $\pm$ & $33$ & $-198.6$ & $9.93$  & -       & -       & $25.24$ & $19.87$ & $18.75$ & $17.47$ \\
AGB289376 & RHeB   & mct6D & 39  & $00:44:36.30$ & $+41:24:13.0$ & $-1.76$  & $\pm$ & $0.17$ & $48$  & $\pm$ & $8$  & $-252.5$ & $8.11$  & -       & $28.15$ & $22.87$ & $19.48$ & $18.33$ & $17.30$ \\
MS332202  & MS     & mct6D & 57  & $00:44:36.37$ & $+41:25:19.5$ & $-0.63$  & $\pm$ & $0.02$ & $25$  & $\pm$ & $3$  & $-271.6$ & $7.82$  & $17.69$ & $17.97$ & $19.37$ & $19.21$ & $19.40$ & $19.32$ \\
MS412514  & MS     & mct6D & 80  & $00:44:32.07$ & $+41:27:06.9$ & $-1.95$  & $\pm$ & $0.12$ & $52$  & $\pm$ & $6$  & $-264.1$ & $7.03$  & $22.04$ & $21.54$ & $22.24$ & $21.20$ & $20.92$ & $20.82$ \\
MS118058  & MS     & mct6E & 79  & $00:44:04.57$ & $+41:15:47.6$ & $-14.25$ & $\pm$ & $1.26$ & $195$ & $\pm$ & $22$ & $-193.6$ & $7.94$  & $21.19$ & $21.07$ & $22.36$ & $21.78$ & $21.74$ & $21.47$ \\
SG127353  & T-MS   & mct6E & 84  & $00:44:04.80$ & $+41:16:21.2$ & $-8.93$  & $\pm$ & $0.63$ & $225$ & $\pm$ & $15$ & $-310.6$ & $7.75$  & $23.32$ & $21.96$ & $21.39$ & $20.88$ & $20.78$ & $20.31$ \\
MS181868  & AGB    & mct6E & 123 & $00:44:02.67$ & $+41:19:28.4$ & $-17.40$ & $\pm$ & $1.49$ & $206$ & $\pm$ & $24$ & $-414.0$ & $6.42$  & -       & -       & $26.25$ & $21.83$ & $20.29$ & $19.27$ \\
MS223912  & T-MS   & mct6E & 161 & $00:43:52.05$ & $+41:21:47.4$ & $-9.25$  & $\pm$ & $0.93$ & $83$  & $\pm$ & $13$ & $-104.6$ & $4.74$  & $22.04$ & $21.32$ & $21.54$ & $21.20$ & $21.15$ & $21.00$ \\
MS310616  & MS     & mct6E & 204 & $00:43:57.06$ & $+41:24:47.3$ & $-15.64$ & $\pm$ & $2.37$ & $115$ & $\pm$ & $31$ & $-221.8$ & $4.56$  & $21.82$ & $21.25$ & $21.81$ & $21.06$ & $20.99$ & $20.81$ \\
MS341793  & MS     & mct6E & 214 & $00:43:56.05$ & $+41:25:32.7$ & $-25.33$ & $\pm$ & $1.49$ & $135$ & $\pm$ & $18$ & $-114.6$ & $4.38$  & $19.55$ & $19.86$ & $21.50$ & $21.39$ & $21.62$ & $21.24$ \\
AGB362536 & AGB C* & mct6E & 218 & $00:43:58.57$ & $+41:26:00.6$ & $-3.49$  & $\pm$ & $0.51$ & $65$  & $\pm$ & $14$ & $-252.9$ & $4.52$  & -       & -       & $26.10$ & $19.83$ & $17.68$ & $16.66$ \\
AGB543679 & RHeB   & mct6F & 14  & $00:45:53.07$ & $+41:34:35.4$ & $-4.47$  & $\pm$ & $0.19$ & $69$  & $\pm$ & $5$  & $-259.0$ & $12.37$ & -       & $25.97$ & $22.98$ & $20.02$ & $19.49$ & $18.62$ \\
AGB568802 & RHeB   & mct6F & 34  & $00:45:56.13$ & $+41:36:10.2$ & $-5.35$  & $\pm$ & $0.17$ & $55$  & $\pm$ & $3$  & $-173.6$ & $12.31$ & -       & -       & $22.45$ & $19.61$ & $19.28$ & $18.57$ \\
AGB646278 & AGB    & mct6F & 70  & $00:45:59.91$ & $+41:39:06.3$ & $-7.93$  & $\pm$ & $1.50$ & $72$  & $\pm$ & $18$ & $-36.5$  & $12.10$ & -       & -       & $24.83$ & $20.02$ & $18.63$ & $17.62$ \\
AGB686015 & RHeB   & mct6F & 93  & $00:45:57.69$ & $+41:41:09.9$ & $-1.94$  & $\pm$ & $0.07$ & $48$  & $\pm$ & $4$  & $-333.2$ & $11.66$ & -       & -       & $22.40$ & $19.56$ & $19.46$ & $18.79$ \\
AGB692878 & AGB    & mct6F & 98  & $00:45:47.09$ & $+41:41:37.0$ & $-0.55$  & $\pm$ & $0.10$ & $11$  & $\pm$ & $7$  & $-202.9$ & $10.87$ & -       & $26.36$ & $24.08$ & $20.56$ & $19.31$ & $18.22$ \\
MS692952  & MS     & mct6F & 99  & $00:45:54.30$ & $+41:41:37.3$ & $-20.90$ & $\pm$ & $1.75$ & $138$ & $\pm$ & $16$ & $-220.6$ & $11.37$ & $22.51$ & $22.32$ & $22.96$ & $22.83$ & $22.80$ & $22.76$ \\
MS700544  & MS     & mct6F & 108 & $00:45:53.87$ & $+41:42:09.7$ & $-22.96$ & $\pm$ & $0.33$ & $121$ & $\pm$ & $4$  & $-184.2$ & $11.29$ & $19.81$ & $19.95$ & $21.21$ & $21.05$ & $21.11$ & $20.94$ \\
AGB708605 & AGB    & mct6F & 118 & $00:45:51.83$ & $+41:42:49.9$ & $-4.20$  & $\pm$ & $1.02$ & $40$  & $\pm$ & $28$ & $-67.0$  & $11.11$ & -       & $27.26$ & $25.84$ & $20.38$ & $18.22$ & $17.17$ \\
AGB709905 & AGB    & mct6F & 121 & $00:46:00.65$ & $+41:42:56.7$ & $-3.22$  & $\pm$ & $0.23$ & $69$  & $\pm$ & $7$  & $-304.0$ & $11.70$ & -       & $27.13$ & $24.40$ & $20.77$ & $19.60$ & $18.64$ \\
MS761954  & MS     & mct6F & 194 & $00:46:08.86$ & $+41:47:58.5$ & $-15.85$ & $\pm$ & $0.61$ & $112$ & $\pm$ & $9$  & $-186.3$ & $12.01$ & $20.46$ & $20.55$ & $21.88$ & $21.46$ & $21.51$ & $21.28$ \\
MS763520  & MS     & mct6F & 200 & $00:46:11.14$ & $+41:48:07.6$ & $-25.73$ & $\pm$ & $1.40$ & $170$ & $\pm$ & $11$ & $-174.8$ & $12.15$ & $21.71$ & $21.60$ & $22.86$ & $22.30$ & $22.23$ & $21.87$ \\
MS775105  & MS     & mct6F & 210 & $00:46:07.52$ & $+41:49:14.2$ & $-7.81$  & $\pm$ & $0.16$ & $212$ & $\pm$ & $4$  & $-66.4$  & $11.95$ & $19.40$ & $19.72$ & $21.25$ & $21.14$ & $21.24$ & $21.06$ \\
MS575747  & MS     & mct6G & 11  & $00:45:16.67$ & $+41:36:28.9$ & $-28.81$ & $\pm$ & $1.22$ & $169$ & $\pm$ & $8$  & $-153.2$ & $9.16$  & $19.87$ & $19.94$ & $21.34$ & $20.78$ & $20.65$ & $20.34$ \\
MS661443  & AGB    & mct6G & 59  & $00:45:36.47$ & $+41:39:45.1$ & $-21.63$ & $\pm$ & $0.54$ & $118$ & $\pm$ & $5$  & $-153.6$ & $10.27$ & -       & -       & $25.99$ & $20.45$ & $18.81$ & $17.78$ \\
MS683060  & MS     & mct6G & 74  & $00:45:27.79$ & $+41:40:58.2$ & $-3.53$  & $\pm$ & $0.12$ & $99$  & $\pm$ & $4$  & $-80.7$  & $9.64$  & $20.12$ & $20.25$ & $21.29$ & $21.19$ & $21.19$ & $21.12$ \\\bottomrule
\end{tabular}}
\end{table*}

\newpage
\begin{table*}
\centering
\resizebox{1.35\textwidth}{!}{%
\begin{tabular}{lcccllrllrllcccccccc} 
\toprule
\multicolumn{1}{|c|}{Object} & \multicolumn{1}{c|}{Class.} & \multicolumn{1}{c|}{Mask} & \multicolumn{1}{c|}{Slit No.} & \multicolumn{1}{c|}{RA} & \multicolumn{1}{c|}{Dec} & \multicolumn{3}{c|}{H$\alpha$ EW ($\text{\AA}$)} & \multicolumn{3}{c|}{$\sigma_{v}$ (km s$^{-1}$)} & \multicolumn{1}{c|}{v$_{Hel}$ (km s$^{-1}$)} & \multicolumn{1}{c|}{R$_{deproj}$ (kpc)} & \multicolumn{1}{c|}{F275W} & \multicolumn{1}{c|}{F336W} & \multicolumn{1}{c|}{F475W} & \multicolumn{1}{c|}{F814W} & \multicolumn{1}{c|}{F110W} & \multicolumn{1}{c|}{F160W}\\ \midrule
MS708596   & AGB    & mct6G & 105 & $00:45:32.67$ & $+41:42:49.8$ & $-6.26$  & $\pm$ & $0.28$  & $166$ & $\pm$ & $8$  & $-140.4$ & $9.91$  & -       & $30.65$ & $25.43$ & $21.02$ & $19.23$ & $18.19$ \\
MS716320   & MS     & mct6G & 113 & $00:45:44.32$ & $+41:43:32.5$ & $-1.92$  & $\pm$ & $0.02$  & $72$  & $\pm$ & $1$  & $-156.2$ & $10.6$  & $17.10$ & $17.18$ & $18.50$ & $18.11$ & $18.14$ & $18.02$ \\
MS722500   & MS     & mct6G & 124 & $00:45:41.94$ & $+41:44:08.4$ & $-36.40$ & $\pm$ & $0.63$  & $109$ & $\pm$ & $4$  & $-170.8$ & $10.45$ & $19.26$ & $19.31$ & $20.75$ & $20.32$ & $20.36$ & $20.06$ \\
MS723461   & MS     & mct6G & 126 & $00:45:39.20$ & $+41:44:14.1$ & $-23.02$ & $\pm$ & $0.67$  & $87$  & $\pm$ & $4$  & $-85.4$  & $10.29$ & $20.51$ & $20.33$ & $21.26$ & $20.85$ & $20.92$ & $20.77$ \\
MS729737   & MS     & mct6G & 138 & $00:45:47.66$ & $+41:44:52.1$ & $-26.98$ & $\pm$ & $0.99$  & $158$ & $\pm$ & $11$ & $-199.0$ & $10.78$ & $20.63$ & $20.66$ & $21.92$ & $21.46$ & $21.43$ & $21.18$ \\
MS742327   & MS     & mct6G & 161 & $00:45:56.06$ & $+41:46:07.0$ & $-33.49$ & $\pm$ & $3.58$  & $108$ & $\pm$ & $14$ & $-228.4$ & $11.27$ & $21.63$ & $21.54$ & $22.34$ & $21.87$ & $21.76$ & $21.44$ \\
MS763598   & MS     & mct6G & 193 & $00:46:08.20$ & $+41:48:08.0$ & $-5.83$  & $\pm$ & $0.27$  & $268$ & $\pm$ & $7$  & $-191.4$ & $11.97$ & $20.51$ & $20.55$ & $21.63$ & $21.56$ & $21.59$ & $21.56$ \\
MS767158   & MS     & mct6G & 198 & $00:46:07.10$ & $+41:48:28.6$ & $-4.05$  & $\pm$ & $0.08$  & $162$ & $\pm$ & $3$  & $-212.4$ & $11.91$ & $18.79$ & $18.99$ & $20.36$ & $20.18$ & $20.26$ & $20.23$ \\
MS767515   & MS     & mct6G & 199 & $00:45:56.11$ & $+41:48:30.7$ & $-11.75$ & $\pm$ & $0.36$  & $143$ & $\pm$ & $7$  & $-231.6$ & $11.32$ & $20.36$ & $20.45$ & $21.82$ & $21.33$ & $21.31$ & $21.08$ \\
MS779790   & MS     & mct6G & 211 & $00:45:48.81$ & $+41:49:35.8$ & $-28.62$ & $\pm$ & $0.30$  & $104$ & $\pm$ & $5$  & $-83.9$  & $11.06$ & $17.96$ & $18.22$ & $19.94$ & $19.50$ & $19.54$ & $19.27$ \\
MS597687   & MS     & mct6H & 21  & $00:45:06.06$ & $+41:37:18.9$ & $-10.37$ & $\pm$ & $0.18$  & $110$ & $\pm$ & $4$  & $-196.2$ & $8.38$  & $18.01$ & $18.30$ & $19.84$ & $19.78$ & $19.97$ & $19.81$ \\
MS608785   & MS     & mct6H & 29  & $00:45:14.69$ & $+41:37:43.3$ & $-2.89$  & $\pm$ & $0.03$  & $90$  & $\pm$ & $1$  & $-185.3$ & $8.92$  & $17.65$ & $17.88$ & $19.32$ & $19.20$ & $19.38$ & $19.33$ \\
MS634466   & MS     & mct6H & 42  & $00:45:15.29$ & $+41:38:39.0$ & $-47.31$ & $\pm$ & $1.95$  & $113$ & $\pm$ & $11$ & $-139.4$ & $8.92$  & $19.88$ & $20.02$ & $21.62$ & $21.40$ & $21.42$ & $21.17$ \\
AGB637838  & RHeB   & mct6H & 46  & $00:45:08.79$ & $+41:38:46.7$ & $-2.71$  & $\pm$ & $0.17$  & $41$  & $\pm$ & $6$  & $-368.2$ & $8.52$  & -       & $25.60$ & $22.37$ & $19.39$ & $18.60$ & $17.92$ \\
MS698062   & MS     & mct6H & 92  & $00:45:19.91$ & $+41:41:58.5$ & $-2.22$  & $\pm$ & $0.10$  & $103$ & $\pm$ & $5$  & $-164.4$ & $9.19$  & $20.86$ & $20.66$ & $21.59$ & $20.91$ & $20.67$ & $20.49$ \\
MS703326   & MS     & mct6H & 101 & $00:45:26.09$ & $+41:42:22.7$ & $-6.77$  & $\pm$ & $0.43$  & $174$ & $\pm$ & $12$ & $-127.2$ & $9.54$  & $21.10$ & $21.11$ & $21.95$ & $21.75$ & $21.87$ & $21.77$ \\
MS710772   & MS     & mct6H & 107 & $00:45:22.30$ & $+41:43:01.6$ & $-49.61$ & $\pm$ & $10.16$ & $162$ & $\pm$ & $31$ & $-223.8$ & $9.36$  & $21.08$ & $21.24$ & $22.49$ & $22.32$ & $22.63$ & $22.66$ \\
AGB727603  & MS     & mct6H & 131 & $00:45:28.91$ & $+41:44:39.2$ & $-2.24$  & $\pm$ & $0.13$  & $45$  & $\pm$ & $5$  & $-158.8$ & $9.77$  & $21.99$ & $21.49$ & $22.11$ & $20.85$ & $20.68$ & $20.44$ \\
MS755421   & MS     & mct6H & 170 & $00:45:29.94$ & $+41:47:20.9$ & $-6.19$  & $\pm$ & $0.29$  & $382$ & $\pm$ & $9$  & $-148.9$ & $10.07$ & $20.89$ & $20.82$ & $21.94$ & $21.34$ & $21.15$ & $20.88$ \\
AGB799233  & RHeB   & mct6H & 216 & $00:45:41.72$ & $+41:50:45.6$ & $-2.53$  & $\pm$ & $0.17$  & $81$  & $\pm$ & $7$  & $-92.7$  & $10.94$ & -       & $25.69$ & $23.15$ & $20.48$ & $19.55$ & $18.67$ \\
MS249482   & MS     & mct6I & 17  & $00:43:39.02$ & $+41:22:55.1$ & $-22.27$ & $\pm$ & $0.52$  & $154$ & $\pm$ & $6$  & $-218.0$ & $3.41$  & $19.09$ & $19.27$ & $20.86$ & $20.53$ & $20.35$ & $20.08$ \\
MS257350   & MS     & mct6I & 27  & $00:43:33.43$ & $+41:23:12.9$ & $-23.29$ & $\pm$ & $0.53$  & $153$ & $\pm$ & $8$  & $-138.2$ & $2.97$  & $19.36$ & $19.62$ & $21.09$ & $20.80$ & $20.83$ & $20.50$ \\
MS267013   & MS     & mct6I & 45  & $00:43:43.99$ & $+41:23:32.8$ & $-13.47$ & $\pm$ & $0.42$  & $140$ & $\pm$ & $6$  & $-177.5$ & $3.71$  & $19.85$ & $19.98$ & $21.17$ & $20.78$ & $20.74$ & $20.33$ \\
MS272171   & T-MS   & mct6I & 51  & $00:43:57.55$ & $+41:23:42.8$ & $-12.81$ & $\pm$ & $0.47$  & $52$  & $\pm$ & $8$  & $-332.4$ & $4.79$  & $21.62$ & $20.65$ & $21.09$ & $20.51$ & $20.34$ & $20.21$ \\
serendip1  & AGB    & mct6I & 54  & $00:44:17.25$ & $+41:23:45.8$ & $-0.78$  & $\pm$ & $0.07$  & $45$  & $\pm$ & $4$  & $-246.7$ & $6.50$  & -       & -       & $25.40$ & $20.17$ & $18.27$ & $17.13$ \\
MS275061   & T-MS   & mct6I & 57  & $00:43:27.94$ & $+41:23:48.0$ & $-7.91$  & $\pm$ & $0.25$  & $50$  & $\pm$ & $5$  & $-252.8$ & $2.64$  & $21.34$ & $20.47$ & $20.74$ & $20.25$ & $20.31$ & $20.08$ \\
MS277924   & T-MS   & mct6I & 63  & $00:43:56.91$ & $+41:23:53.2$ & $-6.73$  & $\pm$ & $0.24$  & $124$ & $\pm$ & $6$  & $-231.0$ & $4.70$  & $22.11$ & $21.58$ & $21.67$ & $20.95$ & $20.52$ & $19.97$ \\
MS283118   & MS     & mct6I & 71  & $00:43:36.25$ & $+41:24:02.5$ & $-20.78$ & $\pm$ & $0.83$  & $118$ & $\pm$ & $7$  & $-127.6$ & $3.12$  & $19.49$ & $19.73$ & $21.22$ & $21.03$ & $21.28$ & $21.10$ \\
serendip1  & MS     & mct6I & 71  & $00:43:36.38$ & $+41:24:02.6$ & $-6.00$  & $\pm$ & $0.75$  & $46$  & $\pm$ & $14$ & $-104.5$ & $3.13$  & $22.69$ & $22.07$ & $22.73$ & $21.92$ & $21.73$ & $21.67$ \\
MS285212   & T-MS   & mct6I & 73  & $00:43:42.35$ & $+41:24:06.1$ & $-1.18$  & $\pm$ & $0.05$  & $74$  & $\pm$ & $4$  & $-114.8$ & $3.53$  & $22.49$ & $21.35$ & $20.97$ & $20.44$ & $20.48$ & $20.19$ \\
MS285376   & MS     & mct6I & 76  & $00:44:34.37$ & $+41:24:06.4$ & $-3.54$  & $\pm$ & $0.11$  & $199$ & $\pm$ & $6$  & $-264.8$ & $7.97$  & $18.89$ & $19.15$ & $20.62$ & $20.69$ & $20.99$ & $21.03$ \\
AGB287947  & T-MS   & mct6I & 80  & $00:44:31.62$ & $+41:24:10.7$ & $-9.62$  & $\pm$ & $0.18$  & $121$ & $\pm$ & $6$  & $-254.8$ & $7.69$  & $22.02$ & $21.27$ & $20.64$ & $19.15$ & $18.67$ & $18.21$ \\
AGB295890  & RHeB   & mct6I & 91  & $00:44:25.36$ & $+41:24:24.0$ & $-4.91$  & $\pm$ & $0.17$  & $75$  & $\pm$ & $4$  & $-288.8$ & $7.07$  & -       & $25.85$ & $23.65$ & $19.94$ & $19.11$ & $18.25$ \\
MS310626   & MS     & mct6I & 118 & $00:43:53.31$ & $+41:24:47.3$ & $-1.33$  & $\pm$ & $0.01$  & $134$ & $\pm$ & $1$  & $-189.4$ & $4.27$  & $16.81$ & $17.21$ & $18.84$ & $18.89$ & $18.89$ & $18.37$ \\
AGB325404  & AGB C* & mct6I & 134 & $00:44:22.14$ & $+41:25:09.8$ & $-4.64$  & $\pm$ & $0.35$  & $187$ & $\pm$ & $9$  & $-236.6$ & $6.60$  & -       & -       & $24.03$ & $19.65$ & $18.03$ & $16.94$ \\
MS338932   & MS     & mct6I & 146 & $00:44:34.43$ & $+41:25:28.7$ & $-50.33$ & $\pm$ & $0.55$  & $87$  & $\pm$ & $7$  & $-223.4$ & $7.61$  & $18.72$ & $18.65$ & $19.75$ & $19.28$ & $19.25$ & $19.01$ \\
AGB344610  & AGB C* & mct6I & 151 & $00:44:15.78$ & $+41:25:36.7$ & $-2.37$  & $\pm$ & $0.21$  & $59$  & $\pm$ & $6$  & $-225.2$ & $5.95$  & -       & -       & $24.62$ & $19.76$ & $18.18$ & $16.95$ \\
MS355973   & MS     & mct6I & 159 & $00:43:40.77$ & $+41:25:51.9$ & $-16.83$ & $\pm$ & $1.23$  & $52$  & $\pm$ & $11$ & $-390.5$ & $3.40$  & $20.78$ & $20.52$ & $21.64$ & $21.20$ & $21.07$ & $20.87$ \\
serendip1  & AGB    & mct6I & 168 & $00:44:39.22$ & $+41:25:59.2$ & $-10.95$ & $\pm$ & $1.83$  & $88$  & $\pm$ & $43$ & $-216.0$ & $7.91$  & -       & -       & $26.58$ & $21.76$ & $20.01$ & $18.81$ \\
MS373085   & T-MS   & mct6I & 179 & $00:44:13.36$ & $+41:26:14.3$ & $-1.85$  & $\pm$ & $0.12$  & $38$  & $\pm$ & $5$  & $-93.7$  & $5.63$  & -       & $21.06$ & $21.02$ & $20.75$ & $20.66$ & $20.57$ \\
MS380470   & MS     & mct6I & 188 & $00:44:20.53$ & $+41:26:23.9$ & $-9.01$  & $\pm$ & $0.22$  & $210$ & $\pm$ & $6$  & $-302.2$ & $6.20$  & $19.89$ & $19.99$ & $21.38$ & $20.96$ & $21.05$ & $20.78$ \\
MS392298   & MS     & mct6I & 197 & $00:44:52.59$ & $+41:26:39.3$ & $-4.08$  & $\pm$ & $0.20$  & $89$  & $\pm$ & $8$  & $-339.8$ & $8.93$  & $21.31$ & $21.35$ & $22.38$ & $22.05$ & $21.98$ & $21.89$ \\
MS578990   & MS     & mct6K & 60  & $00:44:32.21$ & $+41:36:37.0$ & $-14.86$ & $\pm$ & $0.27$  & $128$ & $\pm$ & $4$  & $-75.6$  & $6.60$  & $18.48$ & $18.78$ & $20.39$ & $20.17$ & $20.25$ & $20.04$ \\
MS579231   & MS     & mct6K & 61  & $00:43:58.04$ & $+41:36:37.6$ & $-23.38$ & $\pm$ & $2.59$  & $119$ & $\pm$ & $11$ & $-166.7$ & $6.11$  & $20.73$ & $20.85$ & $21.89$ & $21.78$ & $21.96$ & $21.75$ \\
MS587961   & MS     & mct6K & 75  & $00:44:27.40$ & $+41:36:57.6$ & $-17.85$ & $\pm$ & $0.86$  & $211$ & $\pm$ & $10$ & $-175.6$ & $6.49$  & $20.26$ & $20.41$ & $21.45$ & $21.38$ & $21.55$ & $21.56$ \\
MS594977   & T-MS   & mct6K & 84  & $00:45:06.13$ & $+41:37:13.0$ & $-4.53$  & $\pm$ & $1.22$  & $93$  & $\pm$ & $16$ & $-73.4$  & $8.39$  & $23.60$ & $22.96$ & $22.56$ & $22.06$ & $22.14$ & $21.89$ \\
serendip1  & MS     & mct6K & 93  & $00:45:04.96$ & $+41:37:26.9$ & $-1.14$  & $\pm$ & $0.02$  & $78$  & $\pm$ & $2$  & $-112.1$ & $8.31$  & $16.70$ & $16.93$ & $18.43$ & $18.34$ & $18.49$ & $18.46$ \\
MS602034   & MS     & mct6K & 97  & $00:45:19.73$ & $+41:37:28.4$ & $-9.32$  & $\pm$ & $0.37$  & $179$ & $\pm$ & $9$  & $-183.2$ & $9.29$  & $21.31$ & $21.19$ & $21.98$ & $21.42$ & $21.41$ & $21.17$ \\
serendip1  & RHeB   & mct6K & 98  & $00:44:53.13$ & $+41:37:28.8$ & $-2.88$  & $\pm$ & $0.52$  & $53$  & $\pm$ & $12$ & $-173.5$ & $7.63$  & -       & $32.97$ & $25.04$ & $20.88$ & -       & -       \\
MS607506   & T-MS   & mct6K & 113 & $00:45:20.77$ & $+41:37:40.5$ & $-5.07$  & $\pm$ & $1.08$  & $65$  & $\pm$ & $16$ & $-155.5$ & $9.34$  & $22.26$ & $21.80$ & $21.53$ & $21.19$ & $21.09$ & $21.03$ \\
MS613247   & MS     & mct6K & 123 & $00:44:43.41$ & $+41:37:53.0$ & $-40.88$ & $\pm$ & $1.62$  & $155$ & $\pm$ & $8$  & $-136.5$ & $7.19$  & $20.66$ & $20.69$ & $21.95$ & $21.48$ & $21.51$ & $21.33$ \\
MS623672   & AGB    & mct6K & 155 & $00:45:12.73$ & $+41:38:15.2$ & $-3.29$  & $\pm$ & $0.55$  & $55$  & $\pm$ & $14$ & $-152.8$ & $8.77$  & -       & -       & $25.79$ & $21.46$ & $19.60$ & $18.62$ \\
MS642835   & MS     & mct6K & 181 & $00:45:17.06$ & $+41:38:58.2$ & $-1.71$  & $\pm$ & $0.03$  & $96$  & $\pm$ & $3$  & $-92.1$  & $9.02$  & $17.05$ & $16.56$ & $17.76$ & $16.80$ & $16.69$ & $16.62$ \\
MS918576   & MS     & mct6L & 53  & $00:45:59.77$ & $+41:57:14.9$ & $-6.48$  & $\pm$ & $1.23$  & $58$  & $\pm$ & $11$ & $-124.0$ & $12.55$ & $21.60$ & $21.70$ & $22.61$ & $22.65$ & $22.82$ & $22.81$ \\
MS966060   & MS     & mct6L & 104 & $00:46:10.38$ & $+42:01:02.9$ & $-6.72$  & $\pm$ & $0.60$  & $92$  & $\pm$ & $15$ & $-94.1$  & $13.56$ & $23.06$ & $22.65$ & $22.93$ & $22.68$ & $22.68$ & $22.60$ \\
AGB1008239 & AGB    & mct6L & 210 & $00:45:39.04$ & $+42:07:30.9$ & $-1.80$  & $\pm$ & $0.37$  & $48$  & $\pm$ & $8$  & $-150.2$ & $15.53$ & -       & -       & $25.29$ & $20.30$ & $18.55$ & $17.45$ \\
MS1018382  & AGB    & mct6L & 227 & $00:45:52.62$ & $+42:08:46.4$ & $-11.97$ & $\pm$ & $0.73$  & $158$ & $\pm$ & $7$  & $-186.1$ & $15.76$ & -       & $27.28$ & $25.41$ & $22.35$ & $21.27$ & $20.29$ \\
MS388051   & MS     & mct6M & 35  & $00:44:45.17$ & $+41:26:33.8$ & $-8.49$  & $\pm$ & $0.33$  & $195$ & $\pm$ & $8$  & $-231.3$ & $8.29$  & $20.44$ & $20.38$ & $21.31$ & $20.78$ & $20.65$ & $20.48$ \\
SG406459   & T-MS   & mct6M & 47  & $00:44:46.06$ & $+41:26:58.4$ & $-3.82$  & $\pm$ & $0.38$  & $71$  & $\pm$ & $10$ & $-190.7$ & $8.27$  & -       & $23.12$ & $22.25$ & $21.22$ & $20.88$ & $20.68$ \\
MS434008   & MS     & mct6M & 55  & $00:44:41.15$ & $+41:27:38.8$ & $-9.75$  & $\pm$ & $0.73$  & $149$ & $\pm$ & $14$ & $-247.7$ & $7.69$  & $21.06$ & $21.07$ & $21.92$ & $21.73$ & $21.91$ & $21.85$ \\
MS440713   & MS     & mct6M & 57  & $00:44:33.60$ & $+41:27:49.1$ & $-11.33$ & $\pm$ & $0.58$  & $166$ & $\pm$ & $10$ & $-297.3$ & $7.02$  & $22.39$ & $21.81$ & $22.41$ & $21.19$ & $20.93$ & $20.62$ \\
MS454607   & MS     & mct6M & 61  & $00:44:47.22$ & $+41:28:11.7$ & $-2.67$  & $\pm$ & $0.09$  & $56$  & $\pm$ & $4$  & $-192.3$ & $8.10$  & $19.40$ & $19.43$ & $20.49$ & $20.17$ & $20.21$ & $20.17$ \\
MS507660   & MS     & mct6M & 83  & $00:44:49.18$ & $+41:30:03.5$ & $-6.12$  & $\pm$ & $0.49$  & $62$  & $\pm$ & $8$  & $-215.9$ & $7.90$  & $19.97$ & $20.26$ & $21.67$ & $21.61$ & $21.67$ & $21.43$ \\\bottomrule
\end{tabular}}
\end{table*}

\newpage
\begin{table*}
\centering
\resizebox{1.35\textwidth}{!}{%
\begin{tabular}{lcccllrllrllcccccccc} 
\toprule
\multicolumn{1}{|c|}{Object} & \multicolumn{1}{c|}{Class.} & \multicolumn{1}{c|}{Mask} & \multicolumn{1}{c|}{Slit No.} & \multicolumn{1}{c|}{RA} & \multicolumn{1}{c|}{Dec} & \multicolumn{3}{c|}{H$\alpha$ EW ($\text{\AA}$)} & \multicolumn{3}{c|}{$\sigma_{v}$ (km s$^{-1}$)} & \multicolumn{1}{c|}{v$_{Hel}$ (km s$^{-1}$)} & \multicolumn{1}{c|}{R$_{deproj}$ (kpc)} & \multicolumn{1}{c|}{F275W} & \multicolumn{1}{c|}{F336W} & \multicolumn{1}{c|}{F475W} & \multicolumn{1}{c|}{F814W} & \multicolumn{1}{c|}{F110W} & \multicolumn{1}{c|}{F160W}\\ \midrule
MS514797   & MS     & mct6M & 90  & $00:44:46.83$ & $+41:30:27.9$ & $-3.08$  & $\pm$ & $0.27$  & $47$  & $\pm$ & $8$  & $-191.2$ & $7.64$  & $19.32$ & $19.47$ & $20.72$ & $20.48$ & $20.58$ & $20.53$ \\
MS527247   & MS     & mct6M & 105 & $00:44:37.98$ & $+41:31:36.2$ & $-5.10$  & $\pm$ & $0.45$  & $67$  & $\pm$ & $9$  & $-229.1$ & $6.83$  & $20.97$ & $20.90$ & $21.99$ & $21.75$ & $21.72$ & $21.68$ \\
MS540379   & MS     & mct6M & 137 & $00:44:41.60$ & $+41:34:17.3$ & $-6.20$  & $\pm$ & $0.42$  & $93$  & $\pm$ & $7$  & $-211.6$ & $6.94$  & $20.24$ & $20.25$ & $21.47$ & $21.00$ & $20.92$ & $20.66$ \\
MS548831   & MS     & mct6M & 142 & $00:44:23.37$ & $+41:35:00.1$ & $-14.35$ & $\pm$ & $0.69$  & $99$  & $\pm$ & $7$  & $-189.7$ & $6.07$  & $20.37$ & $20.39$ & $21.45$ & $21.09$ & $21.47$ & $21.18$ \\
AGB571711  & T-MS   & mct6M & 156 & $00:44:30.54$ & $+41:36:18.4$ & $-6.05$  & $\pm$ & $0.88$  & $50$  & $\pm$ & $11$ & $-135.9$ & $6.50$  & -       & $23.55$ & $22.47$ & $21.07$ & $20.62$ & $20.15$ \\
MS579806   & MS     & mct6M & 160 & $00:44:36.52$ & $+41:36:39.0$ & $-8.74$  & $\pm$ & $0.38$  & $82$  & $\pm$ & $7$  & $-244.3$ & $6.78$  & $19.63$ & $19.92$ & $21.09$ & $21.04$ & $21.23$ & $21.18$ \\
MS582186   & MS     & mct6M & 161 & $00:44:19.15$ & $+41:36:44.7$ & $-48.07$ & $\pm$ & $11.23$ & $83$  & $\pm$ & $30$ & $-182.6$ & $6.25$  & $21.37$ & $21.45$ & $22.47$ & $22.35$ & $22.41$ & $22.23$ \\
RGB635965  & RHeB?  & mct6M & 185 & $00:44:20.76$ & $+41:38:42.4$ & $-3.90$  & $\pm$ & $0.36$  & $50$  & $\pm$ & $11$ & $-63.1$  & $6.73$  & -       & $24.78$ & $23.07$ & $20.84$ & $20.13$ & $19.54$ \\
MS673940   & MS     & mct6M & 208 & $00:44:23.36$ & $+41:40:24.9$ & $-10.28$ & $\pm$ & $0.26$  & $120$ & $\pm$ & $5$  & $-137.0$ & $7.21$  & $20.38$ & $20.40$ & $21.53$ & $21.32$ & $21.48$ & $21.33$ \\
SG770695   & T-MS   & mct6O & 13  & $00:45:36.17$ & $+41:48:49.2$ & $-2.77$  & $\pm$ & $0.03$  & $114$ & $\pm$ & $2$  & $-94.5$  & $10.48$ & $21.33$ & $19.48$ & $18.40$ & $16.71$ & $16.26$ & $15.88$ \\
MS790086   & MS     & mct6O & 40  & $00:45:38.25$ & $+41:50:15.7$ & $-23.21$ & $\pm$ & $5.12$  & $51$  & $\pm$ & $51$ & $-71.0$  & $10.75$ & $22.48$ & $22.08$ & $22.97$ & $22.22$ & $22.07$ & $21.89$ \\
serendip1  & RHeB   & mct6O & 76  & $00:45:25.91$ & $+41:51:16.1$ & $-27.81$ & $\pm$ & $1.91$  & $152$ & $\pm$ & $17$ & $-142.7$ & $10.62$ & -       & -       & $24.94$ & $22.73$ & $22.06$ & $21.27$ \\
MS829713   & MS     & mct6O & 123 & $00:44:52.83$ & $+41:52:17.3$ & $-0.55$  & $\pm$ & $0.01$  & $63$  & $\pm$ & $1$  & $-58.8$  & $10.82$ & $18.17$ & $18.19$ & $19.41$ & $18.90$ & $18.87$ & $18.77$ \\
MS829767   & MS     & mct6O & 124 & $00:44:32.23$ & $+41:52:17.5$ & $-7.63$  & $\pm$ & $0.86$  & $60$  & $\pm$ & $15$ & $-186.3$ & $11.26$ & $21.96$ & $21.57$ & $22.10$ & $21.31$ & $21.03$ & $20.81$ \\
MS833627   & MS     & mct6O & 138 & $00:45:38.26$ & $+41:52:28.5$ & $-23.1$  & $\pm$ & $1.35$  & $64$  & $\pm$ & $10$ & $-105.7$ & $11.13$ & $19.36$ & $19.60$ & $21.03$ & $20.94$ & $21.08$ & $21.09$ \\
MS838996   & MS     & mct6O & 154 & $00:44:52.25$ & $+41:52:43.4$ & $-21.19$ & $\pm$ & $0.98$  & $194$ & $\pm$ & $11$ & $-193.6$ & $10.98$ & $19.97$ & $20.06$ & $21.46$ & $20.97$ & $20.97$ & $20.70$ \\
MS843459   & MS     & mct6O & 162 & $00:45:13.19$ & $+41:52:56.6$ & $-11.51$ & $\pm$ & $0.92$  & $198$ & $\pm$ & $17$ & $-76.4$  & $10.92$ & $20.42$ & $20.53$ & $21.82$ & $21.67$ & $21.98$ & $21.89$ \\
MS854403   & MS     & mct6O & 186 & $00:45:19.69$ & $+41:53:28.7$ & $-19.66$ & $\pm$ & $1.31$  & $114$ & $\pm$ & $26$ & $-212.3$ & $11.09$ & $19.99$ & $20.22$ & $21.68$ & $21.51$ & $21.67$ & $21.46$ \\
MS864633   & MS     & mct6O & 207 & $00:45:04.81$ & $+41:53:59.3$ & $-43.46$ & $\pm$ & $1.67$  & $97$  & $\pm$ & $9$  & $-98.9$  & $11.27$ & $19.93$ & $19.88$ & $21.09$ & $20.49$ & $20.43$ & $20.14$ \\
serendip1  & AGB    & mct6O & 208 & $00:44:36.21$ & $+41:54:00.2$ & $-33.94$ & $\pm$ & $7.84$  & $77$  & $\pm$ & $79$ & $-173.9$ & $11.83$ & -       & -       & $26.40$ & $21.22$ & $19.28$ & $18.17$ \\
serendip1  & AGB    & mct6O & 236 & $00:44:49.90$ & $+41:52:42.0$ & $-14.99$ & $\pm$ & $0.68$  & $62$  & $\pm$ & $6$  & $-52.3$  & $11.01$ & -       & $27.27$ & $27.43$ & $22.51$ & $18.49$ & $18.03$ \\
AGB844835  & AGB    & mct6P & 32  & $00:46:10.16$ & $+41:53:00.7$ & $-9.84$  & $\pm$ & $2.29$  & $52$  & $\pm$ & $43$ & $-82.8$  & $12.29$ & -       & $28.99$ & $26.00$ & $20.58$ & $18.39$ & $17.35$ \\
MS854153   & MS     & mct6P & 46  & $00:45:35.18$ & $+41:53:27.9$ & $-6.04$  & $\pm$ & $0.52$  & $51$  & $\pm$ & $35$ & $-78.0$  & $11.27$ & $21.36$ & $21.42$ & $22.34$ & $22.21$ & $22.26$ & $22.20$ \\
MS862789   & T-MS   & mct6P & 66  & $00:45:41.73$ & $+41:53:53.7$ & $-12.75$ & $\pm$ & $0.22$  & $123$ & $\pm$ & $7$  & $-94.7$  & $11.49$ & $19.85$ & $19.22$ & $19.19$ & $18.84$ & $18.76$ & $18.48$ \\
MS883034   & MS     & mct6P & 115 & $00:45:31.87$ & $+41:55:02.4$ & $-11.11$ & $\pm$ & $0.48$  & $99$  & $\pm$ & $6$  & $-148.7$ & $11.60$ & $17.99$ & $18.50$ & $20.19$ & $20.45$ & $20.73$ & $20.81$ \\
MS890980   & MS     & mct6P & 137 & $00:44:53.13$ & $+41:55:32.4$ & $-0.71$  & $\pm$ & $0.17$  & $15$  & $\pm$ & $7$  & $-113.0$ & $12.00$ & $20.13$ & $20.03$ & $20.98$ & $20.40$ & $20.32$ & $20.23$ \\
MS892450   & MS     & mct6P & 142 & $00:46:05.18$ & $+41:55:37.7$ & $-15.50$ & $\pm$ & $1.44$  & $168$ & $\pm$ & $11$ & $-152.8$ & $12.41$ & $20.16$ & $20.25$ & $21.60$ & $21.46$ & $21.50$ & $21.39$ \\
MS897001   & MS     & mct6P & 155 & $00:44:56.71$ & $+41:55:54.8$ & $-17.38$ & $\pm$ & $2.34$  & $113$ & $\pm$ & $37$ & $-88.2$  & $12.06$ & $22.25$ & $22.17$ & $22.85$ & $22.70$ & $22.70$ & $22.64$ \\
MS898676   & MS     & mct6P & 159 & $00:45:42.08$ & $+41:56:01.1$ & $-2.14$  & $\pm$ & $0.01$  & $297$ & $\pm$ & $2$  & $-79.3$  & $11.96$ & $16.11$ & -       & $17.26$ & $16.97$ & $17.03$ & $16.94$ \\
MS921236   & MS     & mct6P & 197 & $00:45:10.16$ & $+41:57:25.8$ & $-28.27$ & $\pm$ & $1.36$  & $171$ & $\pm$ & $8$  & $-156.8$ & $12.38$ & $19.98$ & $20.19$ & $21.75$ & $21.46$ & $21.44$ & $21.21$ \\
MS925371   & MS     & mct6P & 199 & $00:45:02.67$ & $+41:57:41.9$ & $-29.68$ & $\pm$ & $1.78$  & $100$ & $\pm$ & $6$  & $-111.0$ & $12.61$ & $19.62$ & $19.84$ & $21.11$ & $20.91$ & $21.20$ & $21.01$ \\
MS924898   & MS     & mct6Q & 31  & $00:45:11.82$ & $+41:57:40.0$ & $-26.03$ & $\pm$ & $2.36$  & $147$ & $\pm$ & $20$ & $-110.7$ & $12.44$ & $20.66$ & $20.88$ & $22.26$ & $22.17$ & $22.31$ & $22.17$ \\
AGB928293  & AGB C* & mct6Q & 36  & $00:46:16.55$ & $+41:57:54.0$ & $-6.48$  & $\pm$ & $0.47$  & $62$  & $\pm$ & $6$  & $-132.3$ & $13.12$ & -       & -       & $25.30$ & $19.80$ & $17.76$ & $16.65$ \\
MS933068   & MS     & mct6Q & 48  & $00:45:54.98$ & $+41:58:14.4$ & $-5.18$  & $\pm$ & $0.25$  & $175$ & $\pm$ & $9$  & $-190.7$ & $12.67$ & $21.24$ & $21.15$ & $21.88$ & $21.43$ & $21.48$ & $21.33$ \\
MS936504   & MS     & mct6Q & 62  & $00:45:28.26$ & $+41:58:29.2$ & $-12.55$ & $\pm$ & $1.25$  & $144$ & $\pm$ & $17$ & $-167.0$ & $12.57$ & $22.14$ & $21.94$ & $22.86$ & $22.20$ & $21.75$ & $21.68$ \\
MS938750   & MS     & mct6Q & 72  & $00:45:36.02$ & $+41:58:39.2$ & $-12.12$ & $\pm$ & $0.31$  & $174$ & $\pm$ & $7$  & $-185.8$ & $12.61$ & $19.95$ & $20.13$ & $21.48$ & $21.28$ & $21.37$ & $21.20$ \\
MS942221   & MS     & mct6Q & 83  & $00:45:43.63$ & $+41:58:54.7$ & $-14.37$ & $\pm$ & $0.32$  & $220$ & $\pm$ & $4$  & $-141.6$ & $12.71$ & $20.17$ & $20.08$ & $21.28$ & $20.60$ & $20.54$ & $20.29$ \\
MS942322   & MS     & mct6Q & 84  & $00:45:28.28$ & $+41:58:55.2$ & $-25.95$ & $\pm$ & $0.78$  & $158$ & $\pm$ & $6$  & $-100.3$ & $12.71$ & $20.69$ & $20.51$ & $21.59$ & $20.93$ & $20.88$ & $20.52$ \\
MS944360   & MS     & mct6Q & 95  & $00:45:44.20$ & $+41:59:04.8$ & $-17.33$ & $\pm$ & $0.62$  & $141$ & $\pm$ & $8$  & $-106.8$ & $12.76$ & $20.17$ & $20.19$ & $21.39$ & $21.02$ & $21.07$ & $20.88$ \\
MS944642   & MS     & mct6Q & 96  & $00:45:03.54$ & $+41:59:06.1$ & $-7.24$  & $\pm$ & $0.24$  & $111$ & $\pm$ & $4$  & $-70.2$  & $13.11$ & $18.84$ & $19.09$ & $20.55$ & $20.48$ & $20.56$ & $20.52$ \\
serendip1  & AGB    & mct6Q & 96  & $00:45:03.44$ & $+41:59:06.3$ & $-13.05$ & $\pm$ & $2.81$  & $68$  & $\pm$ & $37$ & $-95.4$  & $13.12$ & -       & -       & $26.24$ & $20.82$ & $18.53$ & $17.36$ \\
MS950027   & MS     & mct6Q & 116 & $00:45:01.60$ & $+41:59:30.4$ & $-3.70$  & $\pm$ & $0.27$  & $47$  & $\pm$ & $7$  & $-118.7$ & $13.31$ & $21.08$ & $20.76$ & $21.74$ & $20.77$ & $20.51$ & $20.32$ \\
MS960668   & MS     & mct6Q & 148 & $00:45:18.17$ & $+42:00:26.5$ & $-25.09$ & $\pm$ & $1.32$  & $100$ & $\pm$ & $7$  & $-153.8$ & $13.33$ & $20.88$ & $20.98$ & $21.81$ & $21.73$ & $21.93$ & $21.81$ \\
MS963896   & MS     & mct6Q & 155 & $00:44:56.74$ & $+42:00:47.7$ & $-21.35$ & $\pm$ & $5.58$  & $102$ & $\pm$ & $43$ & $-129.6$ & $13.95$ & $21.76$ & $21.71$ & $22.48$ & $22.30$ & $22.36$ & $22.24$ \\
MS967879   & MS     & mct6Q & 172 & $00:44:59.89$ & $+42:01:16.1$ & $-21.77$ & $\pm$ & $1.98$  & $123$ & $\pm$ & $14$ & $-181.8$ & $14.05$ & $21.13$ & $21.15$ & $22.59$ & $22.53$ & $22.30$ & $22.14$ \\
MS1023626  & MS     & mct6R & 21  & $00:47:23.76$ & $+42:09:28.4$ & $-13.82$ & $\pm$ & $0.80$  & $91$  & $\pm$ & $10$ & $-198.1$ & $16.99$ & $21.75$ & $21.68$ & $22.47$ & $22.33$ & $22.42$ & $22.37$ \\
AGB1025281 & AGB    & mct6R & 27  & $00:47:13.76$ & $+42:09:42.4$ & $-1.47$  & $\pm$ & $0.26$  & $51$  & $\pm$ & $9$  & $-340.2$ & $16.71$ & -       & $26.10$ & $27.53$ & $21.72$ & $18.50$ & $17.57$ \\
MS1026522  & MS     & mct6R & 32  & $00:47:17.84$ & $+42:09:53.1$ & $-15.45$ & $\pm$ & $1.72$  & $82$  & $\pm$ & $19$ & $-95.2$  & $16.86$ & $22.41$ & $22.06$ & $22.64$ & $22.32$ & $22.25$ & $22.22$ \\
MS1027514  & MS     & mct6R & 41  & $00:47:34.26$ & $+42:10:01.5$ & $-38.32$ & $\pm$ & $0.75$  & $101$ & $\pm$ & $5$  & $-141.6$ & $17.44$ & $19.58$ & $19.80$ & $21.22$ & $21.05$ & $21.35$ & $21.06$ \\
MS1032330  & MS     & mct6R & 70  & $00:46:39.30$ & $+42:10:43.4$ & $-6.83$  & $\pm$ & $0.58$  & $63$  & $\pm$ & $13$ & $-94.0$  & $16.26$ & $20.44$ & $20.50$ & $22.14$ & $21.67$ & $21.19$ & $21.00$ \\
MS1035086  & MS     & mct6R & 91  & $00:46:47.68$ & $+42:11:06.4$ & $-3.84$  & $\pm$ & $0.2$   & $63$  & $\pm$ & $9$  & $-201.8$ & $16.45$ & $21.36$ & $21.35$ & $21.92$ & $21.73$ & $21.79$ & $21.71$ \\
MS1036341  & MS     & mct6R & 100 & $00:46:49.93$ & $+42:11:16.4$ & $-30.42$ & $\pm$ & $0.35$  & $128$ & $\pm$ & $3$  & $-134.0$ & $16.52$ & $18.97$ & $19.22$ & $20.82$ & $20.56$ & $20.62$ & $20.40$ \\
MS1037476  & MS     & mct6R & 109 & $00:46:59.50$ & $+42:11:25.2$ & $-18.66$ & $\pm$ & $0.22$  & $165$ & $\pm$ & $2$  & $-161.3$ & $16.70$ & $19.39$ & $19.51$ & $20.81$ & $20.65$ & $20.69$ & $20.51$ \\
MS1037726  & MS     & mct6R & 111 & $00:47:29.15$ & $+42:11:27.4$ & $-27.72$ & $\pm$ & $0.54$  & $152$ & $\pm$ & $3$  & $-111.8$ & $17.45$ & $19.72$ & $19.76$ & $21.09$ & $20.71$ & $20.79$ & $20.50$ \\
MS1037909  & T-MS   & mct6R & 115 & $00:47:10.11$ & $+42:11:28.8$ & $-3.81$  & $\pm$ & $0.11$  & $105$ & $\pm$ & $5$  & $-146.2$ & $16.93$ & $20.97$ & $20.77$ & $21.01$ & $20.81$ & $20.79$ & $20.71$ \\
MS1039213  & MS     & mct6R & 128 & $00:46:20.78$ & $+42:11:39.4$ & $-29.94$ & $\pm$ & $1.92$  & $115$ & $\pm$ & $14$ & $-121.3$ & $16.48$ & $22.08$ & $21.81$ & $22.81$ & $22.03$ & $21.89$ & $21.60$ \\
MS1040439  & MS     & mct6R & 139 & $00:46:44.66$ & $+42:11:49.2$ & $-29.16$ & $\pm$ & $1.86$  & $135$ & $\pm$ & $14$ & $-77.6$  & $16.59$ & $21.32$ & $21.38$ & $22.69$ & $22.24$ & $22.16$ & $21.89$ \\
MS1042778  & MS     & mct6R & 157 & $00:47:13.78$ & $+42:12:10.4$ & $-10.06$ & $\pm$ & $0.41$  & $120$ & $\pm$ & $12$ & $-82.0$  & $17.15$ & $20.32$ & $20.44$ & $21.81$ & $21.57$ & $21.69$ & $21.43$ \\
MS1043257  & MS     & mct6R & 162 & $00:46:20.07$ & $+42:12:15.4$ & $-5.40$  & $\pm$ & $0.20$  & $273$ & $\pm$ & $7$  & $-188.6$ & $16.67$ & $19.81$ & $20.07$ & $21.20$ & $21.27$ & $21.37$ & $21.39$ \\
MS1045217  & MS     & mct6R & 177 & $00:46:36.83$ & $+42:12:34.9$ & $-23.91$ & $\pm$ & $0.74$  & $112$ & $\pm$ & $8$  & $-87.2$  & $16.75$ & $21.27$ & $21.11$ & $22.02$ & $21.32$ & $21.21$ & $20.94$ \\
MS1045832  & MS     & mct6R & 184 & $00:46:58.01$ & $+42:12:40.9$ & $-2.25$  & $\pm$ & $0.58$  & $67$  & $\pm$ & $16$ & $-37.4$  & $16.96$ & $21.08$ & $21.35$ & $22.63$ & $22.60$ & $22.64$ & $22.61$ \\
MS1048995  & MS     & mct6R & 197 & $00:46:36.74$ & $+42:13:16.4$ & $-16.60$ & $\pm$ & $1.54$  & $104$ & $\pm$ & $21$ & $-62.4$  & $16.95$ & $22.64$ & $22.49$ & $22.79$ & $22.39$ & $22.23$ & $22.14$ \\
MS1008068  & MS     & mct6S & 43  & $00:47:05.00$ & $+42:07:29.7$ & $-60.13$ & $\pm$ & $15.08$ & $91$  & $\pm$ & $41$ & $-55.5$  & $16.09$ & $21.76$ & $21.64$ & $22.88$ & $22.57$ & $22.61$ & $22.42$\\\bottomrule
\end{tabular}}
\end{table*}

\newpage
\begin{table*}
\centering
\resizebox{1.35\textwidth}{!}{%
\begin{tabular}{lcccllrllrllcccccccc} 
\toprule
\multicolumn{1}{|c|}{Object} & \multicolumn{1}{c|}{Class.} & \multicolumn{1}{c|}{Mask} & \multicolumn{1}{c|}{Slit No.} & \multicolumn{1}{c|}{RA} & \multicolumn{1}{c|}{Dec} & \multicolumn{3}{c|}{H$\alpha$ EW ($\text{\AA}$)} & \multicolumn{3}{c|}{$\sigma_{v}$ (km s$^{-1}$)} & \multicolumn{1}{c|}{v$_{Hel}$ (km s$^{-1}$)} & \multicolumn{1}{c|}{R$_{deproj}$ (kpc)} & \multicolumn{1}{c|}{F275W} & \multicolumn{1}{c|}{F336W} & \multicolumn{1}{c|}{F475W} & \multicolumn{1}{c|}{F814W} & \multicolumn{1}{c|}{F110W} & \multicolumn{1}{c|}{F160W}\\ \midrule
MS1009654  & MS     & mct6S & 53  & $00:46:47.40$ & $+42:07:41.4$ & $-35.02$ & $\pm$ & $1.78$ & $110$ & $\pm$ & $27$ & $-76.6$  & $15.68$ & $20.38$ & $20.33$ & $21.74$ & $21.13$ & $20.65$ & $20.23$ \\
AGB1011804 & AGB C* & mct6S & 62  & $00:47:36.12$ & $+42:07:57.5$ & $-7.76$  & $\pm$ & $0.44$ & $63$  & $\pm$ & $5$  & $-189.8$ & $17.31$ & -       & -       & $23.53$ & $19.39$ & $17.92$ & $16.88$ \\
MS1016968  & MS     & mct6S & 90  & $00:46:26.88$ & $+42:08:36.0$ & $-21.65$ & $\pm$ & $0.92$ & $251$ & $\pm$ & $15$ & $-59.9$  & $15.60$ & $19.16$ & $19.37$ & $20.88$ & $20.53$ & $20.55$ & $20.31$ \\
MS1024667  & MS     & mct6S & 145 & $00:46:29.69$ & $+42:09:37.2$ & $-17.66$ & $\pm$ & $0.63$ & $110$ & $\pm$ & $7$  & $-67.3$  & $15.90$ & $19.80$ & $19.88$ & $20.98$ & $20.72$ & $20.79$ & $20.60$ \\
MS1029029  & MS     & mct6S & 174 & $00:46:36.91$ & $+42:10:14.6$ & $-54.09$ & $\pm$ & $2.71$ & $94$  & $\pm$ & $11$ & $-110.4$ & $16.12$ & $20.30$ & $20.32$ & $21.54$ & $21.07$ & $21.23$ & $20.95$ \\
AGB925447  & RHeB   & mct6T & 71  & $00:45:59.19$ & $+41:57:42.2$ & $-3.67$  & $\pm$ & $0.17$ & $49$  & $\pm$ & $5$  & $-73.7$  & $12.63$ & -       & $27.24$ & $24.28$ & $20.13$ & $18.99$ & $18.18$ \\
MS934267   & MS     & mct6T & 83  & $00:45:54.98$ & $+41:58:19.4$ & $-14.80$ & $\pm$ & $1.01$ & $65$  & $\pm$ & $7$  & $-116.0$ & $12.69$ & $21.29$ & $21.34$ & $22.09$ & $21.94$ & $21.96$ & $21.92$ \\
MS947307   & MS     & mct6T & 97  & $00:46:05.58$ & $+41:59:18.1$ & $-22.08$ & $\pm$ & $0.52$ & $91$  & $\pm$ & $7$  & $-118.0$ & $13.09$ & $19.22$ & $19.53$ & $21.06$ & $20.93$ & $21.10$ & $20.90$ \\
AGB952727  & AGB    & mct6T & 105 & $00:45:49.67$ & $+41:59:43.6$ & $-2.01$  & $\pm$ & $0.10$ & $77$  & $\pm$ & $5$  & $-134.6$ & $12.97$ & -       & $28.84$ & $24.42$ & $20.25$ & $18.93$ & $17.88$ \\
MS959219   & MS     & mct6T & 113 & $00:45:48.10$ & $+42:00:17.7$ & $-11.76$ & $\pm$ & $0.29$ & $226$ & $\pm$ & $5$  & $-66.3$  & $13.12$ & $19.91$ & $20.16$ & $21.61$ & $21.45$ & $21.65$ & $21.54$ \\
AGB989533  & AGB C* & mct6T & 202 & $00:45:24.03$ & $+42:04:31.2$ & $-0.35$  & $\pm$ & $0.09$ & $57$  & $\pm$ & $5$  & $-126.3$ & $14.71$ & -       & $27.48$ & $24.31$ & $19.94$ & $18.47$ & $17.28$ \\
AGB1001394 & RHeB   & mct6T & 230 & $00:45:34.31$ & $+42:06:37.9$ & $-6.92$  & $\pm$ & $0.13$ & $77$  & $\pm$ & $3$  & $-633.6$ & $15.29$ & -       & $29.71$ & $22.19$ & $19.40$ & $19.16$ & $18.52$ \\
MS1002698  & MS     & mct6T & 234 & $00:45:25.11$ & $+42:06:48.6$ & $-52.43$ & $\pm$ & $6.04$ & $124$ & $\pm$ & $29$ & $-194.6$ & $15.56$ & $20.93$ & $21.10$ & $22.33$ & $22.17$ & $22.28$ & $22.10$ \\
AGB1005555 & AGB    & mct6T & 240 & $00:45:34.70$ & $+42:07:11.2$ & $-6.17$  & $\pm$ & $0.49$ & $68$  & $\pm$ & $8$  & $-311.4$ & $15.49$ & -       & $26.36$ & $24.65$ & $20.66$ & $19.25$ & $18.20$ \\
MS1031822  & MS     & mct6U & 41  & $00:45:48.03$ & $+42:10:38.5$ & $-5.62$  & $\pm$ & $0.73$ & $57$  & $\pm$ & $13$ & $-103.7$ & $16.51$ & $20.02$ & $20.21$ & $21.61$ & $21.60$ & $21.66$ & -       \\
MS1041161  & MS     & mct6U & 60  & $00:45:52.07$ & $+42:11:55.4$ & $-39.32$ & $\pm$ & $1.19$ & $130$ & $\pm$ & $9$  & $-134.2$ & $16.90$ & $21.23$ & $20.88$ & $21.84$ & $20.81$ & $20.55$ & $20.19$ \\
MS1042767  & MS     & mct6U & 64  & $00:46:06.20$ & $+42:12:10.3$ & $-23.87$ & $\pm$ & $6.48$ & $51$  & $\pm$ & $28$ & $-82.2$  & $16.77$ & $21.34$ & $21.43$ & $22.60$ & $22.45$ & $22.51$ & $22.56$ \\
MS1050499  & MS     & mct6U & 98  & $00:45:45.04$ & $+42:13:33.8$ & $-21.17$ & $\pm$ & $1.91$ & $106$ & $\pm$ & $12$ & $-142.8$ & $17.68$ & $20.87$ & $21.04$ & $22.11$ & $21.99$ & $22.06$ & $21.99$ \\
AGB1052705 & AGB C* & mct6U & 113 & $00:45:55.27$ & $+42:13:59.7$ & $-3.58$  & $\pm$ & $0.33$ & $53$  & $\pm$ & $9$  & $-228.5$ & $17.61$ & -       & -       & $24.55$ & $20.07$ & $18.73$ & $17.67$ \\
MS1059485  & MS     & mct6U & 153 & $00:46:19.31$ & $+42:15:36.9$ & $-3.34$  & $\pm$ & $0.21$ & $88$  & $\pm$ & $6$  & $-123.0$ & $17.79$ & $18.89$ & $19.21$ & $20.60$ & $20.71$ & $20.87$ & $20.91$ \\
AGB1060489 & AGB C* & mct6U & 160 & $00:46:32.55$ & $+42:15:55.5$ & $-1.42$  & $\pm$ & $0.13$ & $54$  & $\pm$ & $7$  & $-145.2$ & $17.77$ & -       & $28.06$ & $23.80$ & $19.71$ & $18.57$ & $17.50$ \\
MS1012419  & MS     & mct6V & 14  & $00:46:32.68$ & $+42:08:02.4$ & $-11.02$ & $\pm$ & $0.64$ & $87$  & $\pm$ & $9$  & $-84.9$  & $15.51$ & $21.44$ & $21.24$ & $22.19$ & $21.57$ & $21.58$ & $21.32$ \\
MS1036579  & MS     & mct6V & 66  & $00:46:40.08$ & $+42:11:18.2$ & $-25.44$ & $\pm$ & $0.41$ & $85$  & $\pm$ & $5$  & $-93.4$  & $16.42$ & $20.00$ & $19.97$ & $21.08$ & $20.49$ & $20.46$ & $20.26$ \\
MS1055498  & MS     & mct6V & 130 & $00:46:35.27$ & $+42:14:32.2$ & $-6.89$  & $\pm$ & $0.12$ & $173$ & $\pm$ & $4$  & $-85.9$  & $17.33$ & $19.44$ & $19.70$ & $21.16$ & $20.82$ & $20.92$ & $20.72$ \\
serendip1  & MS     & mct6V & 138 & $00:46:37.09$ & $+42:14:43.8$ & $-7.32$  & $\pm$ & $0.25$ & $159$ & $\pm$ & $7$  & $-170.5$ & $17.38$ & $20.99$ & $20.82$ & $21.99$ & $21.41$ & $21.15$ & $20.89$ \\
SG1013574  & T-MS   & mct6W & 33  & $00:46:47.93$ & $+42:08:10.4$ & $-5.18$  & $\pm$ & $0.59$ & $61$  & $\pm$ & $14$ & $-138.0$ & $15.79$ & $24.09$ & $23.17$ & $22.68$ & $21.87$ & $21.61$ & $21.40$ \\
AGB1016791 & AGB C* & mct6W & 38  & $00:46:19.04$ & $+42:08:34.8$ & $-1.15$  & $\pm$ & $0.22$ & $52$  & $\pm$ & $9$  & $-128.6$ & $15.56$ & -       & $28.77$ & $24.44$ & $20.13$ & $18.62$ & $17.50$ \\
MS1021720  & MS     & mct6W & 50  & $00:46:24.93$ & $+42:09:12.8$ & $-15.45$ & $\pm$ & $0.89$ & $150$ & $\pm$ & $11$ & $-62.1$  & $15.76$ & $20.17$ & $20.30$ & $21.84$ & $21.53$ & $21.63$ & $21.36$ \\
MS1023445  & MS     & mct6W & 52  & $00:46:35.34$ & $+42:09:27.0$ & $-21.48$ & $\pm$ & $0.19$ & $121$ & $\pm$ & $3$  & $-92.0$  & $15.90$ & $18.40$ & $18.53$ & $19.88$ & $19.66$ & $19.71$ & $19.52$ \\
MS1028129  & MS     & mct6W & 58  & $00:46:27.51$ & $+42:10:06.6$ & $-14.21$ & $\pm$ & $0.49$ & $98$  & $\pm$ & $6$  & $-71.4$  & $16.02$ & $20.12$ & $20.19$ & $21.17$ & $20.93$ & $21.00$ & $20.85$ \\
MS1037135  & MS     & mct6W & 82  & $00:46:42.87$ & $+42:11:22.7$ & $-14.41$ & $\pm$ & $0.38$ & $107$ & $\pm$ & $5$  & $-97.5$  & $16.46$ & $20.65$ & $20.55$ & $21.59$ & $21.12$ & $21.16$ & $20.90$ \\
MS1041718  & MS     & mct6W & 99  & $00:46:43.96$ & $+42:12:00.4$ & $-8.01$  & $\pm$ & $1.09$ & $58$  & $\pm$ & $15$ & $-106.6$ & $16.64$ & $20.75$ & $20.99$ & $22.27$ & $22.24$ & $22.36$ & $22.33$ \\
MS1047936  & MS     & mct6W & 130 & $00:46:45.26$ & $+42:13:04.3$ & $-9.6$   & $\pm$ & $3.33$ & $74$  & $\pm$ & $59$ & $-230.7$ & $16.93$ & $22.40$ & $22.21$ & $22.98$ & $22.58$ & $22.32$ & $22.11$ \\
MS1053266  & MS     & mct6W & 154 & $00:47:20.23$ & $+42:14:05.4$ & $-40.51$ & $\pm$ & $1.09$ & $123$ & $\pm$ & $7$  & $-144.5$ & $17.67$ & $19.82$ & $19.95$ & $21.18$ & $20.99$ & $21.07$ & $20.93$ \\
AGB1065524 & RHeB   & mct6W & 200 & $00:47:23.69$ & $+42:17:40.7$ & $-3.40$  & $\pm$ & $0.19$ & $61$  & $\pm$ & $6$  & $-326.9$ & $18.50$ & -       & $26.80$ & $22.78$ & $20.52$ & $19.76$ & $19.02$ \\
RGB1068713 & AGB C* & mct6W & 216 & $00:47:21.37$ & $+42:18:44.2$ & $-10.13$ & $\pm$ & $1.88$ & $46$  & $\pm$ & $21$ & $-83.9$  & $18.72$ & -       & -       & $25.96$ & $21.43$ & $19.57$ & $18.62$ \\
MS922906   & MS     & mct6X & 25  & $00:46:52.42$ & $+41:57:32.3$ & $-5.24$  & $\pm$ & $0.51$ & $62$  & $\pm$ & $10$ & $-166.7$ & $14.58$ & $21.84$ & $21.87$ & $22.75$ & $22.57$ & $22.45$ & $22.38$ \\
MS931088   & MS     & mct6X & 37  & $00:46:54.63$ & $+41:58:06.0$ & $-32.62$ & $\pm$ & $0.82$ & $131$ & $\pm$ & $5$  & $-193.1$ & $14.72$ & $19.28$ & $19.56$ & $21.10$ & $20.81$ & $20.91$ & $20.70$ \\
MS935185   & MS     & mct6X & 44  & $00:46:39.24$ & $+41:58:23.4$ & $-30.85$ & $\pm$ & $0.92$ & $156$ & $\pm$ & $7$  & $-136.0$ & $14.02$ & $19.64$ & $19.85$ & $21.48$ & $21.24$ & $21.26$ & $21.03$ \\
MS935435   & MS     & mct6X & 46  & $00:46:40.07$ & $+41:58:24.4$ & $-1.63$  & $\pm$ & $0.08$ & $48$  & $\pm$ & $4$  & $-213.5$ & $14.05$ & $19.95$ & $19.99$ & $21.24$ & $20.92$ & $20.95$ & $20.91$ \\
MS941562   & MS     & mct6X & 59  & $00:46:28.54$ & $+41:58:51.6$ & $-30.36$ & $\pm$ & $1.02$ & $111$ & $\pm$ & $8$  & $-159.0$ & $13.65$ & $20.23$ & $20.36$ & $21.66$ & $21.52$ & $21.61$ & $21.53$ \\
MS945202   & MS     & mct6X & 68  & $00:46:34.22$ & $+41:59:08.9$ & $-11.54$ & $\pm$ & $0.29$ & $97$  & $\pm$ & $7$  & $-117.7$ & $13.90$ & $20.03$ & $20.19$ & $21.22$ & $21.17$ & $21.27$ & $21.24$ \\
MS950839   & MS     & mct6X & 79  & $00:46:29.69$ & $+41:59:34.3$ & $-8.43$  & $\pm$ & $1.23$ & $58$  & $\pm$ & $17$ & $-92.3$  & $13.79$ & $20.81$ & $21.04$ & $22.34$ & $22.30$ & $22.47$ & $22.48$ \\
serendip1  & RHeB   & mct6X & 87  & $00:46:27.68$ & $+41:59:57.9$ & $-11.23$ & $\pm$ & $3.34$ & $36$  & $\pm$ & $34$ & $-426.1$ & $13.78$ & -       & $26.73$ & $25.06$ & $23.30$ & $22.77$ & $22.20$ \\
MS981758   & MS     & mct6X & 129 & $00:46:14.43$ & $+42:03:03.6$ & $-19.76$ & $\pm$ & $2.20$ & $105$ & $\pm$ & $25$ & $-118.5$ & $14.08$ & $21.37$ & $21.54$ & $22.66$ & $22.73$ & $22.94$ & $22.98$ \\
AGB989244  & RHeB   & mct6X & 147 & $00:46:21.80$ & $+42:04:28.0$ & $-5.32$  & $\pm$ & $0.11$ & $48$  & $\pm$ & $3$  & $-24.7$  & $14.51$ & -       & $24.90$ & $21.94$ & $17.63$ & $16.23$ & $15.50$ \\
MS1004075  & MS     & mct6X & 193 & $00:46:16.42$ & $+42:06:59.8$ & $-49.04$ & $\pm$ & $3.36$ & $157$ & $\pm$ & $13$ & $-80.1$  & $15.11$ & $21.48$ & $21.58$ & $22.67$ & $22.37$ & $22.38$ & $22.21$\\\bottomrule
\end{tabular}}
\end{table*}
\end{landscape}

\clearpage

\section{2D SPLASH Spectra}
\label{sec:appB}

Although the 2D SPLASH spectra were not used for analysis in this paper, we did use them for verifying the efficiency of our selection algorithm. Here we present a sample of four 2D SPLASH spectra, two of which are classified as H$\alpha$ stars. The spectra and a description of what can be seen are given in Fig.~\ref{fig:2D}1, the details of each of the four objects are given in Table \ref{table:2Dspec}1.
\vspace{5mm}
\begin{center}
(a)\includegraphics[width=0.42\textwidth , trim=217 206 53 141,clip]{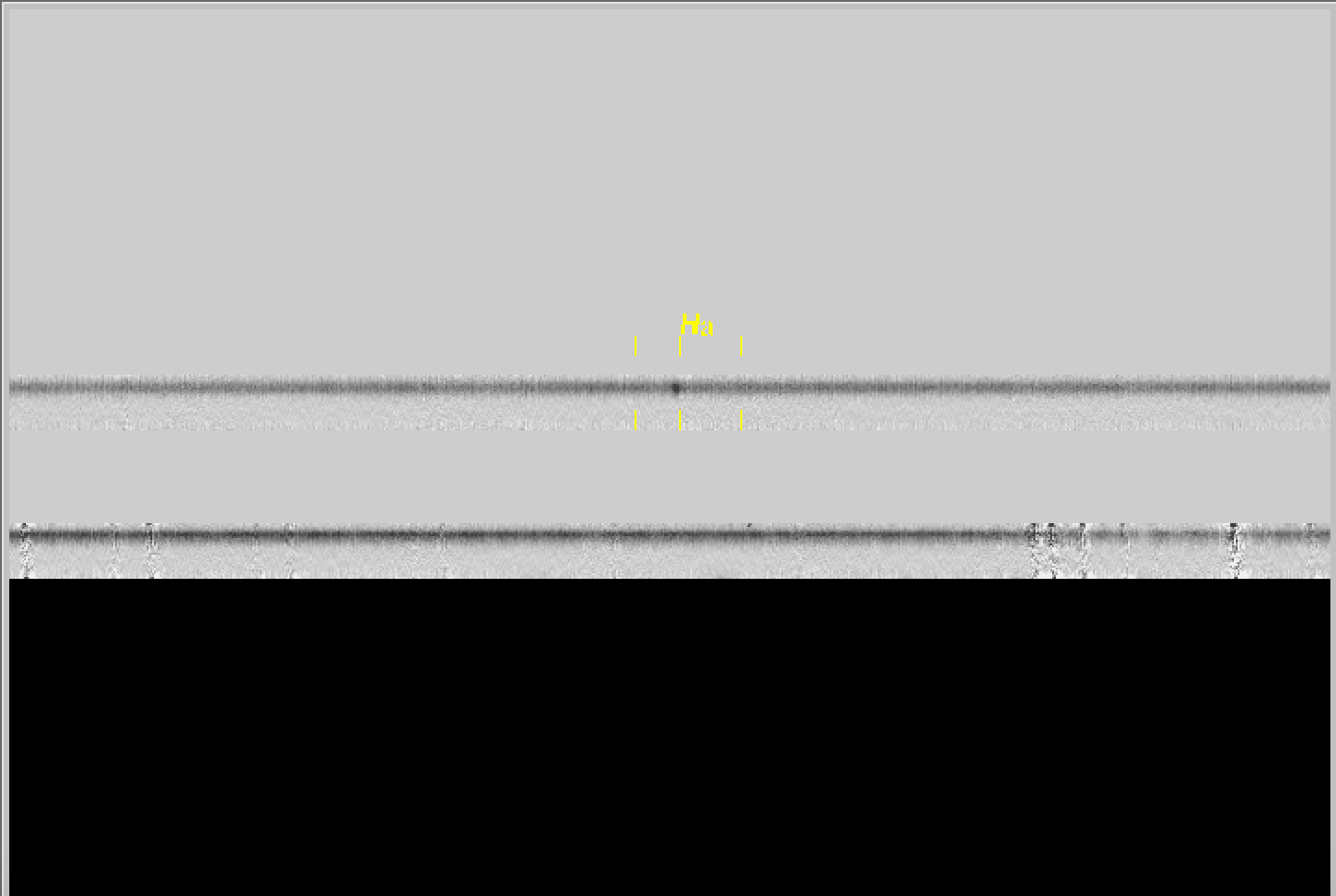}
\hspace{-0.7mm}
(b)\includegraphics[width=0.42\textwidth , trim=167 220 123 84,clip]{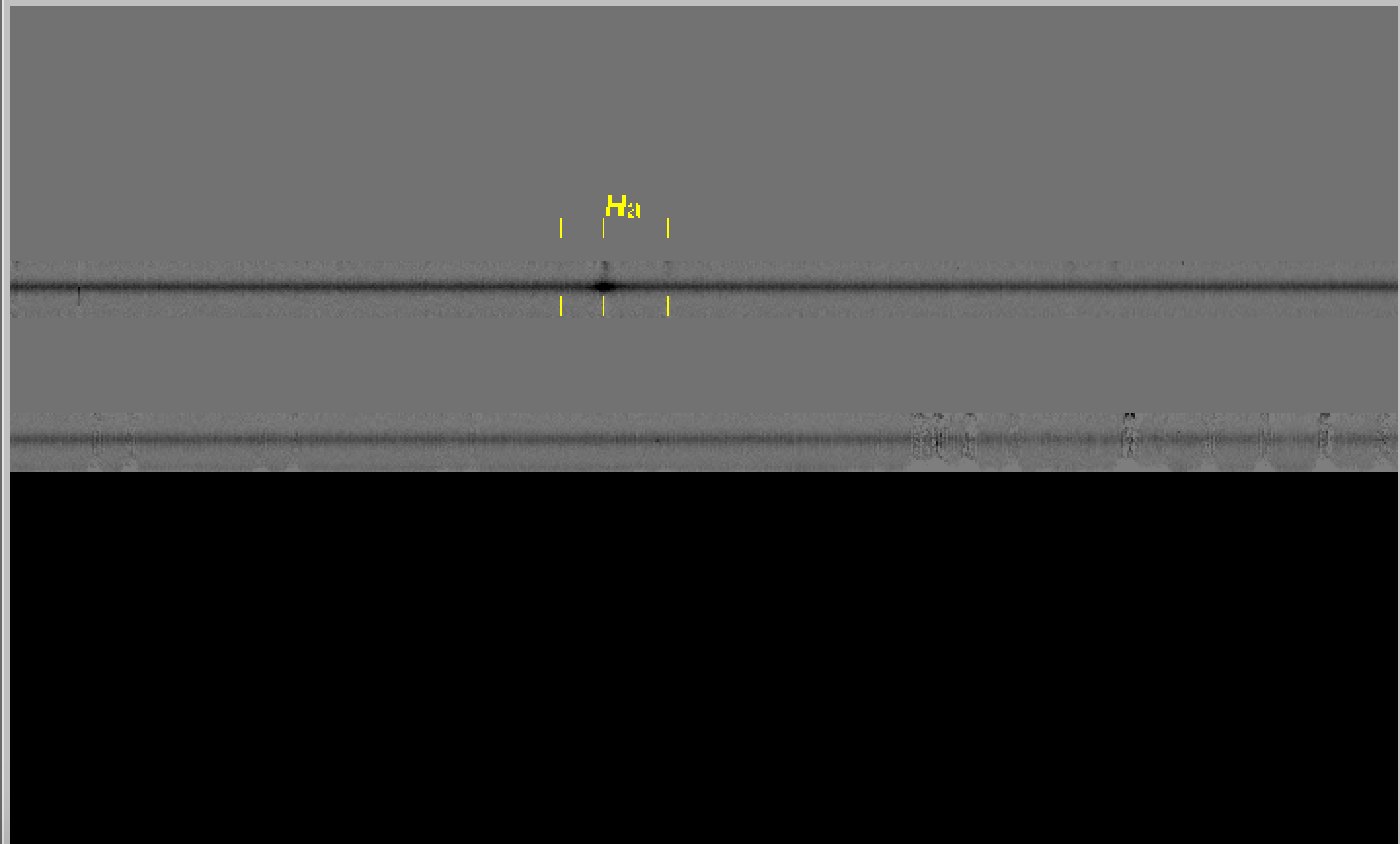}\\
\vspace{1mm}
(c)\includegraphics[width=0.42\textwidth , trim=187 213 93 143,clip]{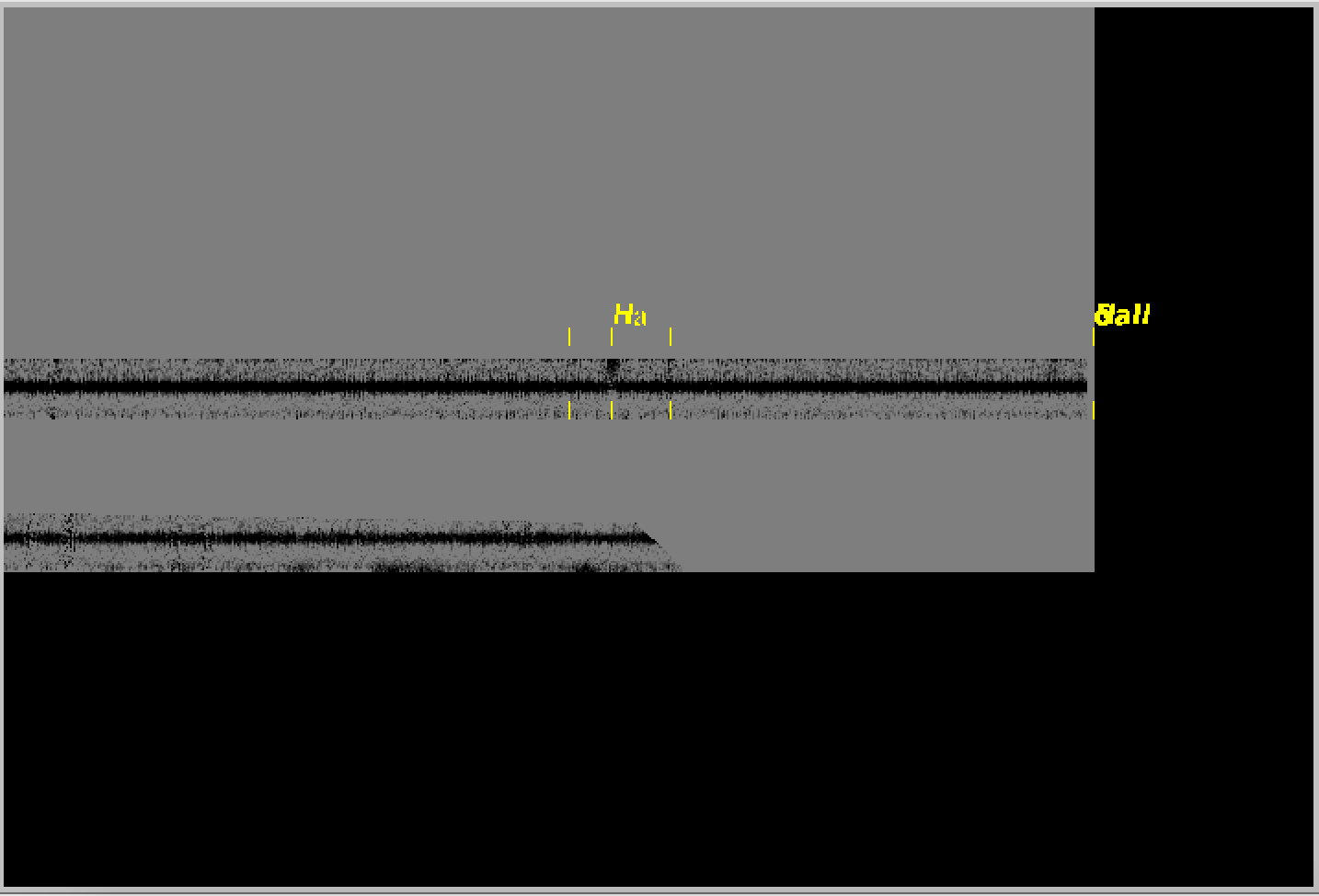}
(d)\includegraphics[width=0.42\textwidth , trim=200 275 80 80,clip]{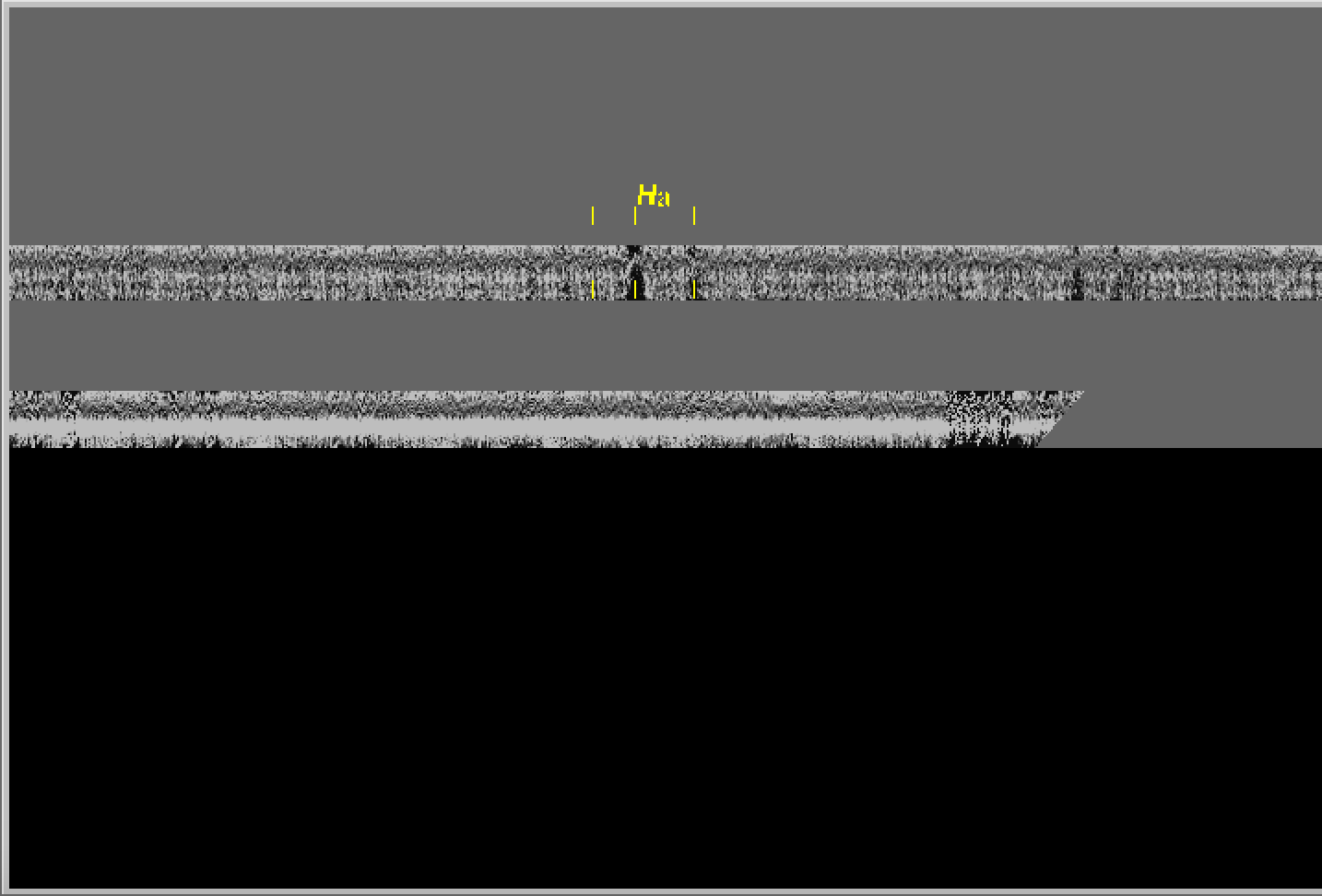}\\
\label{fig:2D}
\end{center}
\vspace{1mm}
{\textbf{Figure B1:} Four examples of 2D spectra from SPLASH to demonstrate the difference between emission-line stars (top panels) and stars surrounded by diffuse nebulous emission (bottom panels). The three yellow lines in every panel mark the wavelength of the [NII]a, H$\alpha$, and [NII]b lines respectively. The spatial extent of emission in each slit is vertical, while wavelength runs horizontally. Information on these four stars is shown in Table \ref{table:2Dspec}1.\\ 
(a) An emission-line star within the H$\alpha$ star catalogue. The stellar continuum is the thin dark horizontal line. H$\alpha$ emission can be seen as a dark spot at the central yellow mark. The indication that this H$\alpha$ emission is from the star and not the surrounding material is that it is the width of the stellar light and does not extend vertically from it.\\ 
(b) Another emission-line star that is also in the H$\alpha$ star catalogue. This star shows the dark spot of H$\alpha$ emission that is the width of the stellar light. However, this star also shows some faint spatially extended H$\alpha$ emission, brightest above the horizontal band of stellar light in the image. This star has nebulous H$\alpha$ emission but does not have detectable [NII] or [SII] emission and therefore has been selected as an H$\alpha$ star. However, the information on the origin of H$\alpha$ emission, either from the star or surrounding material, is lost in the 1D spectra. This is one of the reasons we use the visually selected list of emission-line stars, identified from the 2D spectra, in order to verify the efficiency of our selection algorithm.\\ 
(c) An example of a star that does not emit H$\alpha$ but that is surrounded by nebulous H$\alpha$ emission. This is shown by the spatially extended emission above the thin horizontal band of stellar light. There are also some fainter [NII] emission lines and fainter still [SII] emission lines, the second of which has been cut off due to reaching the end of the blue half of the 2D spectrum. This nebulous emission has resulted in this star being removed from the H$\alpha$ star catalogue.\\
(d) A slightly noisier spectrum of a non-emission-line star. Here, spatially extended [NII] and [SII] emission is visible along with H$\alpha$, causing this star to be removed from our H$\alpha$ star catalogue.\small}

\vspace{10mm}
\noindent {\textbf{Table B1:} Properties for the four objects whose 2D SPLASH spectra are shown in Fig.~\ref{fig:2D}1. The columns show the relevant panel in Fig.~\ref{fig:2D}1, object name, DEIMOS mask name and slit number, RA, and Dec of each object. The last column indicates whether the displayed star is in our H$\alpha$ catalogue. For those that are in the catalogue, we give additional properties on the first page of Appendix \ref{sec:appA}. These objects' names are enclosed in boxes in Table \ref{sec:appA}1.\small}\\

\begin{center}
\resizebox{0.8\textwidth}{!}{
\label{table:2Dspec} 
\begin{tabular}{clccllc} 
\hline
\multicolumn{1}{|c|}{Panel} & \multicolumn{1}{|c|}{Object} & \multicolumn{1}{c|}{Mask} & \multicolumn{1}{c|}{Slit No.} &\multicolumn{1}{c|}{RA} & \multicolumn{1}{c|}{Dec} & \multicolumn{1}{c|}{H$\alpha$ Star}\\ \hline
(a) & AGB154104 & mct6B & 193& $00:44:48.54$ & $+41:17:55.9$ & \cmark \\
(b) & MS179275  & mct6C & 64 & $00:44:31.66$ & $+41:19:19.9$ & \cmark \\
(c) & AGB178257 & mct6B & 208& $00:44:32.95$ & $+41:19:16.8$ & \xmark	\\
(d) & SG151962  & mct6C & 40 & $00:44:23.25$ & $+41:17:48.8$ & \xmark \\\hline
\end{tabular}}
\end{center}

\end{document}